\documentclass{aa}
\usepackage{txfonts}
\usepackage{natbib}

\usepackage{graphicx,multirow}
\usepackage{color}
\bibpunct{(}{)}{;}{a}{}{,}

\begin{document}

\titlerunning{The properties of the extended warm ionised gas around low-redshift QSOs}
\title{The properties of the extended warm ionised gas around low-redshift QSOs and the lack of extended high-velocity outflows\thanks{Based on observations collected at the Centro Astron\'omico
Hispano Alem\'an (CAHA) at Calar Alto, operated jointly by the Max-
Planck-Institut f\"ur Astronomie and the Instituto de Astrof\'isica de
Andaluc\'ia (CSIC) }\fnmsep\thanks{Table 3, 4 and 7 are available in electronic form at http://www.aanda.org}}
\author{B. Husemann\inst{1}
   \and L. Wisotzki\inst{1} 
   \and S. F. S\'anchez\inst{2,3} 
   \and K. Jahnke\inst{4}
}

\institute{Leibniz-Institut f\"ur Astrophysik Potsdam (AIP), An der Sternwarte 16, 14482 Potsdam, Germany\\
\email{bhusemann@aip.de}
\and
Instituto de Astrof\'isica de Andaluc\'ia (CSIC), Camino Bajo de Hu\'etor s/n Aptdo. 3004, E18080-Granada, Spain
\and
Centro Astron\'omico Hispano Alem\'an de Calar Alto (CSIC-MPIA), E-4004 Almer\'ia, Spain
\and 
Max-Planck-Institut f\"ur Astronomie, K\"onigsstuhl 17, D-69117 Heidelberg, Germany}

\abstract{We present a detailed analysis of a large sample of 31 low-redshift, mostly radio-quiet type 1 quasi-stellar objects (QSOs) observed with integral field spectroscopy to study their extended emission-line regions (EELRs). We focus on the ionisation state of the gas, size and luminosity of extended narrow line regions (ENLRs), which corresponds to those parts of the EELR dominated by ionisation from the QSO, as well as the kinematics of the ionised gas.  We detect EELRs around 19 of our 31 QSOs (61\%) after deblending the unresolved QSO emission and the extended host galaxy light in the integral field data with a new dedicated algorithm. Based on standard emission-line diagnostics we identify 13 EELRs to be entirely ionised by the QSO radiation, 3 EELRs are composed of \mbox{\ion{H}{ii}}\ regions and 3 EELRs display signatures of both ionisation mechanisms at different locations. The typical size of the ENLR is $\sim$10\,kpc at a median nuclear \mbox{[\ion{O}{iii}]}\ luminosity of $\log( L([\ion{O}{iii}])/[\mathrm{erg}\mathrm{s}^{-1}])= 42.7\pm0.15$. We show  that the ENLR sizes are least a factor of $\sim$2 larger than determined with the \textit{Hubble} Space Telescope, but are consistent with those of recently reported type 2 QSOs at matching \mbox{[\ion{O}{iii}]}\ luminosities. The ENLR of type 1 and type 2 QSOs therefore appear to follow the same size-luminosity relation. Furthermore, we show for the first time that the ENLR size is much better correlated with the QSO continuum luminosity than with the total/nuclear \mbox{[\ion{O}{iii}]}\ luminosity. We show that ENLR luminosity and radio luminosity are correlated, and argue that radio jets even in radio-quiet QSOs are  important for  shaping the properties of the ENLR. 
Strikingly, the kinematics of the ionised gas is quiescent and likely gravitationally driven in the majority of cases and we find only 3 objects with radial gas velocities exceeding $>400\,\mathrm{km}\,\mathrm{s}^{-1}$ in specific regions of the EELR that can be associate with radio jets. In general, these are significantly lower outflow velocities and detection rates compared to starburst galaxies or radio-loud QSOs. This represent a challenge for some theoretical feedback models in which luminous QSOs are expected to radiatively drive an outflow out to scales of the entire host galaxy.
}

\keywords{Galaxies: active - quasars: emission-lines - Galaxies: ISM }

\maketitle

\section{Introduction}
Extended warm ionised gas with sizes of several tens of kpc around luminous radio-loud quasi-stellar objects (QSOs) was initially discovered by \citet{Wampler:1975} and \citet{Stockton:1976}. Follow-up longslit spectroscopy presented by \citet{Boroson:1984} and \citet{Boroson:1985} confirmed that a substantial part of the extended ionised gas is predominantly photoionised by the hard radiation of the active galactic nucleus (AGN) as indicated by the similar line ratios compared to the compact narrow-line region (NLR). \citet{Stockton:1987} conducted a large ground-based narrow-band imaging survey of 47 luminous QSOs at $z<0.5$ in the \mbox{[\ion{O}{iii}]}\,$\lambda5007$ line (shortly \mbox{[\ion{O}{iii}]}\ hereafter) and detected ionised gas beyond 10\,kpc around 25\% of the objects, predominantly radio-loud QSOs.  These nebulae are often referred to as extended emission line regions (EELRs) when their \mbox{[\ion{O}{iii}]}\ line luminosity exceeds $5\times10^{41}\,\mathrm{erg}\,\mathrm{s}^{-1}$. A strong alignment effect between the major EELR axis and the radio axis was found for RLQs at $z>0.6$ \citep{McCarthy:1987}, suggesting that radio jets have a strong impact on the EELR properties and may directly enhance their brightnesses. Furthermore, EELRs are preferentially found around radio-loud QSOs with a systematically lower broad line region (BLR) metallicity \citep{Fu:2007b}. Thus, the nature of EELRs and the origin of the gas are not yet clear and remain strongly debated.

The properties of the extended ionised gas around radio-quiet QSOs has been studied much less intensively. \citet{Bennert:2002} imaged a sample of 7\,objects in the \mbox{[\ion{O}{iii}]}\ line  with the \textit{Hubble} Space Telescope (HST) and detected ionised gas on scales of up to 10\,kpc, significantly smaller than the EELR around radio-loud QSOs. Including the NLR sizes of low-luminosity Seyfert galaxies, \citet{Bennert:2002} established a correlation between the \mbox{[\ion{O}{iii}]}\ luminosity and the size of the NLR, often referred to as the extended NLR (ENLR) when its size exceeds 1\,kpc. The adopted distinction between the terms ENLR and EELR throughout the paper is that an (E)NLR designates \emph{only} the AGN photoionised phase of the emission-line region, whereas an EELR refer to the entire emission-line region, independently of the ionisation mechanism. Another large HST narrow-band snapshot survey targeting Seyfert galaxies confirmed the existence of such a relation \citep{Schmitt:2003b}, but the exact slope and zero-point still remain controversial. Potential caveats of the HST images are the low sensitivity and the lack of spectroscopic information to verify the ionisation mechanism of the gas to be AGN photoionization. For example, longslit spectroscopy of Seyfert galaxies allowed \citet{Bennert:2006b,Bennert:2006a} to estimate a characteristic distance of 0.7--4\,kpc from the nucleus at which the photoionization of star forming region started to dominate against the AGN photoionisation. 

The study of extended ionised gas around luminous QSOs is significantly hampered by the severe contamination from the emission of the bright nucleus.  Recent studies therefore focused on rare type 2 QSOs for which the emission of the  nucleus is thought to be obscured by an optically thick torus following in the Unified Model of AGN \citep{Antonucci:1993}. For example, \citet{Villar-Martin:2010} detected a huge EELR (180\,kpc) around a luminous type 2 QSO ($z=0.399$) and determined an ENLR size of $\sim$25\,kpc from longslit spectra using emission line diagnostics. Furthermore, small samples of type 2 QSOs were recently studied by means of longslit \citep{Greene:2011,Villar-Martin:2011a} and integral field spectroscopy \citep{Humphrey:2010}. Typical ENLR sizes of $\sim$10\,kpc were reported, almost 2 times larger than the ENLRs of unobscured (type 1) QSOs observed with HST. 

The extended warm ionised gas around AGN has attracted further attention because it may allow to directly study the impact of an AGN on its surrounding host galaxy. Complex kinematics in the EELR are found around radio-loud QSOs and radio galaxies with line widths up to $1000\,\mathrm{km}\,\mathrm{s}^{-1}$ and high velocity clouds that exceed the escape velocity of the galaxy \citep[e.g.][]{McCarthy:1996b,Villar-Martin:1999,Tadhunter:2001,Holt:2003,Holt:2006,Nesvadba:2006,Fu:2009}. These QSOs are therefore expected to expel a large fraction of gas from the host galaxy. This is usually interpreted as a clear signature for AGN feedback which has been invoked to explain the observed scaling relation between the black hole (BH) mass and host galaxy properties \citep[e.g.][]{Magorrian:1998,Gebhardt:2000,Tremaine:2002,Haering:2004,Gueltekin:2009}, and to efficiently quench star formation. This is required to prevent the excess of massive young galaxies that is often seen in cosmological simulations \citep[e.g.][]{Croton:2006,Khalatyan:2008}. Given the short duty cycle of radio-loud AGN and the low radio-loud fraction of the order of 10\%, it remains still unclear whether the radio-loud mode of AGN activity can be the primary channel for AGN feedback. Furthermore, the coupling  of kinetic jet energy to the ambient medium may be significantly less efficient than currently required by numerical simulation \citep{Holt:2011}. Extremely complex extended gas kinematics similar to those in radio-loud counterparts have also been confirmed in individual radio-quiet QSOs like the ultra luminous infrared galaxy (ULIRG) Mrk~231 \citep{Hamilton:1987,Lipari:2009a,Rupke:2011,Aalto:2012}, but the majority of radio-quiet type 2 QSOs appears to have more  quiescent gas kinematics \citep{Humphrey:2010, Greene:2011}. An exception are double-peaked \mbox{[\ion{O}{iii}]}\ AGN, which were initially thought to be binary AGN candidates, but \citet{Fu:2012} reported that the majority of them are simply AGN with a luminous ENLR based on integral-field spectroscopy. The line splitting of several hundred to thousand $\mathrm{km}\,\mathrm{s}^{-1}$ suggests the presence of either significant outflows driven by the AGN or severe inflows as a result of merging processes. Anyway, the fraction of double-peaked \mbox{[\ion{O}{iii}]}\ AGN is $\lesssim$1\% \citep{Wang:2009,Liu:2010a,Smith:2010}.

Given that type 1 QSOs are much more frequent and allow to infer basic AGN properties from the spectrum of the nucleus, we took the challenge to deal with the deblending of the QSO and extended host galaxy emission. In our early work  \citep[][hereafter Hu08]{Husemann:2008}, we studied a sample of 20 mainly radio-quiet QSOs observed with an integral field unit (IFU) and found that 8 of the 20 QSOs are surrounded by an EELR. We emphasised that its presence is linked to the spectral properties of the nucleus itself, in particular the equivalent width (EW) of the \ion{Fe}{ii} emission and the width of the broad H$\beta$ line. Since then, we further extended our sample with additional observations of 11 radio-quiet QSOs. In this paper we present a detailed analysis of the EELR ionisation mechanisms, the ENLR size and luminosity as well as the ionised gas kinematics for our entire sample. 

The paper is organised as follows. In section~\ref{sect_pmas:sample_data} we report on the characteristics of the sample including its selection and host galaxy properties based on broad-band imaging. The IFU observations, data reduction and QSO-host deblending are described in Section~\ref{sect_pmas:IFU_data}.  Basic properties of the AGN are determined from the QSO spectra in Section~\ref{sect_pmas:QSO_spec} and we apply standard emission-line diagnostics to infer the ionisation mechanism of the extended gas in Section~\ref{sect_pmas:EELR_charac}. We describe the properties of the ENLR in Section~\ref{sect_pmas:ENLR_prop} and discuss them in comparison to previous results in the literature. Finally, we present the kinematics of the extended gas in Section~\ref{sect_pmas:EELR_kin}.

Throughout the paper we assume a standard $\Lambda$CDM cosmological model with $H_0=70\,\mathrm{km}\,\mathrm{s}^{-1}\,\mathrm{Mpc}$, $\Omega_\mathrm{m}=0.3$, and $\Omega_\Lambda=0.7$.

\section{The QSO sample and characteristics}\label{sect_pmas:sample_data}
\subsection{Sample selection}
\begin{table*}\centering
 \begin{footnotesize}
 \caption{Characteristics of the QSO sample.}
 \label{tbl:sample}
 \begin{tabular}{llccccccccc}\hline\hline\noalign{\smallskip}
Name & Alt. Name & $\alpha$ (J2000) & $\delta$ (J2000) & $z$\tablefootmark{a} & $m_B$\tablefootmark{b} & $M_B$\tablefootmark{c} & $f_{1.4\,\mathrm{GHz}}$\tablefootmark{d} & $f_{5\,\mathrm{GHz}}$\tablefootmark{e} & $R$\tablefootmark{f} & Class.\tablefootmark{g}\\\noalign{\smallskip}\hline\noalign{\smallskip}
PG 0026$+$129 &  & 00${}^\mathrm{h}$29${}^\mathrm{m}$13\fs8 &  13\degr16\arcmin05\farcs0 & $0.142$ & $14.95$ & $-24.24$ & 7.1 & 5.1 & 1.08 & RI \\
PG 0050$+$124 & I Zw 1 & 00${}^\mathrm{h}$53${}^\mathrm{m}$34\fs9 &  12\degr41\arcmin35\farcs0 & $0.061$ & $14.39$ & $-22.86$ & 8.3 & 2.6 & 0.33 & RQ \\
PG 0052$+$251 &  & 00${}^\mathrm{h}$54${}^\mathrm{m}$52\fs2 &  25\degr25\arcmin39\farcs0 & $0.155$ & $15.42$ & $-23.97$ & 7.1 & 0.7 & 0.24 & RQ \\
SDSS J0057$+$1446 &  & 00${}^\mathrm{h}$57${}^\mathrm{m}$09\fs9 &  14\degr46\arcmin10\farcs2 & $0.172$ & $15.58$ & $-24.05$ & 2.4 & $1.3^\dagger$ & 0.48 & RQ \\
HE 0132$-$0441 & RBS 219 & 01${}^\mathrm{h}$35${}^\mathrm{m}$27\fs0 & $-$04\degr26\arcmin36\farcs2 & $0.154$ & $16.25$ & $-23.12$ & 7.4 & 2.9 & 2.03 & RI \\
SDSS J0155$-$0857 &  & 01${}^\mathrm{h}$55${}^\mathrm{m}$30\fs0 & $-$08\degr57\arcmin04\farcs1 & $0.165$ & $16.83$ & $-22.71$ & $<0.9$ & $<0.5^\dagger$ & $<$0.58 & RQ \\
HE 0157$-$0406 &  & 01${}^\mathrm{h}$59${}^\mathrm{m}$49\fs0 & $-$03\degr52\arcmin00\farcs4 & $0.218$ & $17.09$ & $-23.11$ & $<0.9$ & $<0.5^\dagger$ & $<$0.75 & RQ \\
PG 0157$+$001 & Mrk 1014 & 01${}^\mathrm{h}$59${}^\mathrm{m}$50\fs2 &  00\degr23\arcmin41\farcs0 & $0.164$ & $15.20$ & $-24.32$ & 26.2 & 8.0 & 2.12 & RI \\
SDSS J0836$+$4426 &  & 08${}^\mathrm{h}$36${}^\mathrm{m}$58\fs9 &  44\degr26\arcmin02\farcs3 & $0.254$ & $15.54$ & $-25.03$ & 10.0 & $5.3^\dagger$ & 1.93 & RI \\
SDSS J0948$+$4335 &  & 09${}^\mathrm{h}$48${}^\mathrm{m}$59\fs5 &  43\degr35\arcmin19\farcs0 & $0.226$ & $16.81$ & $-23.48$ & $<0.9$ & $<0.5^\dagger$ & $<$0.56 & RQ \\
SDSS J1131$+$2632 &  & 11${}^\mathrm{h}$31${}^\mathrm{m}$09\fs2 &  26\degr32\arcmin07\farcs9 & $0.244$ & $17.41$ & $-23.06$ & 2.7 & $1.4^\dagger$ & 2.90 & RI \\
HE 1217$+$0220 & PKS 1217+023 & 12${}^\mathrm{h}$20${}^\mathrm{m}$11\fs9 &  02\degr03\arcmin41\farcs7 & $0.239$ & $17.23$ & $-23.20$ & 652.0 & 440.0 & 757.18 & RL \\
HE 1226$+$0219 & 3C 273 & 12${}^\mathrm{h}$29${}^\mathrm{m}$06\fs7 &  02\degr03\arcmin07\farcs9 & $0.158$ & $11.42$ & $-28.01$ & 55296.0 & 37000.0 & 302.22 & RL \\
SDSS J1230$+$6621 &  & 12${}^\mathrm{h}$30${}^\mathrm{m}$22\fs2 &  66\degr21\arcmin54\farcs7 & $0.184$ & $16.75$ & $-23.05$ & $<2.5$ & $<1.3^\dagger$ & $<$1.46 & RQ? \\
HE 1228$+$0131 &  & 12${}^\mathrm{h}$30${}^\mathrm{m}$50\fs0 &  01\degr15\arcmin21\farcs8 & $0.117$ & $14.82$ & $-23.92$ & 1.4 & 0.2 & 0.04 & RQ \\
SDSS J1230$+$1100 &  & 12${}^\mathrm{h}$30${}^\mathrm{m}$54\fs1 &  11\degr00\arcmin11\farcs2 & $0.236$ & $16.61$ & $-23.78$ & $<1.0$ & $<0.5^\dagger$ & $<$0.52 & RQ \\
PG 1427$+$480 &  & 14${}^\mathrm{h}$29${}^\mathrm{m}$43\fs1 &  47\degr47\arcmin27\farcs0 & $0.221$ & $16.33$ & $-23.90$ & $<0.9$ & 0.0 & 0.02 & RQ \\
SDSS J1444$+$0633 &  & 14${}^\mathrm{h}$44${}^\mathrm{m}$14\fs7 &  06\degr33\arcmin06\farcs8 & $0.208$ & $17.10$ & $-22.99$ & $<1.0$ & $<0.5^\dagger$ & $<$0.81 & RQ \\
HE 1453$-$0303 &  & 14${}^\mathrm{h}$56${}^\mathrm{m}$11\fs5 & $-$03\degr15\arcmin27\farcs8 & $0.206$ & $15.68$ & $-24.39$ & 2.7 & $1.4^\dagger$ & 0.58 & RQ \\
PG 1545$+$210 & PKS 1545+210 & 15${}^\mathrm{h}$47${}^\mathrm{m}$43\fs5 &  20\degr52\arcmin17\farcs0 & $0.266$ & $16.05$ & $-24.63$ & 1560.0 & 720.0 & 418.15 & RL \\
PG 1612$+$261 & TON 256 & 16${}^\mathrm{h}$14${}^\mathrm{m}$13\fs2 &  26\degr04\arcmin16\farcs0 & $0.131$ & $16.00$ & $-23.00$ & 17.9 & 5.1 & 2.81 & RI \\
SDSS J1655$+$2146 &  & 16${}^\mathrm{h}$55${}^\mathrm{m}$51\fs4 &  21\degr46\arcmin01\farcs8 & $0.154$ & $16.00$ & $-23.37$ & 6.6 & $3.5^\dagger$ & 1.94 & RI \\
PG 1700$+$518 &  & 17${}^\mathrm{h}$01${}^\mathrm{m}$24\fs9 &  51\degr49\arcmin21\farcs0 & $0.292$ & $15.43$ & $-25.48$ & 21.6 & 7.2 & 2.36 & RI \\
PG 2130$+$099 & Mrk 1513 & 21${}^\mathrm{h}$32${}^\mathrm{m}$27\fs8 &  10\degr08\arcmin19\farcs0 & $0.061$ & $14.62$ & $-22.63$ & 6.0 & 2.0 & 0.32 & RQ \\
HE 2152$-$0936 & PHL 1811 & 21${}^\mathrm{h}$55${}^\mathrm{m}$01\fs5 & $-$09\degr22\arcmin24\farcs5 & $0.192$ & $14.26$ & $-25.63$ & 1.2 & $0.6^\dagger$ & 0.07 & RQ \\
HE 2158$-$0107 &  & 22${}^\mathrm{h}$01${}^\mathrm{m}$03\fs1 & $-$00\degr53\arcmin01\farcs0 & $0.213$ & $16.69$ & $-23.45$ & 1.6 & $0.8^\dagger$ & 0.88 & RQ \\
HE 2158$+$0115 & RBS 1812 & 22${}^\mathrm{h}$01${}^\mathrm{m}$29\fs7 &  01\degr29\arcmin56\farcs1 & $0.160$ & $16.63$ & $-22.83$ & $<0.9$ & $<0.5^\dagger$ & $<$0.48 & RQ \\
PG 2214$+$139 & Mrk 304 & 22${}^\mathrm{h}$17${}^\mathrm{m}$11\fs6 &  14\degr14\arcmin28\farcs0 & $0.067$ & $14.98$ & $-22.48$ & $<2.5$ & 0.2 & 0.05 & RQ \\
HE 2307$-$0254 &  & 23${}^\mathrm{h}$10${}^\mathrm{m}$29\fs7 & $-$02\degr38\arcmin13\farcs0 & $0.221$ & $16.90$ & $-23.34$ & $<2.5$ & $<1.3^\dagger$ & $<$1.68 & RQ? \\
HE 2349$-$0125 & PKS 2349$-$014 & 23${}^\mathrm{h}$51${}^\mathrm{m}$56\fs0 & $-$01\degr09\arcmin13\farcs6 & $0.174$ & $16.03$ & $-23.63$ & 1623.2 & 622.7 & 355.46 & RL \\
HE 2353$-$0420 &  & 23${}^\mathrm{h}$56${}^\mathrm{m}$32\fs9 & $-$04\degr04\arcmin00\farcs7 & $0.229$ & $17.23$ & $-23.09$ & $<2.5$ & $<1.3^\dagger$ & $<$2.27 & RQ? \\
\noalign{\smallskip}
\hline
\end{tabular}
\tablefoot{
\tablefoottext{a}{QSO redshift taken from the PG, HES, or SDSS catalogues.}
\tablefoottext{b}{Apparent $B$ band magnitudes taken from the catalogues or synthesised from the SDSS spetrum.}
\tablefoottext{c}{Absolute magnitudes $M_B$ taking into account Galactic foreground extinction and an appropriate $k$-correction \citep{Wisotzki:2000b}.}
\tablefoottext{d}{Radio fluxes at 1.4\,GHz in mJy as measured by the NVSS all-sky survey or the FIRST survey when available.}
\tablefoottext{e}{Radio fluxes at 5\,GHz in mJy taken from \citet{Kellermann:1989} and an unpublished catalog (Gopal-Krishna et al. private communication), if available. Values marked with $\dagger$ were estimated from the radio flux at 1.4\,GHz assuming a spectral index of $\alpha=-0.5$.}
\tablefoottext{f}{Ratio $R$ of the 5\,GHz radio to optical flux density at $4400$\AA. The optical flux density $f_\nu(4400\mathrm{\AA})$ in mJy was estimated from the optical $B$ band magnitude with the relation $m_B = -2.5\log f_\nu(4400\,\mathrm{\AA}) -48.36$ \citep{Schmidt:1983}.}
\tablefoottext{g}{Radio classification of the QSO following \citet{Kellermann:1989} into radio loud (RL) and radio quiet (RQ) and radio intermediate  (RI) QSOs. An unambigous classification for a few object cannot be made due to upper limits in $R$.}
}

\end{footnotesize}
\end{table*}

Our QSO sample consists of 31 objects in the redshift range $0.06<z<0.3$, which we observed with the Potsdam Multi-Aperture Spectrophotometer \citep[PMAS,][]{Roth:2005} mounted to the Cassegrain focus of the 3.5\,m telescope at the Calar Alto observatory. 
The QSOs were drawn from two samples that were constructed with different selection criteria. An initial sample of 21 low-redshift QSOs was randomly selected from the Palomar-Green Survey \citep[PG,][]{Schmidt:1983} and from the Hamburg/ESO survey \citep[HES,][]{Wisotzki:2000} with an absolute magnitude of $M_B\leq-23$ (computed with $H_0=50\,\mathrm{km}\,\mathrm{s}^{-1}\,\mathrm{Mpc}^{-1}$), a commonly used yet arbitrary division line between Seyfert galaxies and QSOs, and a declination of $\delta>-10^\circ$ to be observable from the Calar Alto observatory.  An additional set of 10 QSOs were drawn from the Sloan Digital Sky Survey (SDSS) DR6 \citep{Adelman-McCarthy:2008}. These QSOs were selected to be brighter than  $g<17$ within the redshift range $0.13<z<0.3$ and to have a \mbox{[\ion{O}{iii}]}/\mbox{H$\beta$}\ flux density ratio at their peaks greater than 1.5 as measured from the SDSS spectra. This last criterion is based on early results from the initial sample (Hu08) and should give preference to QSOs surrounded by extended ionised gas. The complete list of our observed objects is given in Table~\ref{tbl:sample} including information on some of their main characteristics.

\begin{figure}
\centering
\resizebox{\hsize}{!}{\includegraphics[clip]{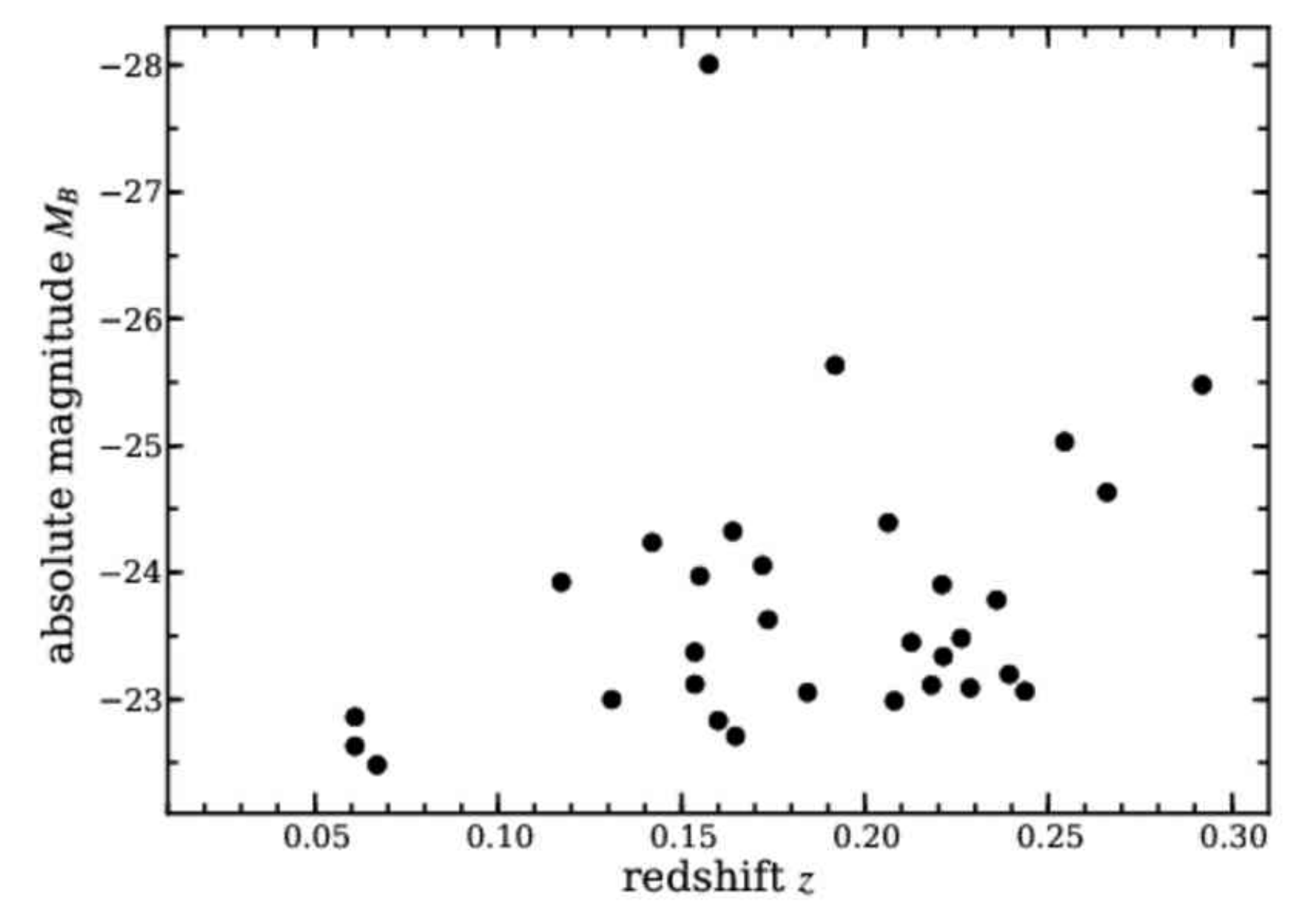}}
\caption{Absolute $B$ band magnitude against the redshift for all objects in our QSO sample.}
\label{fig_pmas:sample}
\end{figure}

The redshift-luminosity diagram of our observed QSO sample is shown in Fig.~\ref{fig_pmas:sample}. We computed absolute magnitudes ($M_B$) from the apparent $B$ band magnitudes taking into account Galactic foreground extinction \citep{Schlegel:1998} and an appropriate $k$-correction \citep{Wisotzki:2000b}. 
 The $B$ band magnitudes and redshifts were either taken from the available catalogues or directly measured from publicly available SDSS spectra. The absolute magnitudes are in the range of  $-26<M_B<-22.3$ with a median value of -23.45. The median redshift of the sample is $z=0.184$. Thus, the majority of our objects are luminous low-redshift QSOs.

\subsection{Radio properties}
Radio-loud and radio-quiet QSOs (RLQs and RQQs, respectively) are often distinguished based on the $R$ parameter, which is defined as the ratio of the radio (at 6\,cm/5\,GHz) to the optical flux density (at 4400\,\AA). We adopt the classification of \citet{Kellermann:1989}, who separated RQQs with $R<1$, RLQs with $R>10$ and a few radio-intermediate objects in between the two classes. Not every object of our sample has been observed at 5\,GHz, but the NRAO VLA Sky Survey \citep[NVSS][]{Condon:1998} and the Faint Images of the Radio Sky at Twenty-centimetres (FIRST) survey \citep{Becker:1995} provide radio measurements  at 1.4\,GHz (21\,cm) for a large fraction of the sky with a completeness level of 2.5\,mJy and 1\,mJy, respectively. We adopt a power-law dependence of the radio flux densities ($f_\nu\propto \nu^{\alpha_\mathrm{r}}$) with a spectral index of $\alpha_\mathrm{r}=-0.5$, at the border line between steep- and flat-spectrum radio sources, to estimate the flux density at 5\,GHz from measurements at 1.4\,GHz when required. Only 4 objects in the sample can be classified as RLQs, 8 objects have intermediate radio properties, and the majority of objects within the sample are clearly RQQs.

\begin{table}\centering
 \begin{footnotesize}
 \caption{Available radio imaging data for some RQQs.}
 \label{tbl:radio_prop}
 \begin{tabular}{llcccc}\hline\hline\noalign{\smallskip}
Name           & $\nu$\tablefootmark{a} &  Beam\tablefootmark{b}   & Size\tablefootmark{c}  & PA\tablefootmark{d} & Reference \\\noalign{\smallskip}\hline\noalign{\smallskip}
PG 0026$+$129  & 8.4\,GHz   &  $0\farcs35$ & $0\farcs60$  & ...        & (1) \\
PG 0052$+$251  & 4.8\,GHz   &  $0\farcs59$ & $<0\farcs09$ & ...        & (1) \\
Mrk 1014       & 8.4\,GHz   &  $0\farcs36$ & $2\farcs6$   & $90\degr$  & (1) \\
PG 1612$+$261  & 8.4\,GHz   &  $0\farcs28$ & $1\farcs0$   & $66\degr$  & (2) \\
PG 1700$+$518  & 8.4\,GHz   &  $0\farcs28$ & $0\farcs5$   & $-42\degr$ & (2) \\
PG 1700$+$518  & 1.6\,GHz   &  $0\farcs02$ & $0\farcs5$   & $-41\degr$ & (3) \\
PG 2130$+$099  & 4.8\,GHz   &  $0\farcs35$ & $2\farcs9$   & $-48\degr$ & (1) \\
\noalign{\smallskip}
\hline
\end{tabular}
\tablefoot{
\tablefoottext{a}{Observing frequency $\nu$ of the radio observations.}
\tablefoottext{b}{Beam size of the interferometric radio observations.}
\tablefoottext{c}{Maximum size of the extended radio emission.}
\tablefoottext{d}{Position angle of the major radio axis. 90\degr\ points to the East and 0\degr\ to the North.}
}
\tablebib{ (1)~\citet{Leipski:2006b}; (2)~\citet{Kukula:1998}; (3)~\citet{Yang:2012}}

\end{footnotesize}
\end{table}

Besides the integrated radio measurements, archival high spatial resolution images for 6 of our RQQs were obtained with the VLA interferometer at subarcsecond resolution and were presented by \citet{Kukula:1998} and \citet{Leipski:2006b}. We retrieved the processed images from the VLA image archive\footnote{https://archive.nrao.edu/archive/archiveimage.html} and some have also been kindly provided by the authors. The characteristics of these images are listed in Table~\ref{tbl:radio_prop}. All except PG~0052$+$251 display extended radio emission with sizes up to $3\arcsec$. A triple radio source as an unambiguous sign of a jet was identified in the RQQs Mrk~1014, in PG~2130$+$099 using the VLA, and in PG~1700$+$518 using the higher spatial resolution of the European VLBI network \citep{Yang:2012}.

\subsection{QSO host morphologies}
\subsubsection{Available broad-band imaging}
High spatial resolution broad-band images were obtained with the HST for most of the PG QSOs over the last decade, except  PG~1427$+$490, PG~1612$+$261 and PG~2214$+$139. These HST images were initially published in various unrelated articles (see Table~\ref{tab_pmas:host_morph}), but a homogeneous analysis of all the public data was presented by \citet{Kim:2008}. Ground-based broad-band images are additionally available for all QSOs covered by the SDSS footprint. The SDSS survey images  with an effective exposure time of 54s were usually obtained at a seeing around $\sim$1\farcs2 ($r$). Yet, it has so far not been attempted to uncover the underlying QSO host  galaxies in these images given the short exposure time and the low spatial resolution of the images. 

No broad-band images have so far been presented for most of the HES QSOs in our sample. Ground-based $H$ band images were obtained for a few QSOs in 2003 with the SOFI IR spectrograph and imaging camera \citep{Moorwood:1998} mounted on the New Technology Telescope (NTT) and are publicly available in the ESO archive, which we reduced with the corresponding standard ESO pipeline. Additionally, we obtained $R$ band images for three HES QSOs with the 60\,cm telescope of the Lightbuckets robotic telescope network (hereafter LB). This small telescope is equipped with an Apogee Alta U42 CCD camera providing a spatial resolution of 0\farcs57\,$\mathrm{pixel}^{-1}$. With an exposure time of $\sim$1\,h, these images are nominally deeper than SDSS, but suffer from a seeing of $\gtrsim$2\arcsec. We reduced these images following the standard procedures including bias subtraction, flat-fielding and combination of the dithered exposures with custom \texttt{Python} scripts.

\subsubsection{QSO-host deblending}
In order to perform a consistent analysis of the available broad-band images we deblended the QSO and host galaxy component using \texttt{GALFIT} \citep{Peng:2002,Peng:2010}. We emphasise that our main aim here is only to recover the stellar light surface brightness distribution of the underlying host galaxies and distinguish between bulge-dominated and disc-dominated host galaxies whenever possible. A more detailed analysis of the images would be beyond our scope and hardly possible given the low spatial resolution  of many images.

\begin{figure}
\resizebox{\hsize}{!}{\includegraphics[clip]{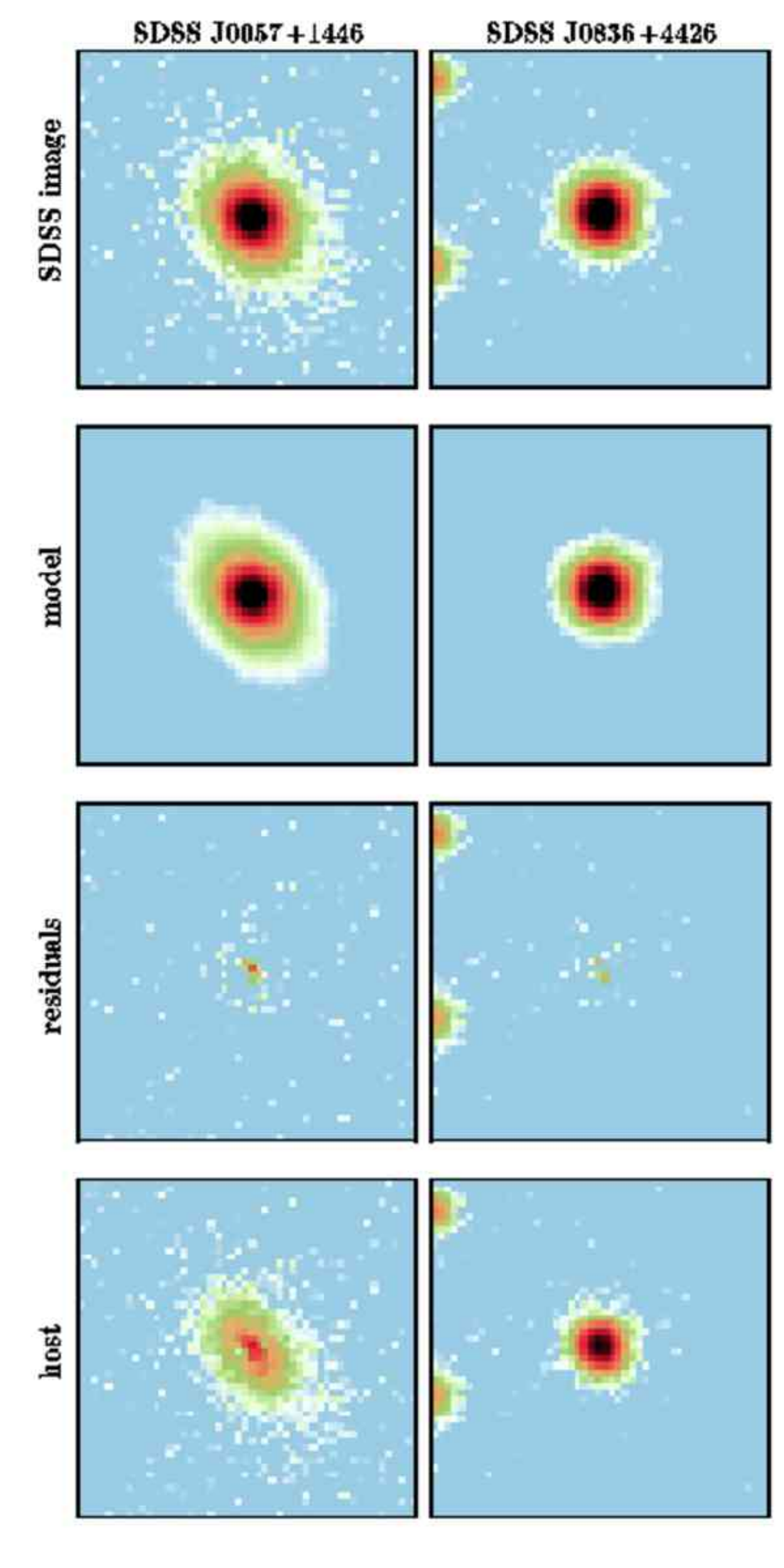}}
\caption{Two examples of the QSO-host deblending process for SDSS $i$ band images. The PSF was constructed from a bright nearby star in the field. A best-fitting Sersi\'c surface brightness model plus a PSF for the nucleus was obtained with \texttt{GALFIT} and subtracted from the original SDSS image to highlight the residuals. The host galaxy is recovered after subtracting the best-fit PSF component from the original image with nucleus-to-host ratios of $\sim$3.5 and $\sim$2.2 for SDSS~J0057$+$1446 and SDSS~J0836$+$4426, respectively.}
\label{fig_pmas:SDSS_host_decomp}
\end{figure}

We created an empirical Point Spread Function (PSF) from bright stars in the field close to the target. Dedicated PSF stars were observed with HST only for a few objects, so that we relied on synthetic PSFs generated with the \texttt{TinyTim} software package \citep{Krist:1995}. First, we used \texttt{GALFIT} to model each object with a PSF component only, to check whether any significant residuals of a host galaxy could actually be detected. In a second step, we modelled the images with a superposition of a Sersi\'c light profile \citep{Sersic:1968} for the host and a PSF for the QSO component. Two examples of the deblending results for the ground-based SDSS image are presented in Fig.~\ref{fig_pmas:SDSS_host_decomp}.

\onltab{
\begin{table*}\centering
\begin{footnotesize}
 \caption{Broad-band imaging analysis results -- Structural parameters and morphological host galaxy classification.}
 \label{tab_pmas:host_morph}
  \begin{tabular}{llcccccccc}\hline\hline\noalign{\smallskip}
Name              & Telescope/Instrument & Ref.\tablefootmark{a} &$\chi^2_\nu(n=1)$ & $\chi^2_\nu (n=4)$ &  $\chi^2_\nu(n)$ & $r_{\mathrm{e}}$\tablefootmark{b} & $b/a$\tablefootmark{c} & PA\tablefootmark{d} &  Class.\tablefootmark{e}  \\\noalign{\smallskip}\hline\noalign{\smallskip}
PG~0026$+$129     & HST NICMOS (F160W) & (1) &	4.400		  &	3.566	      & 3.563 (3.6)     & 1\farcs2 (3.0\,kpc)     & 0.82  & $85\degr$   &  B     \\                                   
PG~0050$+$124     & HST WFPC2 (F814W) & &	2.826 		  &	2.684         & 2.669 (2.8)     & 8\farcs7 (10.2\,kpc)    & 0.73  & $-58\degr$  & B+D      \\ 
PG~0052$+$251     & HST WFPC2 (F606W) & (2) &	2.705		  &	3.368	      & 2.704 (1.0)     & 3\farcs5 (9.4\,kpc)     & 0.65  & $-8\degr$   & D      \\ 
SDSS~J0057$+$1446  & SDSS ($i'$)       &  &	1.002		  &	0.990	      & 0.989 (2.7)     & 3\farcs7 (10.8\,kpc)    & 0.52  & $-50\degr$  & B?      \\
		  & SDSS ($r'$)      &  &	1.018		  &	0.985	      & 0.984 (6.7)     & 2\farcs2 (6.4\,kpc)     & 0.62  & $-49\degr$  &       \\ 
RBS~219    & LB Apogee U42 ($R$) & &	1.505		  &	1.508	      & 1.505 (1.1)     & 3\farcs8 (10.1\,kpc)    & 0.79  & $28\degr$   & D(V)       \\ 
SDSS~J0155$-$0857  & SDSS ($i'$)       &  &	1.180		  &	1.150	      & 1.148 (5.4)     & 1\farcs3 (3.4\,kpc)     & 0.62  & $32\degr$   & B?      \\
		  & SDSS ($r'$)       &  &	1.136		  &	1.139	      & 1.132 (1.7)     & 2\farcs4 (6.8\,kpc)     & 0.57  & $31\degr$   &      \\
PG~0157$+$001     & HST WFPC2 (F675W) & (3) &	6.406		  &	6.471	      & 6.360 (1.5)     & 4\farcs9 (13.8\,kpc)    & 0.89  & $15\degr$   & D/M      \\  
SDSS~J0836$+$4426  & SDSS ($i'$)       &  &	0.984		  &	0.979	      & 0.980 (6.1)     & 0\farcs2 (0.8\,kpc)     & 0.52  & $-83\degr$  & ?      \\
		  & SDSS ($r'$)       &  &	0.994		  &	0.984	      & 0.984 (12.0)    & 0\farcs2 (0.8\,kpc)     & 0.90  & $16\degr$   &       \\
SDSS~J0948$+$4335  & SDSS ($i'$)       &  &	1.244		  &	1.259	      & 1.243 (1.2)     & 2\farcs4 (8.7\,kpc)     & 0.80  & $-31\degr$  & D(V)      \\
		  & SDSS ($r'$)      &   &	1.052		  &	1.083	      & 1.051 (0.8)     & 2\farcs7 (9.8\,kpc)     & 0.79  & $-27\degr$  &       \\
SDSS~J1131$+$2632  & SDSS ($i'$)       &  &	1.213		  &	1.255	      & 1.212 (0.8)     & 3\farcs7 (14.2\,kpc)    & 0.74  & $-70\degr$  & M(V)      \\
		  & SDSS ($r'$)      &   &	1.111		  &	1.174	      & 1.105 (0.7)     & 3\farcs4 (13.1\,kpc)    & 0.75  & $-73\degr$  &       \\
PKS~1217$+$023    & HST WFPC2 (F675W) & (4) &	4.850		  &	4.722	      & 4.719 (3.15)    & 2\farcs5 (9.7\,kpc)     & 0.79  & $85\degr$   & B      \\  
3C~273            & HST WFPC2 (F606W) & (5) &	1.941		  &	1.823	      & 1.814 (2.8)     & 5\farcs0 (13.7\,kpc)    & 0.80  & $78\degr$   & B     \\ 
SDSS~J1230$+$6621  & SDSS ($i'$)       &  &	1.340		  &	1.340	      & 1.339 (1.7)     & 1\farcs8 (5.6\,kpc)     & 0.85  & $15\degr$   & D?      \\
		  & SDSS ($r'$)       &  &	1.047		  &	1.052	      & 1.047 (1.1)     & 1\farcs8 (5.6\,kpc)     & 0.73  & $24\degr$   &       \\
HE~1228$+$0131    & SDSS ($i'$)       &  &	1.038		  &	1.006	      & 0.992 (9.9)     & 0\farcs3  (0.6\,kpc)    & 0.66  & $-48\degr$  & ?      \\
		  & SDSS ($r'$)       &  &	1.122		  &	1.110	      & 1.104 (9.0)     & 0\farcs8  (1.7\,kpc)    & 0.68  & $-63\degr$  &       \\
SDSS~J1230$+$1100  & SDSS ($i'$)       &  &	0.878		  &	0.867	      & 0.864 (6.1)     & 0\farcs2 (0.8\,kpc)     & 0.77  & $-86\degr$  & ?      \\
		  & SDSS ($r'$)       &  &	1.055		  &	1.047	      & 1.043 (8.5)     & 0\farcs2 (0.8\,kpc)     & 0.82  & $-30\degr$  &       \\
SDSS~J1444$+$0633  & SDSS ($i'$)       &  &	0.982		  &	0.975	      & 0.972 (12.1)    & 1\farcs3 (4.4\,kpc)     & 0.68  & $85\degr$   & ?       \\
		  & SDSS ($r'$)      &  &	0.981		  &	0.974	      & 0.974 (7.2)     & 1\farcs7 (5.4\,kpc)     & 0.70  & $90\degr$   &      \\
PG~1427$+$480     & SDSS ($i'$)      &  &	1.438		  &	1.408	      & ...             & 0\farcs9 (3.2\,kpc)     & 0.90  & $34\degr$   & B?     \\
		  & SDSS ($r'$)      &  &	1.039		  &	1.033	      & 1.032 (6.8)     & 0\farcs6 (2.1\,kpc)     & 0.79  & $20\degr$   &      \\
PKS~1545$+$210    & HST WFPC2 (F606W) & (5) &	2.667		  &	2.606	      & 2.603 (3.0)     & 1\farcs8 (7.4\,kpc)     & 0.85  & $77\degr$   & B       \\    
PG~1612$+$261     & SDSS ($i'$)       &  &	1.057		  &	1.053	      & 1.047 (2.2)     & 1\farcs6 (3.7\,kpc)     & 0.37  & $10\degr$   & D(V)       \\
		  & SDSS ($r'$)      &  &	1.057		  &	1.068	      & 1.047 (1.7)     & 1\farcs7 (3.96\,kpc)    & 0.40  & $11\degr$   &        \\
SDSS~J1655$+$2146  & SDSS ($i'$)       &  &	1.795		  &	1.662	      & ...             & 0\farcs2 (0.5\,kpc)     & 0.94  & $51\degr$   & ?       \\
		  & SDSS ($r'$)      &  &	1.188		  &	1.140	      & ...             & 0\farcs6 (1.6\,kpc)     & 0.82  & $30\degr$   &        \\ 
PG~1700$+$518     & HST WFPC2 (F547M) & (1) &	1.800		  &	1.742	      & ...             & 1\farcs4 (6.2\,kpc)     & 0.80  & $23\degr$   & M(V)     \\ 
PG~2130$+$099     & HST WFPC2 (F606W) & (6) &	1.870		  &	1.875	      & 1.833 (1.8)     & 12\farcs0 (14\,kpc)     & 0.36  & $60\degr$   & B+D      \\ 
		  & NTT SOFI ($H$)    &  &	1.417		  &	1.372	      & 1.360 (2.4)     & 7\farcs3 (8.6\,kpc)     & 0.31  & $51\degr$   &       \\ 
PHL~1811    	  & HST ACS (F625W)   & (7) &	1.591		  &	1.552	      & ...	        & 1\farcs7 (5.3\,kpc)     & 0.52  & $72\degr$   & D(V)      \\
		  & NTT SOFI ($H$)    &  &	1.156		  &	1.128	      & 1.091 (9.6)     & 0\farcs7 (2.2\,kpc)     & 0.48  & $59\degr$   &       \\  
		  & LB Apogee U42 ($R$)  &  &  1.263	  	  &	1.248	      & 1.243 (8.5)     & 1\farcs7 (5.3\,kpc)     & 0.59  & $60\degr$   &      \\
HE~2158$-$0107\tablefootmark{f} & SDSS ($z$)       &  &	0.884		  &	0.880	      & 0.880 (6.2)     & 0\farcs2 (0.7\,kpc)     & 0.93   & $43\degr$   & B? \\
		  & NTT SOFI ($H$)    &  &	...		  &	...	      & ...             &  ...            & ...   & ...&      \\
PG~2214$+$139     & NTT SOFI ($H$)    &  &	1.014		  &	1.063	      & 0.950 (1.8)     & 2\farcs9 (3.7\,kpc)     & 0.97  & $64\degr$   & B      \\ 
		  & SDSS ($i'$)       &  &	1.825		  &	1.380	      & 1.299 (5.4)     & 4\farcs2 (5.4\,kpc)     & 0.94  & $63\degr$   &       \\
		  & SDSS ($r'$)       &  &	2.326		  &	1.814	      & 1.684 (5.8)     & 4\farcs1 (5.3\,kpc)     & 0.94  & $56\degr$   &        \\
HE~2307$-$0254    & NTT SOFI ($H$)    &  &	0.542		  &	0.542	      & 0.542 (6.2)     & 0\farcs2 (0.7\,kpc)    & 0.67  & $35\degr$    &  B?     \\ 
PKS~2349$-$014    & HST WFPC2 (F606W) & (8) &	2.649		  &	2.441	      & 2.436 (3.3)     & 5\farcs6 (16.5\,kpc)    & 0.81  & $83\degr$   & B/M      \\ 
HE~2353$-$0420    & LB Apogee U42 ($R$)  &  &	0.940	  	  &	0.957	      & 0.939 (0.7)     & 3\farcs8 (13.9\,kpc)    & 0.49  & $15\degr$   & D(V)       \\ \noalign{\smallskip}\hline 
\end{tabular}
\tablefoot{
\tablefoottext{a}{References to the articles in which a morphological analysis of these data was initally presented.}
\tablefoottext{b}{Effective radius $r_\mathrm{e}$ of the best-fitting realistic host-galaxy model given in apparent and physical scale.}
\tablefoottext{c}{Axis ratio of the minor to major axis of the best-fitting model.}
\tablefoottext{d}{Position angle of the major axis of the best-fitting model oriented such that 0\degr\ is North and 90\degr\ is East.}
\tablefoottext{e}{Our host galaxy morphological classification into: B (bulge-dominated), D (disc-dominated), B+D (bulge+disc component), M (Merger or strongly interacting), B?/D? (only tentative classifications), ? (unconstrained). We append (V) to the classification, if
it is due to a visual inspection of the image. }
\tablefoottext{f}{unresolved host galaxy}
}
\tablebib{ (1)~\citet{Veilleux:2009}; (2)~\citet{Bahcall:1996}; (3)~\citet{McLure:1999}; (4)~\citet{Dunlop:2003}; (5)~\citet{Bahcall:1997}; (6)~\citet{Kim:2008}; (7)~\citet{Jenkins:2005}; (8)~\citet{Bahcall:1995}}
\end{footnotesize}
\end{table*}
}
Each QSO host was modelled with three different radial Sersi\'c surface brightness profiles: 1) Sersi\'c index fixed to $n=1$ (exponential profile), 2) Sersi\'c index fixed to $n=4$ (de Vaucouleurs profile), or 3) Sersi\'c variable as a free parameter. The resulting $\chi^2$ values and the best-fit structural parameters are reported in Table~\ref{tab_pmas:host_morph}.   The superposition of a bulge and a disc model provides a significantly better fit to the surface brightness distribution of PG~2130$+$099 and I~Zw~1 than a single component. The structural parameters reported in Table~\ref{tab_pmas:host_morph} for these two QSO hosts refer to the disc component of the model.

The QSO hosts observed with HST are all well resolved and we recover the host galaxies also for the majority of the ground-based observations. PSF-subtracted host galaxy images are presented below in Fig.~\ref{fig_pmas:EELR_det} for all objects. The ground-based images clearly suffer from their low spatial resolution, so that the models are strongly affected by PSF mismatches in compact galaxies. Such mismatches sometimes lead to non-physically high Sersi\'c indices. The comparison of the SDSS $i$ and $r$ images allow us to judge the robustness of the results. Strongly different results indicate that the structural parameters are not well constrained. Furthermore, a close apparent companion affected the modelling of SDSS~J1655+2146, so that a meaningful model could only be inferred for the host galaxy with a fixed Sersi\'c index.

\subsubsection{Morphological classification}
Discriminating between a bulge- and a disc-dominated QSO host galaxy is not straightforward, even for the high spatial resolution images obtained with HST. A commonly used criterion is the best-fitting Sersi\'c index $n$, where $n>2.5$ indicates a bulge-dominated and $n<2.5$ indicates a disc-dominated systems \citep[e.g.][]{Sanchez:2004b,Bell:2004b}. Alternatively, the lowest $\chi^2$ value is computed for a bulge and a disc-model to identify the best matching model \citep[e.g.][]{McLeod:2001,Sanchez:2003,Dunlop:2003,Jahnke:2003,Guyon:2006}. However, the statistical significance of the favoured model compared to other options needs to be considered. On the other hand, a visual classification of the host is often preferred to identify spiral structures of disc-dominated systems, or apparently disturbed merging system.

For the HST images, the $\chi^2$ values robustly discriminate between the models. We adopt the criterion by \citet{Guyon:2006}, which demands a significant difference in the reduced $\chi^2$ of $|\chi_\nu^2(n=1)-\chi_\nu^2(n=4)|>0.05$. The $\chi^2$ classification always agreed with the one based solely on the best-fit Sersi\'c index and we achieved good agreement with literature results. The situation is more difficult for the ground-based observations, where the $\chi^2$ values sometimes differ only marginally. A visual inspection of the images, together with large axis ratios ($b/a$), may provide evidence in some cases for disc-dominated hosts even when a bulge-dominated model was formally preferred as in the cases of RBS~219, PG~1612$+$261, PHL~1811 and HE~2353$-$0420. When the best-fitting Sersi\'c index is $n>7$, we doubt that any reliable classification can be made, because flux transfer likely happened between the PSF and the host galaxy component. In the other cases, we refer to the best-fitting Sersi\'c indices for our classification, but consider these classifications as tentative.

From the 28 QSOs with available broad-band imaging, we classified 17 host galaxies into 5 bulge-dominated, 8 disc-dominated and 4 ongoing merging galaxies. 5 of the remaining hosts are likely bulge-dominated systems and one is probably disc-dominated. For the rest of the host galaxies we were not able to constrain the morphology. Our sample seems to be almost evenly split up into bulge- and disc-dominated systems without any preference. A similar balance of morphological types were also reported for completely different samples of QSO host galaxies \citep[e.g.][]{Jahnke:2004,Cisternas:2011}.

\section{Spatially resolved spectroscopy of QSOs}\label{sect_pmas:IFU_data}
\subsection{PMAS IFU observations}
\onltab{
\begin{table*}\centering
\begin{footnotesize}
\caption{PMAS IFU observational log.}
\label{tbl:pmas_obslog}
\begin{tabular}{lcccccc}\hline\hline\noalign{\smallskip}
Object		& Obs. Date	& Exp. Time [s]	& Grating & Wavelength Range
(\AA)  & Seeing		 & Obs. Condition \\\hline\noalign{\smallskip}
PG 0026$+$129	& 2002--09--05  & 1$\times$1200, 1$\times$2400	& V300 & 4600--7800 & 1\farcs1	         & photometric    \\
I Zw 1  	& 2002--09--03  & 1$\times$900, 2$\times$1800	& V300 & 3900--7200 & 1\farcs0--1\farcs4 & mainly clear    \\
PG 0052$+$251	& 2002--09--06  & 2$\times$1800			& V300 & 4600--7800 & 1\farcs0	    	 & photometric    \\
SDSS J0057+1446 & 2008--09--05  & 5$\times$1800                 & V600 & 5100--6600 & 1\farcs5           & part. cloudy   \\ 
RBS 219		& 2002--09--02  & 1$\times$1200, 1$\times$1800	& V300 & 4600--7800 & 1\farcs2--1\farcs8 & photometric    \\
SDSS J0155$-$0857& 2008--09--07 & 4$\times$1800                 & V600 & 5100--6600 & 1\farcs4--1\farcs8 & photometric  \\ 
                 &              & 2$\times$1800                 &      &            & 1\farcs2           & photometric  \\ 
HE 0157$-$0406	& 2002--09--02	& 1$\times$1200, 1$\times$1800  & V300 & 5400--8600 & 1\farcs2--1\farcs3 & photometric    \\
Mrk 1014	& 2002--09--07	& 1$\times$1200, 1$\times$1800  & V300 & 5400--8600 & 1\farcs2--1\farcs3 & photometric    \\
                & 2008--09--06  & 4$\times$1800                 & V600 & 5100--6600 & 1\farcs4           & mainly clear \\ 
SDSS J0836+4426 & 2008--03--16  & 3$\times$1800                 & V600 & 5500--7000 & 1\farcs1           & part. cloudy \\
SDSS J0948+4335 & 2008--03--13  & 3$\times$1800                 & V600 & 5250--6750 & 1\farcs1           & cloudy \\
SDSS J1131+2623 & 2008--03--13  & 3$\times$1800                 & V600 & 5250--6750 & 1\farcs1           & part. cloudy \\
PKS 1217$+$023	& 2003--05--02  & 4$\times$1800			& V300 & 5600--8900 & 1\farcs0--1\farcs2 & mainly clear \\
3C 273  	& 2003--05--03  & 3$\times$1200			& V300 & 4350--7650 & 1\farcs4--1\farcs9 & mainly clear \\
HE 1228$+$0131  & 2003--05--04  & 4$\times$1800	 	        & V300 & 4350--7650 & 1\farcs0--1\farcs3 & part. cloudy    \\
SDSS J1230+6621 & 2008--03--16  & 2$\times$1800                 & V600 & 5000--6500 & 1\farcs0           & mainly clear \\   
SDSS J1230+1100 & 2008--03--13  & 1$\times$1800                 & V600 & 5250--6750 & 1\farcs1           & cloudy \\   
PG 1427+490     & 2003--05--01  & 3$\times$1200                 & V300 & 5600--8900 & 0\farcs9--1\farcs2 & mainly clear \\
SDSS J1444+0633 & 2008--03--16  & 1$\times$1800                 & V600 & 5250--6750 & 1\farcs3--1\farcs5 & cloudy   \\
HE 1453$-$0303  & 2003--05--02  & 2$\times$1800		        & V300 & 5600--8900 & 1\farcs2--1\farcs5 & mainly clear    \\
		& 2003--05--03  & 2$\times$1800			&      &            & 1\farcs3--1\farcs6 & mainyl clear \\
PKS 1545$+$21	& 2003--05--03  & 3$\times$1800			& V300 & 5600--8900 & 1\farcs2--1\farcs5 & mainly clear \\
PG 1612+261	& 2008--09--07  & 3$\times$1800			& V600 & 5100--6600 & 1\farcs1           & mainly clear \\
SDSS J1655+2146 & 2008--03--16  & 3$\times$1200                 & V600 & 5000--6500 & 1\farcs3--1\farcs6 & part. cloudy \\
                &               & 1$\times$1800                 &      &            & 1\farcs0           & photometric  \\
PG 1700$+$518	& 2003--05--02  & 2$\times$1800			& V300 & 5600--8900 & 0\farcs9--1\farcs2 & photometric  \\
		& 2003--05--03  & 2$\times$1800			&      &            & 1\farcs6    	 & mainly clear \\
PG 2130$+$099	& 2002--09--04  & 2$\times$2400			& V300 & 3900--7200 & 1\farcs3           & photometric \\
PHL 1811	& 2002--09--06  & 2$\times$1800			& V300 & 5400--8600 & 1\farcs2	    	 & photometric \\
HE 2158$-$0107  & 2002--09--06  & 2$\times$1800		        & V300 &
5400--8600 & 1\farcs1	         & photometric \\
RBS~1812	& 2002--09--04  & 1$\times$2400                 & V300 & 4600--7800 & 1\farcs5	& photometric   \\
                & 2002--09--05  & 2$\times$2400                 &      &   & 1\farcs6-2\farcs0	& mainly clear   \\
PG 2214$+$139	& 2002--09--02  & 2$\times$2400, 1$\times$1800  & V300 & 3900--7200 & 1\farcs3--1\farcs4 & photometric \\
HE 2307$-$0254	& 2002--09--06  & 2$\times$1800			& V300 & 5400--8600 & 1\farcs2           & photometric \\
PKS 2349$-$014	& 2002--09--06  & 3$\times$900			& V300 & 5400--8600 & 1\farcs0--1\farcs1 & photometric \\
HE 2353$-$0420	& 2002--09--07  & 2$\times$1800			& V300 & 5400--8600 & 1\farcs1--1\farcs4 & photometric \\\noalign{\smallskip}\hline
\end{tabular}

\end{footnotesize}
\end{table*}
}
IFU observations for the initial sample were carried out with PMAS in the nights from the 2th to 7th of April in 2002 and from the 2nd to 4th of May 2003. The QSOs of the extended sample were observed in the nights from the 13th to 17th of March and from the 5th to 8th of September in 2008. In all cases we used the PMAS lens-array with the highest magnification providing a seeing-limited sampling of 0\farcs5$\times$\,0\farcs5 with a field of view (FoV) of 8\arcsec\,$\times$\,8\arcsec\, for the 256 spaxels\footnote{spectral pixel, a single spatial resolution element containing a spectrum along the entire wavelength range}. The host galaxies of the observed QSOs effectively fill the FoV in most cases (see Fig.~\ref{fig_pmas:EELR_det}), but are significantly larger for the three nearest objects ($z<0.10$) and a few extended systems at higher redshift.

The low resolution V300 grating was used for the observations of the initial QSO sample covering a large wavelength range including the major diagnostic spectral features around the H$\beta$ and H$\alpha$ emission lines simultaneously. Higher spectral reso\-lution observations with the V600 grating were taken for the extension of the sample, but then centred on the H$\beta$ emission line only. 
For both setups we estimated the spectral resolution from the \ion{O}{i}\,$\lambda 5577$ night-sky emission line to be 6.1\,\AA\ and 3.2\,\AA\ Full-Width at Half Maximum (FWHM), respectively. At least two exposures were usually taken for each QSO with individual integration times  in the range of 900\,s up to 2400\,s reaching a total integration time of 1--2 hours per object.  Table~\ref{tbl:pmas_obslog} summarises the details of the observations including notes on the ambient observing conditions.

A set of arc lamp (HgNe) and continuum lamp exposures were acquired before or after the target frames for fibre tracing and wavelength calibration. Skyflats were taken during twilight to measure the variations in the fibre-to-fibre transmission, and  spectro-photometric standard stars were regularly observed at the beginning and at the end of the nights with the same instrumental setups as the QSOs.

\subsection{Data reduction}
The basic data reduction was performed with the \texttt{R3D} reduction package \citep{Sanchez:2006a} designed to reduce fibre-fed integral field spectroscopy data. We used the independent modular tasks of the package for the following basic reduction steps: 1. bias subtraction, 2. fibre tracing on the continuum lamp exposure, 3. extraction of the fibre spectra, 4. wavelength calibration with the arc-lamp exposures, 5. fibreflat correction based on observed skyflats, 6. flux calibration of the spectra using spectro-photometric standard star observations, and 7. creation of a data\-cube from the row-stacked spectra.

Prior to the reduction we detected and removed cosmic rays hits from the individual raw frames with the \texttt{DCR} routine \citep{Pych:2004} that was designed originally for long-slit spectroscopic data. We used a  conservative configuration of \texttt{DCR} such that night sky lines or emission-line features of our target itself were not removed. A few prominent cosmic ray hits inevitably remained in the data given the imperfection of the algorithm for IFU data.  Thus, we always required that emission lines must be spatially extended, that they have a line width equal or greater than the instrumental resolution, and that the doublet line ratios are consistent with their theoretical values.

As discussed in \citet{Roth:2005}, the PMAS instrument suffers from flexure depending on the hour angle of the observations. This inevitably leads to shifts between different exposures of the same target in dispersion and/or cross-dispersion direction of the spectra on the charge-coupled device (CCD) detector. Flexure shifts were measured with the \texttt{IRAF} task \emph{imalign} by comparing the position of $\sim$20 sky emission lines distributed over the whole CCD in the raw image. When only one set of calibration frames (continuum+arc lamp) was available for a sequence of observations, we resampled the calibration frames to match with position of the individual exposures.  The science frame taken right after or before the calibration frames was used as the reference for the zero position.

A median sky spectrum was created from spaxels containing no emission lines from the host galaxy and subsequently subtracted from the whole datacube. Given that several QSO hosts fill the entire PMAS FoV we accepted a contamination of the sky spectrum by stellar continuum emission in these cases. This has no strong impact on our emission line analysis, but the Balmer emission lines need to be corrected for the underlying stellar absorption lines. How the inaccurate sky subtraction affects our emission-line measurements will be addressed in Sect.~\ref{sect:sky_subtr}. 

The datacubes were then corrected for differential atmospheric refraction (DAR) by resampling each monochromatic slice such that the position of the QSO is aligned to a common reference point at all wavelengths. All standard star observations were used to create a mean sensitivity function for each run and instrument setup, so that the datacubes are properly flux calibrated taking into account the airmass and the atmospheric extinction law at the Calar Alto observatory \citep{Sanchez:2007b}. Finally, we combined the datacubes using either a variance weighted mean or a clipping method whenever 3 or more exposures were available to reject remaining cosmic rays and artefacts.

Most of the data were taken under clear or photometric conditions. In order to check the quality and uncertainty of the spectrophotometry, we compared the flux-calibrated QSO spectra with available SDSS spectra taking into account the  size of SDSS fibres. The deviation between the QSO spectra in the continuum over the whole wavelength range was $\leq 20\%$, which we adopt as our systematic spectrophotometric uncertainty.

\subsection{QSO--host deblending with \texttt{QDeblend${}^{\mathrm{3D}}$}}
Similar to the deblending of the broad-band images we need to decompose the QSO and host emission also in our IFU data to study and interpret the emission of the host galaxy without the QSO contamination. The three-dimensional nature of the IFU date requires new analysis techniques and software. We developed the software tool \texttt{QDeblend${}^{\mathrm{3D}}$}\ for this particular task\footnote{available for download at http://sf.net/projects/qdeblend/}. At the core of the program is an iterative deblending algorithm which is a significant improvement of a simple technique introduced by \citet{Christensen:2006}.

The basic concept is to treat the spaxels of the IFU datacube as a set of independent spectra rather than as a sequence of  monochromatic images. In the absence of atmospheric dispersion the spectrum of a point source would then be the same in each spaxel, multiplied by a scale factor according to the PSF. In the case of QSO observations, the scale factors of the PSF can be directly determined from the strength of the broad emission lines with respect to the adjacent continuum as described by \citet{Jahnke:2004}. Here we outline the iterative algorithm implemented in \texttt{QDeblend${}^{\mathrm{3D}}$}. A more detailed explanation with illustrative diagrams of the iterative process can be found in the user manual of \texttt{QDeblend${}^{\mathrm{3D}}$}.

In a first step a high S/N nuclear QSO spectrum is extracted from the datacube. After a PSF has been determined from a broad QSO emission line,  the nuclear QSO spectrum is subtracted from each spaxel after multiplied by the corresponding PSF scale factor. However, significant over-subtraction will occur when the nuclear QSO spectrum is contaminated by host galaxy light. In \texttt{QDeblend${}^{\mathrm{3D}}$}\ we minimize this host galaxy contamination by iteratively estimating a \emph{mean} host galaxy spectrum from the residual datacube that is subsequently subtracted from the QSO spectrum. A rectangular aperture and surrounding annulus is used as a default to extract QSO and host galaxy spectra, but any set of spaxels for the QSO and host galaxy spectrum could be specified. Because the host galaxy becomes brighter towards the galaxy centre, the mean host spectrum from the annulus should be scaled in brightness before it is subtracted from the initial nuclear QSO spectrum. That scale factor is unconstrained from the IFU data itself. One may either adopt an a-priori defined analytic surface brightness profile, a constant surface brightness, or a manually set scale factor which fulfils a certain criterion. The iteration is stopped when the decontaminated nuclear QSO spectrum converges.

For our PMAS spectra we use the broad H$\alpha$ or H$\beta$ lines of the QSOs to perform a QSO-host deblending with \texttt{QDeblend${}^{\mathrm{3D}}$}. Because we focus on the emission lines of the ionised gas around H$\beta$ and H$\alpha$ region in this work, we restrict the QSO-host deblending to this narrow wavelength regions around the two Balmer lines. This almost entirely avoids the problem of PSF mismatch due to the wavelength dependence  of the atmospheric seeing. The nuclear QSO spectra are taken from the brightest spaxel in the data, and  the surrounding 8 spaxels are used to estimate a mean host galaxy spectrum from the residuals. Here we assume a constant surface brightness to extrapolate the host galaxy emission towards the central spaxel. The QSO spectra converged already after four iterations at which we stopped the iterative process.

\subsection{Imperfect sky subtraction and Balmer absorption}\label{sect:sky_subtr}
One problem with our PMAS IFU observations is the imperfect subtraction of the sky background. Because no blank sky fields were observed together with the target, we were forced to estimate the sky directly from the target datacubes itself. We used spatial regions free of extended emission lines to avoid any over-subtraction of these lines, but those regions may contain significant emission from stellar continuum if the host galaxy extends across the PMAS FoV. Thus, the stellar continuum is over-subtracted in several cases for which we were unable to reliably infer the ``real'' stellar continuum from our data.

\begin{figure}
\resizebox{\hsize}{!}{\includegraphics[clip]{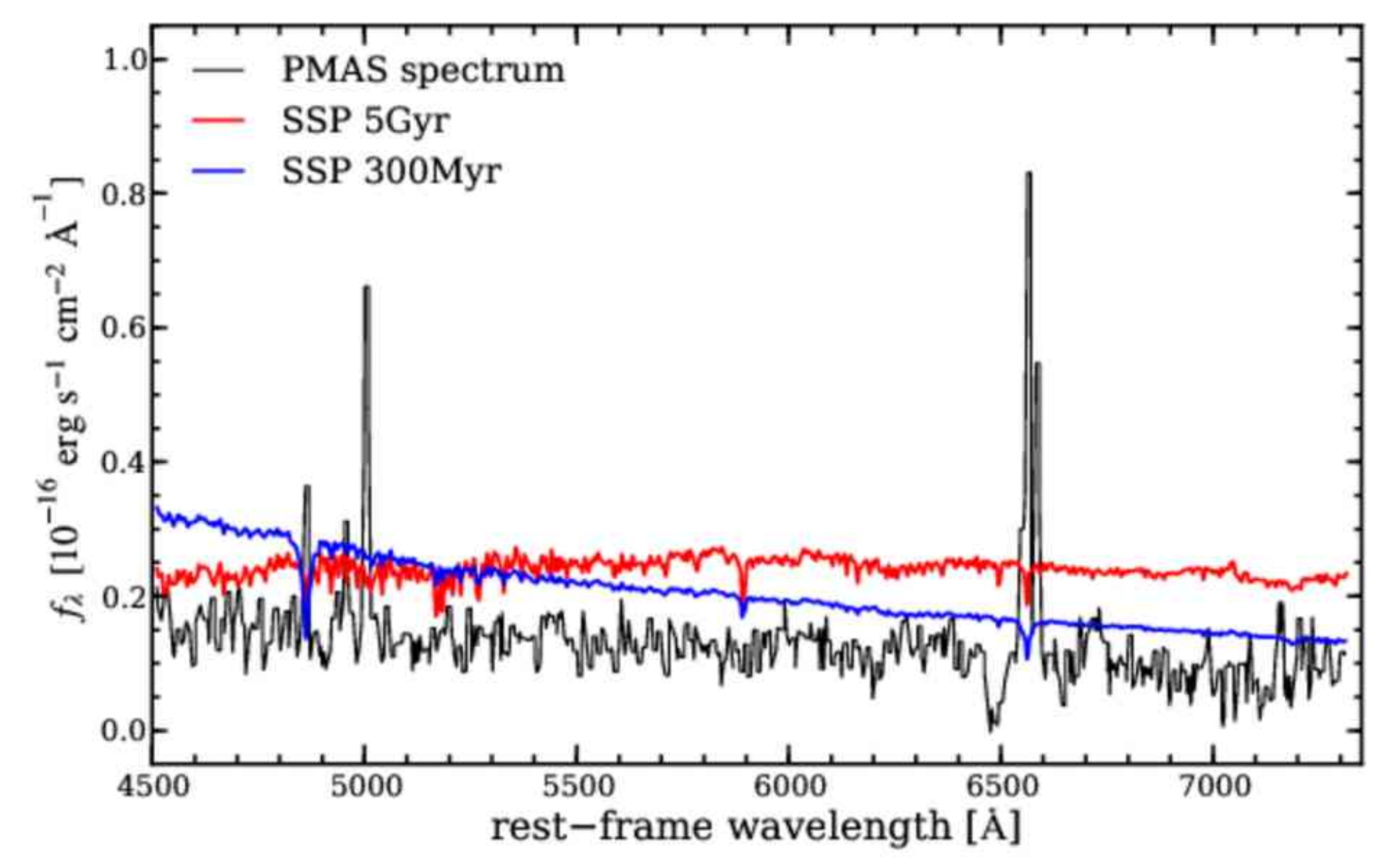}}
\caption{Comparing the PMAS spectra of PKS~2349$-$014 (region A) with two SSP model spectra matched to the HST photometry. The systematic offset between the SSP models and the PMAS spectra is caused by an over-subtraction of the sky background in the PMAS observations.}
\label{fig_pmas:HST_stellar_cont}
\end{figure}
\begin{figure}
\resizebox{\hsize}{!}{\includegraphics[clip]{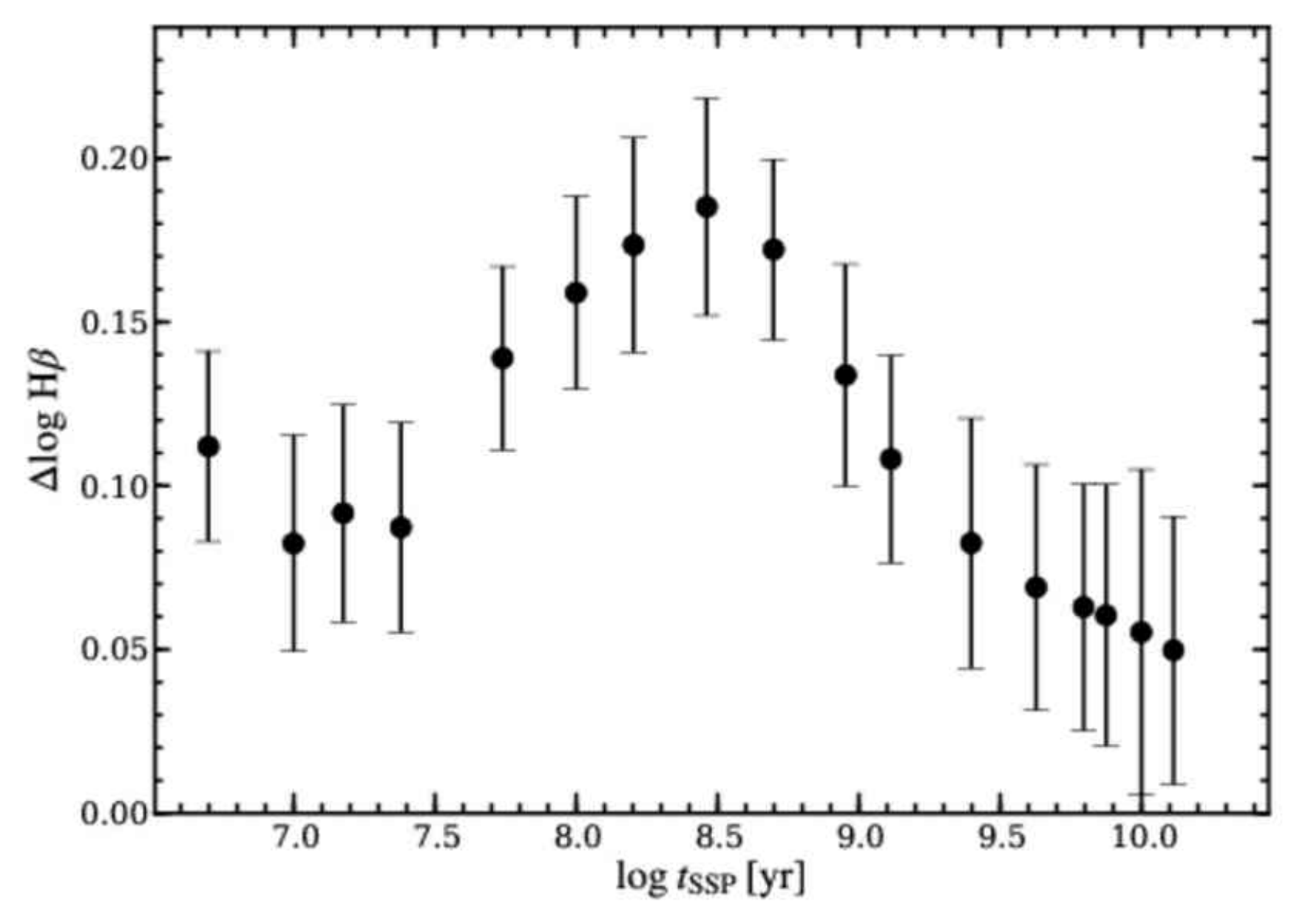}}
\caption{Underestimation of the $\mathrm{H}\beta$ emission line flux ($\Delta \log(\mathrm{H}\beta)$) due to the stellar absorption line as a function of the stellar population age for the PMAS spectrum of PKS~2349$-$014 (region A).}
\label{fig_pmas:Hbeta_cont_age}
\end{figure}

\begin{figure*}
\centering
 \includegraphics[width=0.95\textwidth,clip]{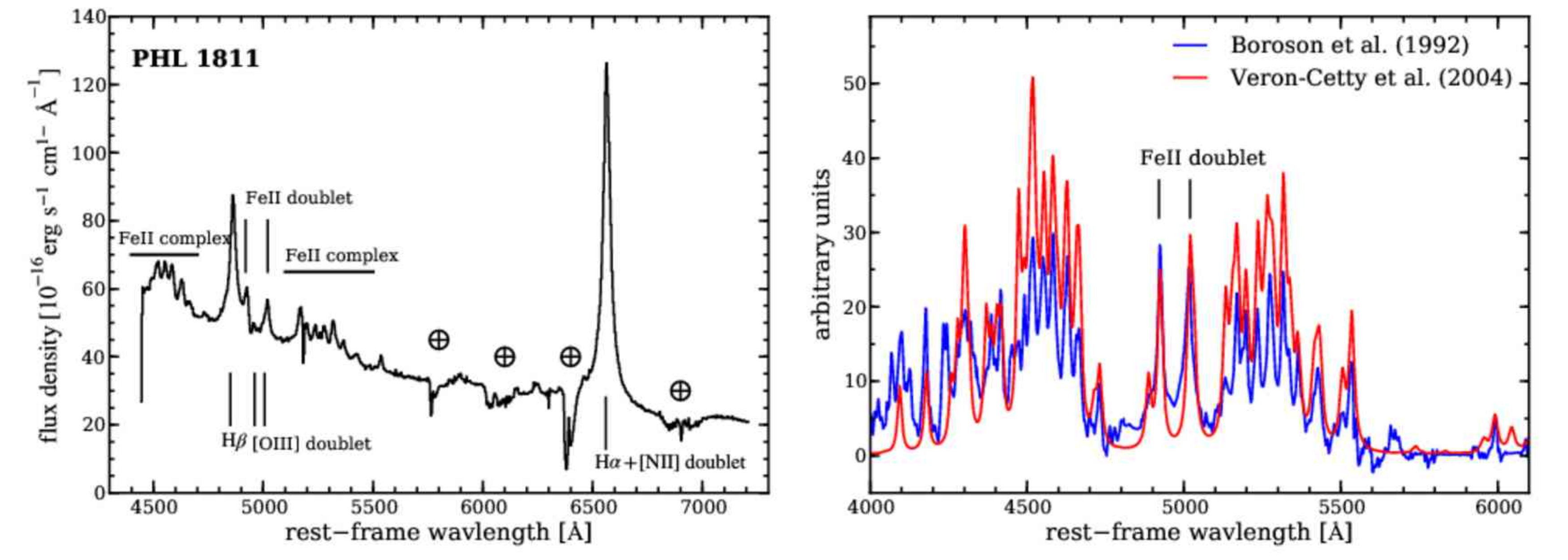}
\caption{\emph{Left panel:} QSO spectrum of PHL~1811 observed with PMAS as a representative example for a strong \ion{Fe}{II} emitter. The available wavelength range of the V300 grating was set such that H$\beta$ and H$\alpha$ would be covered at the redshift of the QSO. The \ion{Fe}{II} complex blueward of H$\beta$ is not fully covered for PHL~1811, as in many other objects. \emph{Right panel:} Comparison of two available \ion{Fe}{ii} templates from \citet{Boroson:1992} and \citet{Veron-Cetty:2004} both based on the proto-typical QSO I~Zw~1. Both templates were scaled to match the isolated \ion{Fe}{ii} $\lambda 5014$ line. The disagreement between the templates is severe given that they were generated from the same QSO.}
\label{fig_pmas:full_QSO_Fe}
\end{figure*}

Our PMAS observations were primarily targeted to map the emission lines of the ionised gas. While it would be also interesting to study the stellar populations of the QSO hosts from their continuum emission, we find that given our exposure times, the S/N is generally too low to quantitatively interpret the stellar absorption patterns. Nevertheless, the measured strength of the Balmer emission lines will depend on the depth of the underlying stellar absorption, which we take into account when using the Balmer lines for diagnostic purposes.
In order to explore the impact of the Balmer absorption line on our measurements, we combined synthetic single stellar population (SSP) models from \citet{Bruzual:2003} with the absolute surface brightness photometry of the HST broad-band images. In a first step, we convolved the nucleus-subtracted HST images with the PSF of the corresponding PMAS observation and rebinned the image to the spatial sampling of our PMAS data. Afterwards, we scaled a suite of SSP spectra to the surface brightness in several specific regions. In Fig.~\ref{fig_pmas:HST_stellar_cont} we compare two SSP models (300\,Myr and 5Gyr) matched to the \emph{HST} photometry with the PMAS spectrum of PKS~2349$-$014 (region A as defined in Fig.~\ref{fig_pmas:EELR_det}) as an example. The PMAS spectrum is well below both SSP model spectra, which can be entirely attributed to the over-subtracted sky background. 

The strengths of the Balmer absorption lines depend on the age of the stellar population.
We measured the differences between the H$\beta$ emission line fluxes of the continuum subtracted and unsubtracted PMAS spectra expressed as $\Delta \log(\mathrm{H}\beta) = \log(\mathrm{H}\beta_{\mathrm{sub.}})-\log(\mathrm{H}\beta_{\mathrm{unsub.}})$ for 18 SSP spectra with ages of the order of $300\,\mathrm{Myr}<t_\mathrm{SSP}< 5\,\mathrm{Gyr}$ assuming solar metallicity. We show the dependence of $\Delta\mathrm{H}\beta$ on $t_\mathrm{SSP}$ for PKS~2349$-$014 (region A) as an example in Fig.~\ref{fig_pmas:Hbeta_cont_age}. We find that the H$\beta$ emission line flux is underestimated due to the corresponding stellar absorption line by 0.07--0.2\,dex in this case of PKS~2349$-$014. The redshift of that object ($z=0.174$) is close to the median redshift of our sample and representative for the majority of objects. The H$\beta$ emission-line flux is more severely affected for the few object at $z=0.06$ given the lower equivalent width of their emission line.

Because of the unknown age of the stellar population, we used the average of $\Delta \log(\mathrm{H}\beta)$ taken over all SSP ages with an error corresponding to the possible range of values. For the host that fill the PMAS FoV and were not observed with \emph{HST}, we corrected all \mbox{H$\beta$}\ lines by an average factor of $\Delta \log(\mathrm{H}\beta)=0.13\pm0.08$\,dex, determined from all objects with reliable \emph{HST} photometry. Although the $\mathrm{H}\alpha$ emission line is also affected by the stellar absorption, we made no correction given that the correction factor ($0.03\pm0.01$\,dex) is negligible for our purposes.

\section{AGN parameters from the QSO spectra}\label{sect_pmas:QSO_spec}
Our PMAS data provide high S/N spectra of the  QSOs without any wavelength-dependent slit losses, because the IFU covers the entire seeing disc of a point source even under bad seeing conditions. We use those spectra to derive some fundamental AGN parameters, like redshifts, black hole masses $M_\mathrm{BH}$, and Eddington ratios $\lambda=L_\mathrm{bol}/L_\mathrm{Edd}$. The contamination of the QSO spectrum by host galaxy light was removed or at least minimised by our QSO--host deblending process described above. For a few QSOs where the SDSS spectra had significantly higher S/N we used these instead. From the SDSS broad-band images we estimated a host contamination of  $\lesssim$15\% in the SDSS spectra and corrected the continuum luminosities correspondingly.

\subsection{Spectral modelling of the \mbox{H$\beta$}--\mbox{[\ion{O}{iii}]}\ region}
All of the AGN parameters of interest were calibrated to spectral features around the broad \mbox{H$\beta$}\ line. This wavelength region is bracketed by two adjacent \mbox{\ion{Fe}{ii}}\ complexes as shown in Fig.~\ref{fig_pmas:full_QSO_Fe} (left panel). Most studies of QSO spectra \citep[e.g.][]{Boroson:1992,Greene:2005,Woo:2006,McGill:2008} used a template generated from the proto-typical QSO I~Zw~1 \citep{Boroson:1992,Veron-Cetty:2004} to subtract the \mbox{\ion{Fe}{ii}}\ complexes over a large wavelength range. We cannot apply this \mbox{\ion{Fe}{ii}}\ template fitting method here in a consistent manner because most of our QSO spectra do not reach enough to the blue to reliably estimate the underlying power-law continuum. We also note that the two available \mbox{\ion{Fe}{ii}}\ templates significantly disagree (Fig.~\ref{fig_pmas:full_QSO_Fe} left panel). However, the dominant \mbox{\ion{Fe}{ii}}\ emission in the proximity of \mbox{H$\beta$}\ and \mbox{[\ion{O}{iii}]}\ originates in both cases from two isolated \mbox{\ion{Fe}{ii}}\ lines, enabling us to choose a more simpler approach to handle the \mbox{\ion{Fe}{ii}}$\,\lambda\lambda4929,5018$ lines over a restricted wavelength range.

We thus modelled only the wavelength region of 4700--5100\AA, which encompasses \mbox{H$\beta$}, \mbox{[\ion{O}{iii}]}\ and almost pure continuum regions on either side. A superposition of Gaussians was used to represent the line profile of the broad and narrow emission lines including the two prominent \mbox{\ion{Fe}{ii}}$\,\lambda\lambda4929,5018$ lines. The full model consists of up to 3 Gaussian components for the \mbox{H$\beta$}\ line, up to 2 Gaussian components for the profile of the \mbox{[\ion{O}{iii}]}\ and \mbox{\ion{Fe}{ii}}\ doublets and a straight line to approximate the local continuum. The two lines of a doublet are coupled as a system in line width, radial velocity and line ratio (a factor of 3 for $\mbox{[\ion{O}{iii}]}\,\lambda\lambda 4960,5007$ and 1.29 for  $\mbox{\ion{Fe}{ii}}\,\lambda\lambda4924,5018$), which significantly increases the robustness of our model and reduces the number of free parameters. If a narrow \mbox{H$\beta$}\ component is present on top of a broad \mbox{H$\beta$}\ emission profile, we couple its kinematic parameters with that of the narrow \mbox{[\ion{O}{iii}]}\ component as these lines originate both from the NLR. In cases where the \mbox{\ion{Fe}{ii}}\ lines were broad and their width could hardly be constrained against the background continuum, we coupled the line width and radial velocity to that of the broad \mbox{H$\beta$}\ line.
\begin{figure}[!ht]
\centering
\includegraphics[clip,height=6cm]{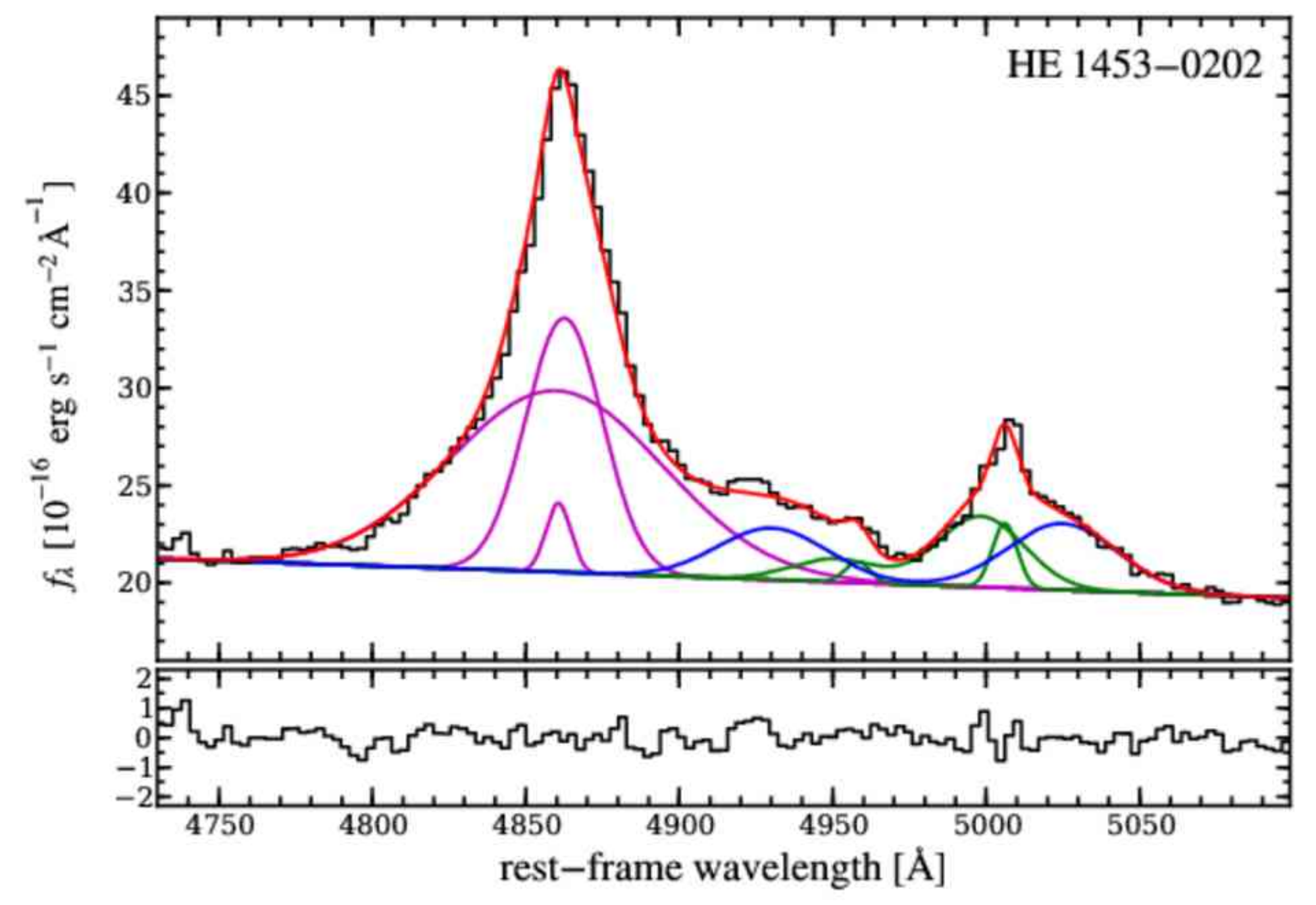}
\centering
\includegraphics[clip,height=6cm]{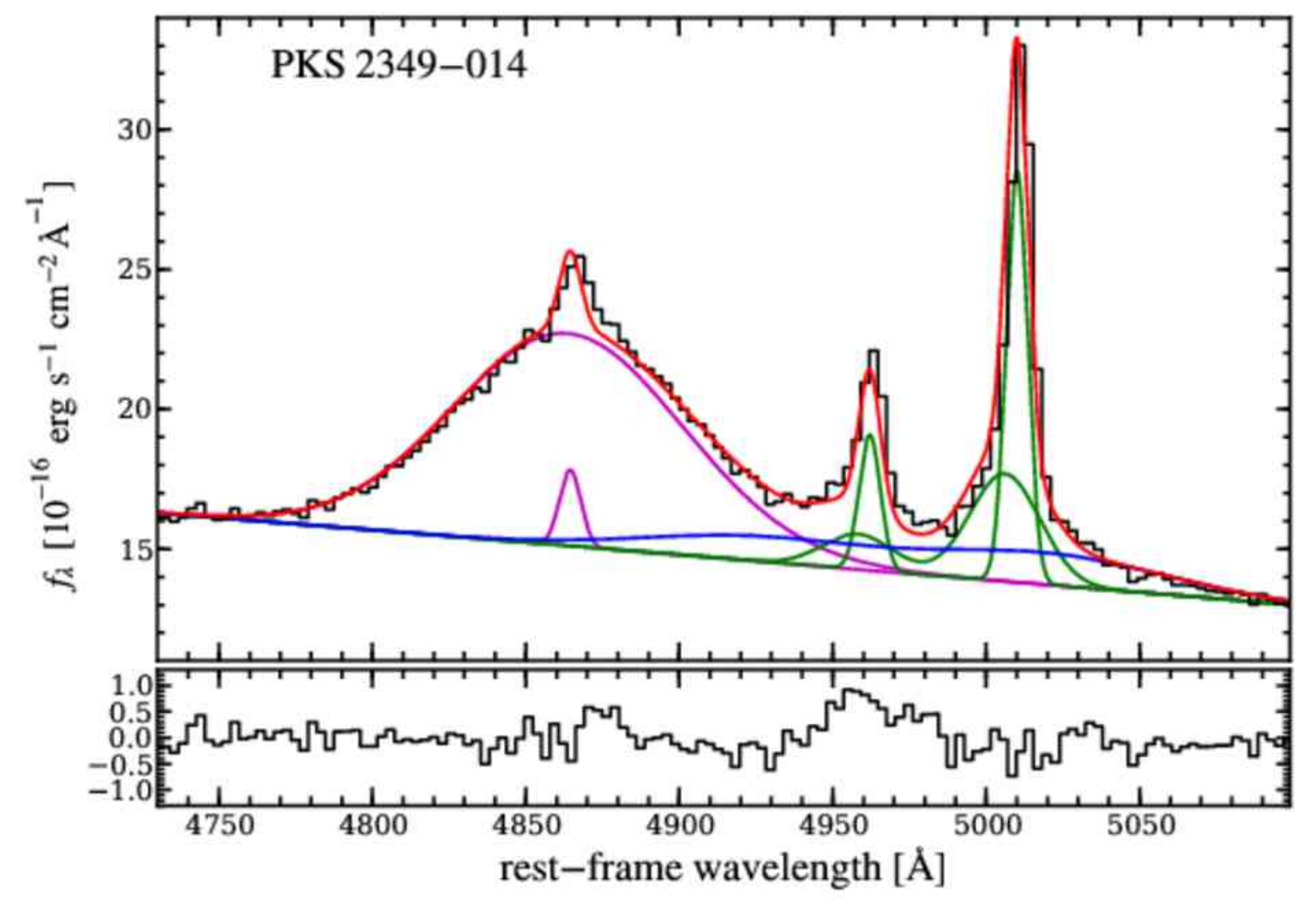}
\caption{Spectral modelling of the \mbox{H$\beta$}--\mbox{[\ion{O}{iii}]}\ region for two representative QSO spectra. The black line correspond to the observed QSO spectra and the red curve represents our best fit model. The model consists of individual Gaussian components for the emission lines (\mbox{H$\beta$}--magenta, \mbox{[\ion{O}{iii}]}--green, \mbox{\ion{Fe}{ii}}--blue) above a straight line for the local continuum. The residual of the model are shown in a dedicated panel below.}
\label{fig_pmas:QSO_fit}
\end{figure}

\begin{table*}\centering
\begin{footnotesize}
\caption{AGN parameters measured or derived from the QSO spectrum}
\label{tbl_pmas:bhmass}
\begin{tabular}{lcccccccc}\hline\hline\noalign{\smallskip}
Object & $f_{\mathrm{NLR},[\ion{O}{iii}]}$ & $\log L_{\mathrm{NLR},[\ion{O}{iii}]}$ & $f_{5100}$  &  $\log L_{5100}$ & $FWHM_{\mathrm{H}\beta}$ & $\sigma_{\mathrm{H}\beta}$ &  $\log\left(M_{\mathrm{BH}}\right)$ & $\log \lambda$\\
  & $\left[10^{-16}\frac{\mathrm{erg}}{\mathrm{s}\,\mathrm{cm}^2}\right]$ & $[\mathrm{erg}\,\mathrm{s}^{-1}]$ & $\left[10^{-16}\frac{\mathrm{erg}}{\mathrm{s}\,\mathrm{cm}^2\,\mathrm{\AA}}\right]$ & $[\mathrm{erg}\,\mathrm{s}^{-1}]$ & $[\mathrm{km}\,\mathrm{s}^{-1}]$ & $[\mathrm{km}\,\mathrm{s}^{-1}]$  & [$M_{\sun}$] &    \\ \noalign{\smallskip}\hline\noalign{\smallskip}
PG 0026$+$129 & $505\pm3$ & $42.54\pm0.09$ &  $27.5\pm 6.7$ & $44.96\pm0.11$ & $2430\pm 28$ & $3471\pm 18$ & $9.0$ & $-1.2$ \\
I Zw 1 & $1667\pm13$ & $42.23\pm0.09$ &  $45.8\pm 11.3$ & $44.31\pm0.11$ & $1111\pm 12$ & $1652\pm 10$ & $8.0$ & $-0.9$ \\
PG 0052$+$251 & $667\pm2$ & $42.69\pm0.09$ &  $25.1\pm 5.8$ & $44.98\pm0.10$ & $4604\pm 24$ & $2088\pm 8$ & $8.6$ & $-0.7$ \\
SDSS J0057$+$1446 & $203\pm14$ & $42.27\pm0.12$ &  $18.6\pm 4.2$ & $44.95\pm0.10$ & $10506\pm 159$ & $4143\pm 39$ & $9.1$ & $-1.3$ \\
HE 0132$-$0441 & $200\pm33$ & $42.16\pm0.16$ &  $12.6\pm 2.9$ & $44.69\pm0.10$ & $1487\pm 21$ & $960\pm 47$ & $7.7$ & $-0.2$ \\
HE 0157$-$0406 & $82\pm2$ & $42.08\pm0.10$ &  $6.2\pm 1.3$ & $44.72\pm0.09$ & $5493\pm 174$ & $2333\pm 77$ & $8.5$ & $-1.0$ \\
SDSS J0155$-$0857 & $211\pm11$ & $42.23\pm0.11$ &  $6.7\pm 1.4$ & $44.47\pm0.09$ & $6650\pm 87$ & $2824\pm 35$ & $8.6$ & $-1.2$ \\
Mrk 1014 & $901\pm10$ & $42.85\pm0.09$ &  $13.7\pm 3.0$ & $44.77\pm0.09$ & $3002\pm 25$ & $1549\pm 16$ & $8.2$ & $-0.6$ \\
SDSS J0846$+$4426 & $1038\pm53$ & $43.34\pm0.11$ &  $17.9\pm 3.9$ & $45.35\pm0.10$ & $4169\pm 128$ & $1771\pm 56$ & $8.6$ & $-0.4$ \\
SDSS J0948$+$4335 & $179\pm5$ & $42.45\pm0.10$ &  $5.4\pm 1.2$ & $44.72\pm0.09$ & $4060\pm 61$ & $1571\pm 30$ & $8.2$ & $-0.6$ \\
SDSS J1131$+$2623 & $252\pm33$ & $42.68\pm0.14$ &  $3.4\pm 0.7$ & $44.59\pm0.09$ & $9370\pm 328$ & $3957\pm 128$ & $8.9$ & $-1.5$ \\
PKS 1217$+$023 & $281\pm2$ & $42.71\pm0.09$ &  $14.6\pm 3.2$ & $45.20\pm0.10$ & $4300\pm 178$ & $2272\pm 686$ & $8.7$ & $-0.7$ \\
3C 273 & $996\pm32$ & $42.87\pm0.10$ &  $272.7\pm 59.2$ & $46.04\pm0.09$ & $2852\pm 7$ & $1999\pm 7$ & $9.1$ & $-0.2$ \\
SDSS J1230$+$1100 & $241\pm7$ & $42.64\pm0.10$ &  $5.1\pm 1.1$ & $44.73\pm0.10$ & $4439\pm 129$ & $2307\pm 30$ & $8.5$ & $-0.9$ \\
HE 1228$+$0131 & $454\pm27$ & $42.23\pm0.11$ &  $44.5\pm 9.7$ & $44.95\pm0.09$ & $1671\pm 11$ & $1572\pm 38$ & $8.3$ & $-0.5$ \\
SDSS J1230$+$6621 & $447\pm7$ & $42.65\pm0.09$ &  $5.9\pm 1.2$ & $44.54\pm0.09$ & $3297\pm 93$ & $2190\pm 22$ & $8.4$ & $-1.0$ \\
PG 1427$+$480 & $240\pm5$ & $42.55\pm0.10$ &  $5.3\pm 1.1$ & $44.67\pm0.09$ & $2455\pm 68$ & $1268\pm 105$ & $8.0$ & $-0.4$ \\
SDSS J1444$+$0633 & $447\pm7$ & $42.65\pm0.09$ &  $5.9\pm 1.2$ & $44.54\pm0.09$ & $3297\pm 93$ & $2190\pm 22$ & $8.4$ & $-1.0$ \\
HE 1453$-$0303 & $195\pm13$ & $42.53\pm0.12$ &  $27.4\pm 7.9$ & $45.31\pm0.12$ & $2513\pm 43$ & $1795\pm 27$ & $8.6$ & $-0.4$ \\
PKS 1545$+$21 & $286\pm2$ & $42.84\pm0.09$ &  $14.1\pm 3.1$ & $45.29\pm0.10$ & $5644\pm 38$ & $2397\pm 15$ & $8.8$ & $-0.7$ \\
PG 1612$+$261 & $2591\pm17$ & $43.14\pm0.09$ &  $21.2\pm 5.1$ & $44.74\pm0.11$ & $2770\pm 26$ & $1177\pm 11$ & $7.9$ & $-0.4$ \\
SDSS J1655$+$2146 & $687\pm7$ & $42.71\pm0.09$ &  $12.0\pm 3.0$ & $44.65\pm0.11$ & $2483\pm 37$ & $1576\pm 27$ & $8.1$ & $-0.6$ \\
PG 1700$+$518 & $145\pm5$ & $42.62\pm0.10$ &  $15.6\pm 3.4$ & $45.43\pm0.09$ & $2067\pm 41$ & $2600\pm 53$ & $9.0$ & $-0.7$ \\
PG 2130$+$099 & $870\pm15$ & $41.98\pm0.09$ &  $48.3\pm 11.1$ & $44.40\pm0.10$ & $2727\pm 10$ & $2144\pm 17$ & $8.3$ & $-1.0$ \\
HE 2152$-$0936 & $122\pm7$ & $42.17\pm0.11$ &  $60.7\pm 13.8$ & $45.59\pm0.10$ & $1877\pm 13$ & $1853\pm 16$ & $8.8$ & $-0.3$ \\
HE 2158$-$0107 & $241\pm1$ & $42.59\pm0.09$ &  $9.6\pm 2.3$ & $44.89\pm0.11$ & $5486\pm 45$ & $2325\pm 19$ & $8.6$ & $-0.9$ \\
HE 2158$+$0115 & $124\pm1$ & $41.92\pm0.09$ &  $4.8\pm 1.1$ & $44.21\pm0.10$ & $2739\pm 33$ & $1637\pm 42$ & $7.9$ & $-0.9$ \\
PG 2214$+$139 & $466\pm11$ & $41.78\pm0.10$ &  $48.3\pm 12.1$ & $44.44\pm0.11$ & $5506\pm 23$ & $3190\pm 77$ & $8.6$ & $-1.4$ \\
HE 2307$-$0254 & $61\pm3$ & $42.00\pm0.11$ &  $8.6\pm 1.9$ & $44.89\pm0.10$ & $2172\pm 27$ & $1823\pm 32$ & $8.4$ & $-0.7$ \\
PKS 2349$-$014 & $279\pm7$ & $42.40\pm0.10$ &  $14.0\pm 3.0$ & $44.85\pm0.09$ & $4041\pm 83$ & $2265\pm 25$ & $8.6$ & $-0.9$ \\
HE 2353$-$0420 & $224\pm1$ & $42.58\pm0.09$ &  $2.7\pm 0.6$ & $44.42\pm0.10$ & $4176\pm 40$ & $1774\pm 17$ & $8.1$ & $-0.9$ \\
\noalign{\smallskip}\hline
\end{tabular}

\end{footnotesize}
\end{table*}

The best-fitting parameters were found by searching for the lowest $\chi^2$ of the residuals with a downhill simplex algorithm \citep[e.g.][]{Press:1992}. Two representative examples of these models are presented in Fig.~\ref{fig_pmas:QSO_fit}. From the best-fitting model, we measured several quantities including the continuum flux at 5100\AA\ ($f_{5100}$) and the width of the broad \mbox{H$\beta$}\ component expressed both in terms of FWHM ($\mathrm{FWHM}_{\mathrm{H}\beta}$) and line dispersion ($\sigma_{\mathrm{H}\beta}$). These values are listed in Table~\ref{tbl_pmas:bhmass} for the entire sample. To check whether these measurements are robust compared to a full \mbox{\ion{Fe}{ii}}\ template fitting approach we performed a simple consistency check.  We generated artificial QSO spectra from our best-fitting models excluding the \mbox{\ion{Fe}{ii}}\ components but adding the \citeauthor{Boroson:1992} iron template matched to the \mbox{\ion{Fe}{ii}}$\lambda$\,5014 line strength and broadened to the FWHM of the broad \mbox{H$\beta$}\ line. These artificial spectra were subsequently analysed in exactly the same way as the original QSO spectra. The recovered values for the FWHM of \mbox{H$\beta$}\ and the continuum flux at 5100\AA\ differ by only 0.02\,dex in the mean, with a maximum offset of 0.05\,dex and 0.08\,dex, respectively.

\subsection{Black hole masses and Eddington ratios}
Black hole masses can be estimated from single epoch broad-line AGN spectra using the so-called virial method \citep{Peterson:2000,Vestergaard:2002}. It combines the empirically derived size-luminosity relation of the BLR as determined via reverberation mapping \citep[e.g.][]{Kaspi:2000,Peterson:2004,Kaspi:2005,Bentz:2006,Bentz:2009a}  and the kinematics of the BLR estimated from the width of the broad emission-lines assuming virial motion of the clouds. We adopted the empirical relation between the BLR size and the continuum luminosity at 5100\,\AA\ ($L_{5100}$) calibrated by \citet{Bentz:2009a}, and the prescription of \citet{Collin:2006} to infer the kinematics of the broad line clouds from the H$\beta$ line dispersion ($\sigma_{\mathrm{H}\beta}$). A virial scale factor of $3.85$ was estimated by \citet{Collin:2006} to match the virial BH masses with the  $M_\mathrm{BH}-\sigma_*$ relation of local inactive galaxies \citep{Tremaine:2002}. This yields a black hole mass calibration of
\begin{equation}
 M_\mathrm{BH}=10^{7.41}\left(\frac{\sigma_{\mathrm{H}\beta}}{1000\,\mathrm{\mathrm{km}\,\mathrm{s}^{-1}}}\right)^2\left(\frac{L_{5100}}{10^{44}\,\mathrm{erg}\,\mathrm{s}^{-1}}\right)^{0.52}\,M_{\sun}\ .
\end{equation}
The error on the estimated BH masses using the virial method is dominated by systematic errors rather than measurement errors from  high S/N QSO spectra \citep[see][]{Denney:2009}. In addition to the intrinsic uncertainties in the adopted scaling relations and the adopted line profile model, the unknown geometry and kinematics of the BLR and the correct determination of the QSO continuum flux are sources of uncertainty. The BH masses estimated in this way (see Table~\ref{tbl_pmas:bhmass}) are assumed to be accurate within a factor of $\sim$2 ($\sim$0.3\,dex). 

Furthermore, we computed the bolometric luminosity of the QSOs adopting a standard bolometric correction of
\begin{equation}
 L_\mathrm{bol} = 9\times \lambda L_\lambda(5100\mathrm{\AA})\ ,
\end{equation}
  following \citet{Kaspi:2000} which is in agreement with the estimates of \citet{Richards:2006}. From the BH masses we also calculated the Eddington luminosities $L_\mathrm{Edd}$ and Eddington ratios $\lambda \equiv L_\mathrm{bol}/L_\mathrm{Edd}$. The derived values for the $M_\mathrm{BH}$ and $\lambda$ are given in Table~\ref{tbl_pmas:bhmass}. 
Almost all of our QSOs have black hole masses of $M_\mathrm{BH}\gtrsim10^{8}M_{\sun}$ and Eddington ratios in between $0.1\lesssim\lambda<1$. None of our QSOs accretes above the Eddington limit, which is in agreement with the results of low-redshift QSO studies \citep[e.g.][]{Schulze:2010,Steinhardt:2010}. The median BH mass and median Eddington ratio for our sample are $M_\mathrm{BH}\simeq4\times 10^{8}M_{\sun}$ and $\lambda\simeq0.15$. In individual cases, $M_\mathrm{BH}$ are found to be 0.5\,dex larger than previous estimates based on the FWHM of H$\beta$ \citep[e.g.][]{Vestergaard:2006}, but the values based on $\sigma_{H\beta}$ should be more robust as argued by \citet{Peterson:2004} and \citet{Collin:2006}.

\section{Emission line diagnostic of extended ionised gas}\label{sect_pmas:EELR_charac}
\subsection{Mapping of extended emission and their spectra}
After subtracting the QSO contribution with \texttt{QDeblend${}^{\mathrm{3D}}$}\ from the observed data we study the properties of ionised gas around the QSOs. Firstly, we extract 40\,\AA\ wide narrow-band images centred on the \mbox{[\ion{O}{iii}]}\ and \mbox{H$\alpha$}\ lines of the QSO after removing any continuum emission estimated the adjacent bands. The synthetic passband of our narrow-band images roughly correspond to $\pm900\,\mathrm{km}\,\mathrm{s}^{-1}$ in the rest frame at the median redshift of the sample, so that extended emission even with high velocities would be captured. We detect extended emission in \mbox{[\ion{O}{iii}]}\ around 16 of 31 QSOs ($\sim$50\%), of which a high fraction have also detectable H$\alpha$ emission (when covered by our observations). Whether the apparent \mbox{[\ion{O}{iii}]}\ emission around 3C~273 is a real feature is currently unclear as discussed below. Additionally, H$\alpha$ emission is detected around three objects without noticeable \mbox{[\ion{O}{iii}]}\ emission, I~Zw~1, RBS~219 and PHL~1811.

\begin{figure*}
\centering
\vspace*{10mm}
\includegraphics[width=0.95\textwidth,clip]{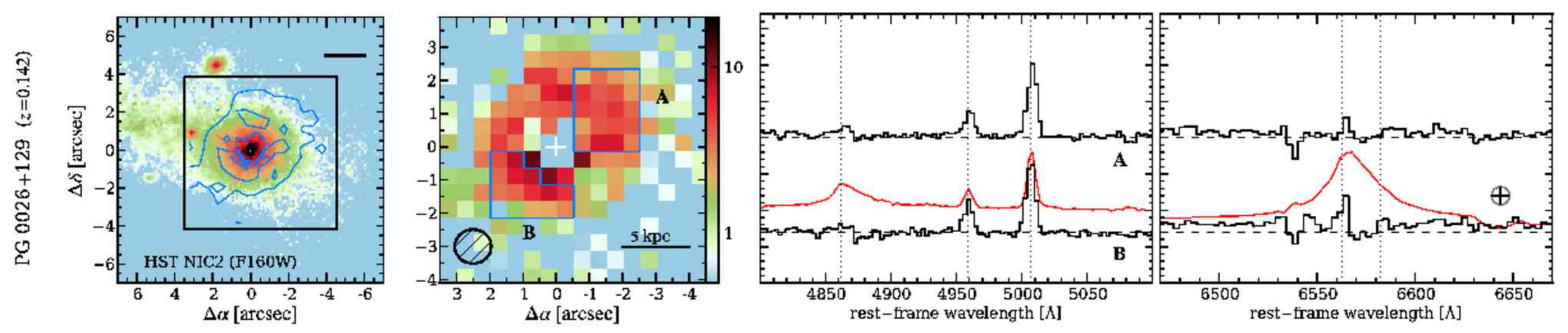}
\includegraphics[width=0.95\textwidth,clip]{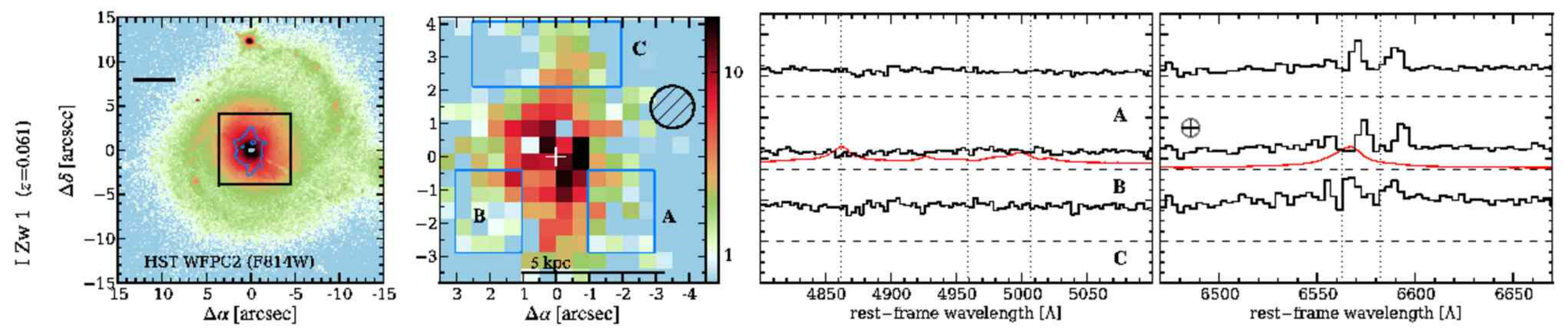}
\includegraphics[width=0.95\textwidth,clip]{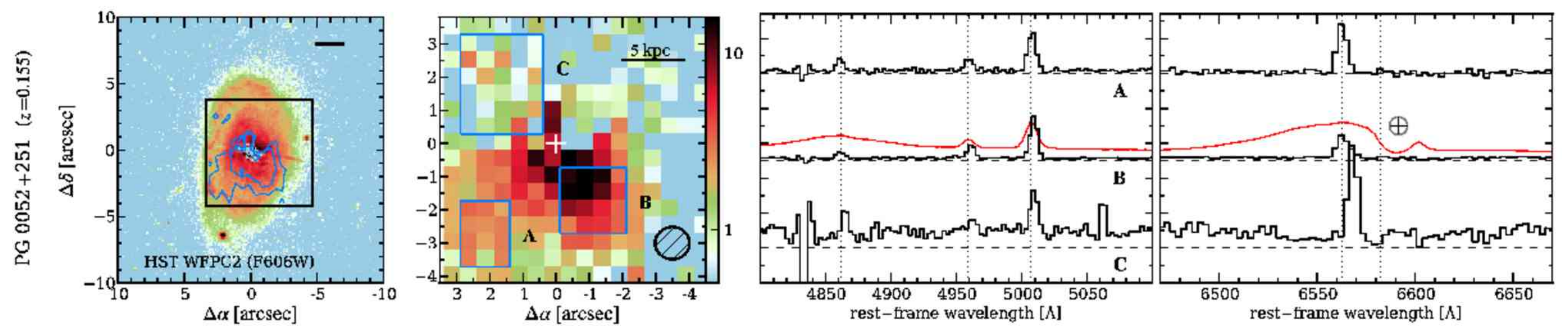}
\includegraphics[width=0.95\textwidth,clip]{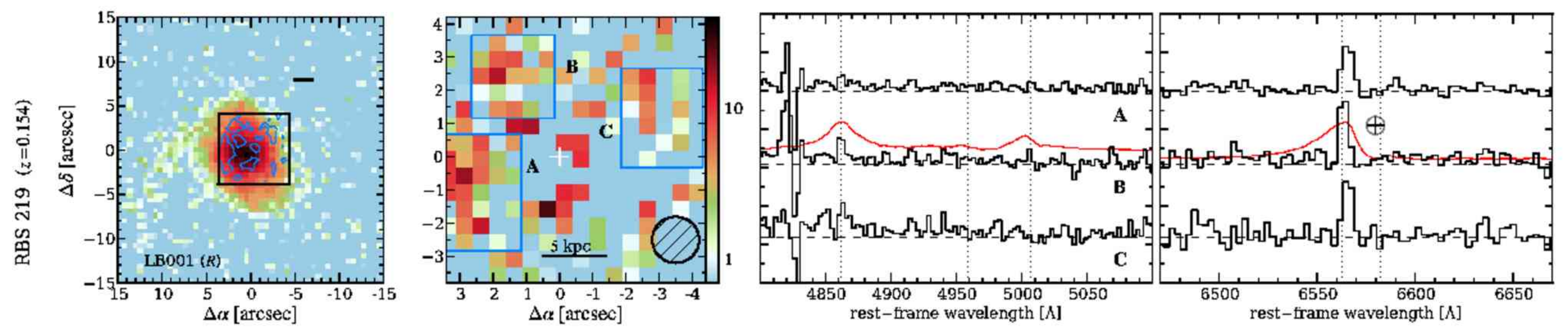}
\includegraphics[width=0.95\textwidth,clip]{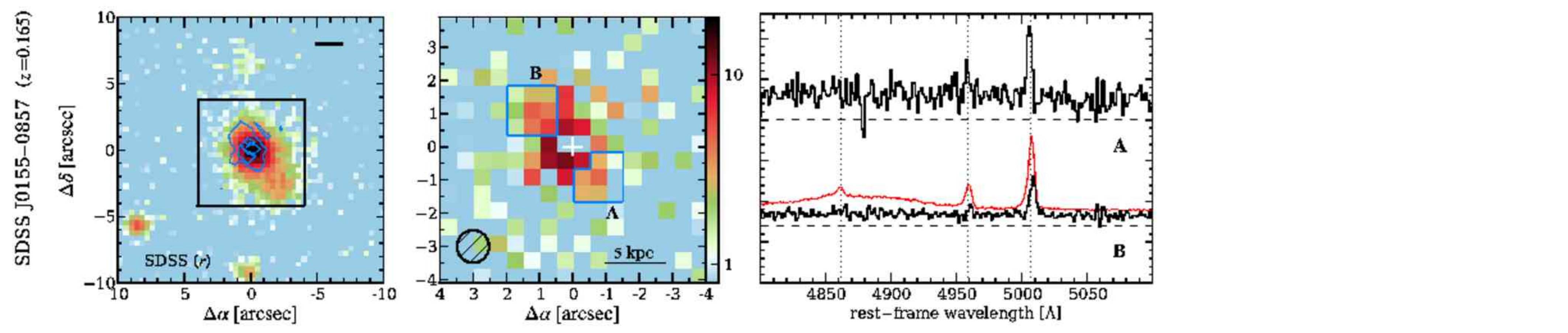}
\caption{Overview of the individual objects with extended emission after the point-like emission of the QSO was subtracted.  \textit{Left panels:} Nucleus-subtracted broad-band images with overplotted contours of extended line emission within the PMAS FoV (black box). A scale bar represents 5\,kpc at the object redshift and the image is oriented such that North is up and East is to the left. \textit{Middle panels:} Nucleus-subtracted \mbox{[\ion{O}{iii}]}\  narrow-band PMAS images  with a logarithmic scaling in units of $10^{-16}\,\mathrm{erg}\,\mathrm{s}^{-1}\mathrm{cm}^{-2}\mathrm{arcsec}^{-2}$. Note that in the case of I~Zw~1, RBS~219 and PHL~1811, the \mbox{H$\alpha$}\ narrow-band images are shown instead. The blue boundaries indicate the spatial regions defined to obtain the co-added spectra shown in the right panels. A white cross marks the position of the QSO and the black ellipse indicates the FWHM of the PSF. \textit{Right panels:} Co-added spectra of the specific spatial regions around the \mbox{H$\beta$}\ and \mbox{H$\alpha$}\ line, if it was covered by the spectral range of the used instrument setup. The different spectra are arbitrarily scaled and offset from each other, for which the dashed lines indicate their corresponding zero flux density. Vertical dotted lines mark the position of emission lines at the QSO redshift, estimated from the peak of the narrow \mbox{[\ion{O}{iii}]}\ line or from the peak of the \mbox{H$\beta$}\ line when the \mbox{[\ion{O}{iii}]}\ line was too weak or exceptionally broad. The arbitrarily scaled QSO spectrum is shown as a red line for comparison. Telluric absorption line features are highlighted by a crossed circle symbol.}

\label{fig_pmas:EELR_det}
\end{figure*}
\begin{figure*}
\centering
\includegraphics[width=0.99\textwidth,clip]{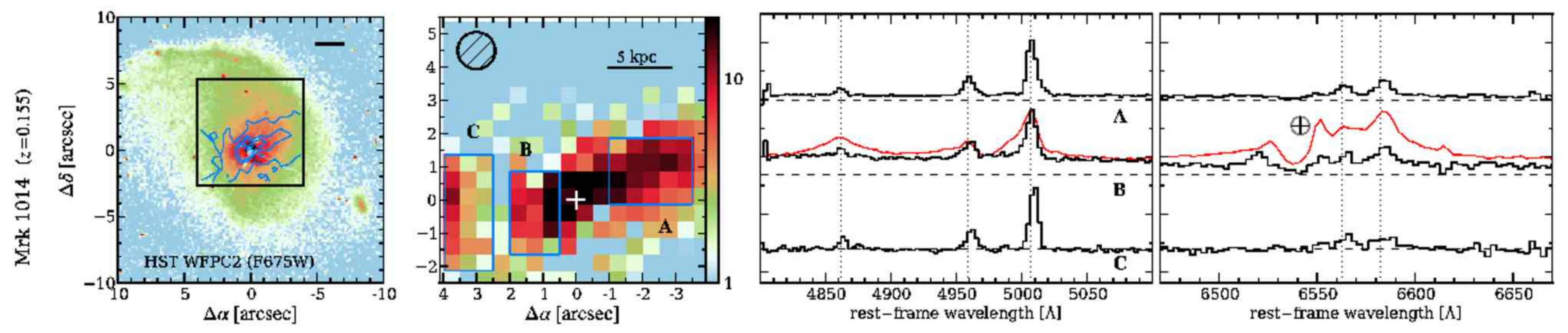}
\includegraphics[width=0.99\textwidth,clip]{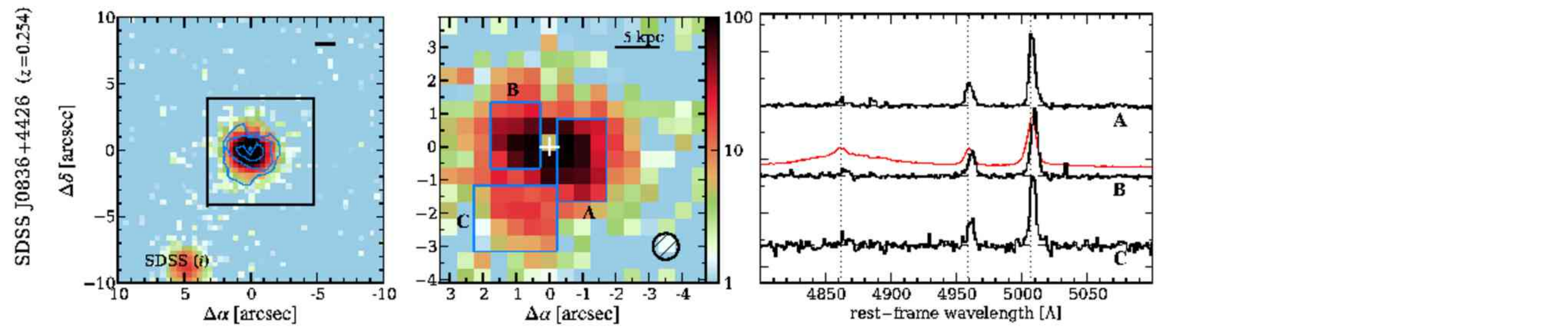}
\includegraphics[width=0.99\textwidth,clip]{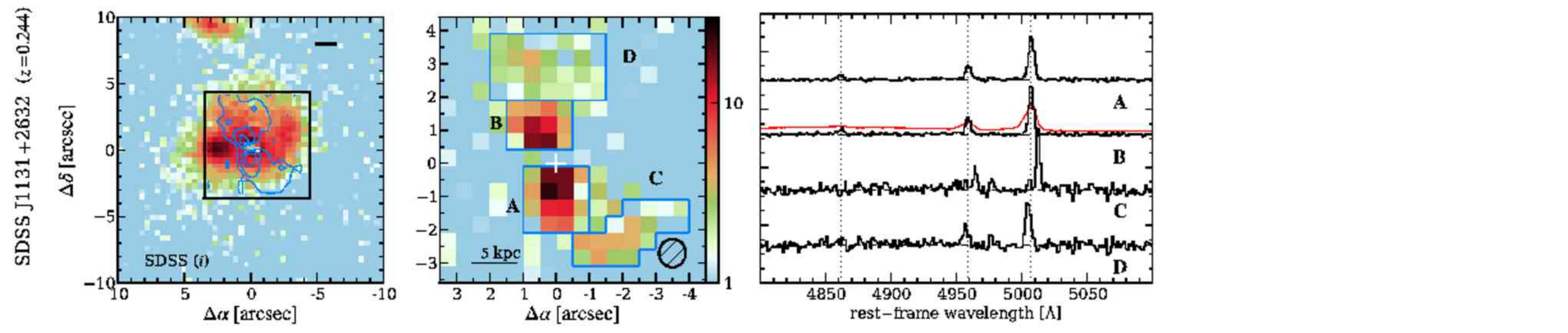}
\includegraphics[width=0.99\textwidth,clip]{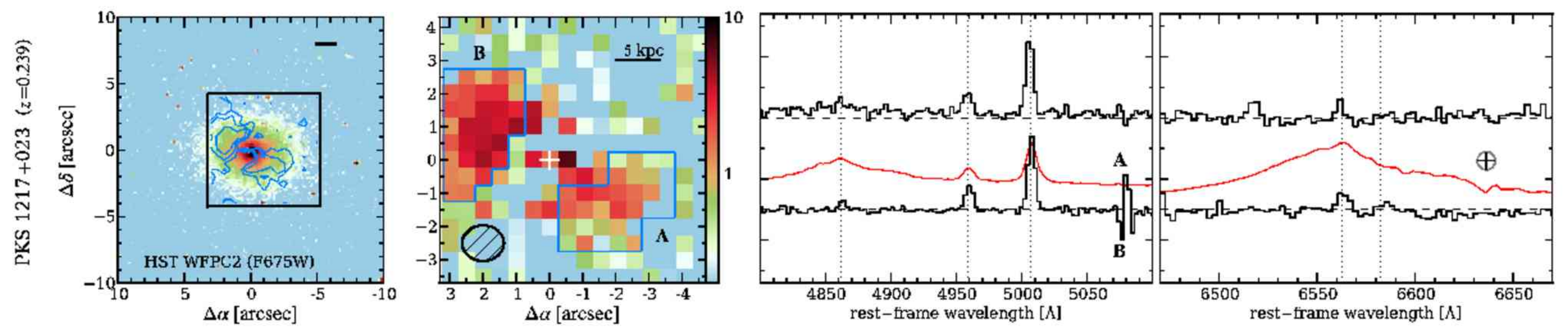}
\includegraphics[width=0.99\textwidth,clip]{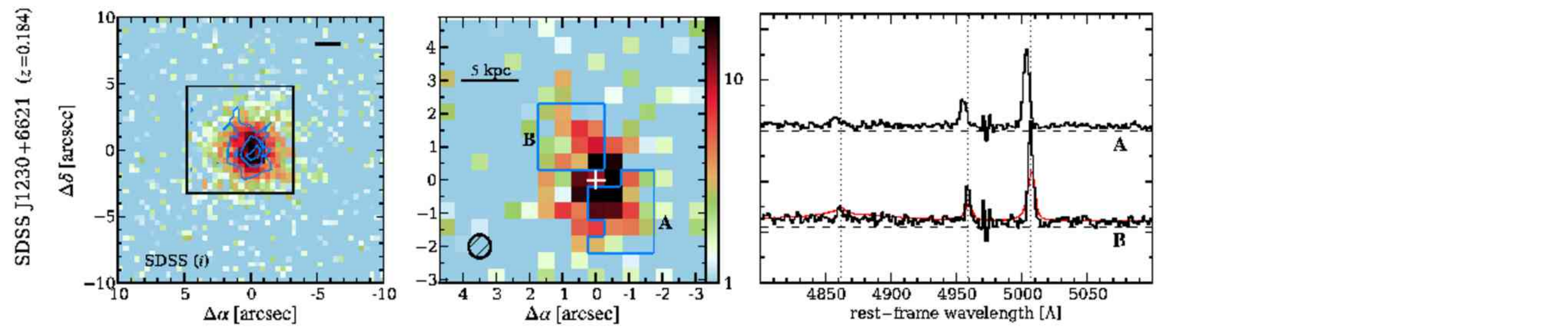}
\includegraphics[width=0.99\textwidth,clip]{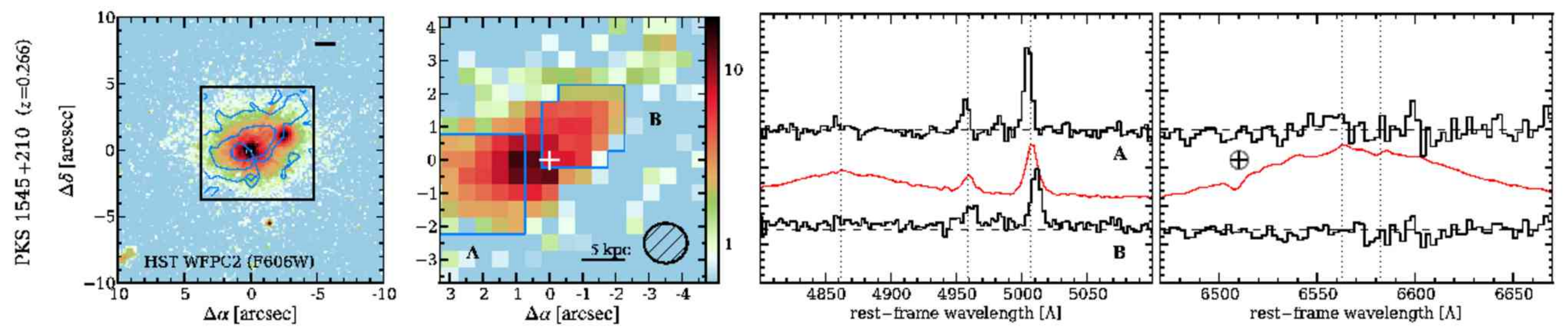}
\addtocounter{figure}{-1}
\caption[test]{Continued.}
\end{figure*}

\begin{figure*}
\centering
\includegraphics[width=0.99\textwidth,clip]{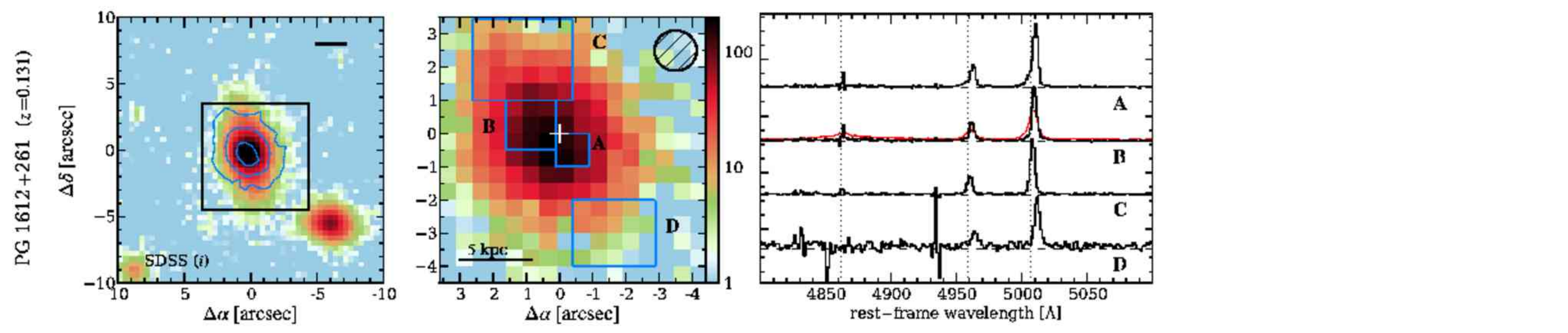}
\includegraphics[width=0.99\textwidth,clip]{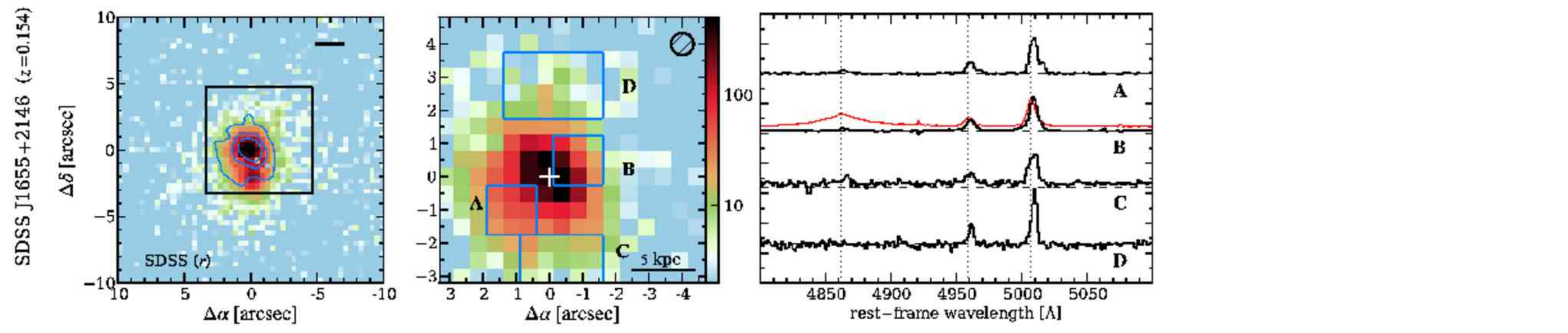}
\includegraphics[width=0.99\textwidth,clip]{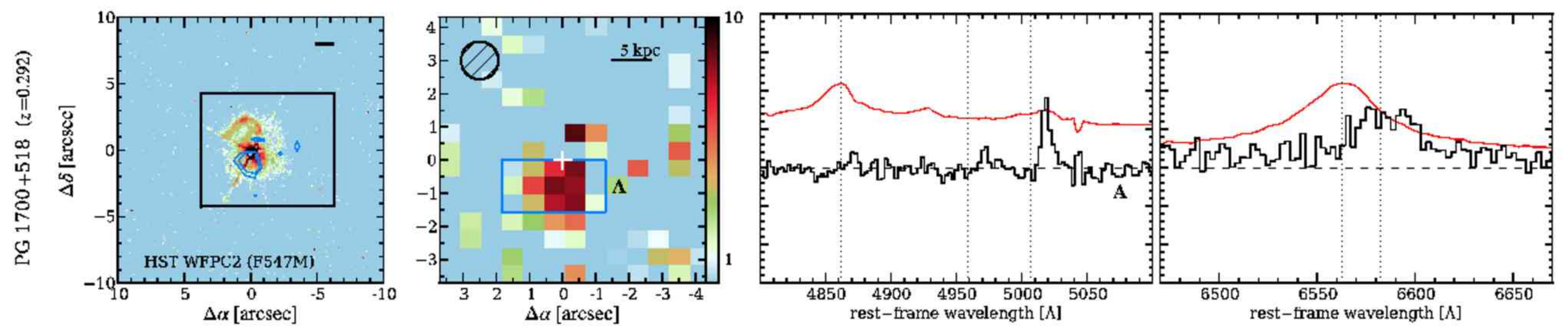}
\includegraphics[width=0.99\textwidth,clip]{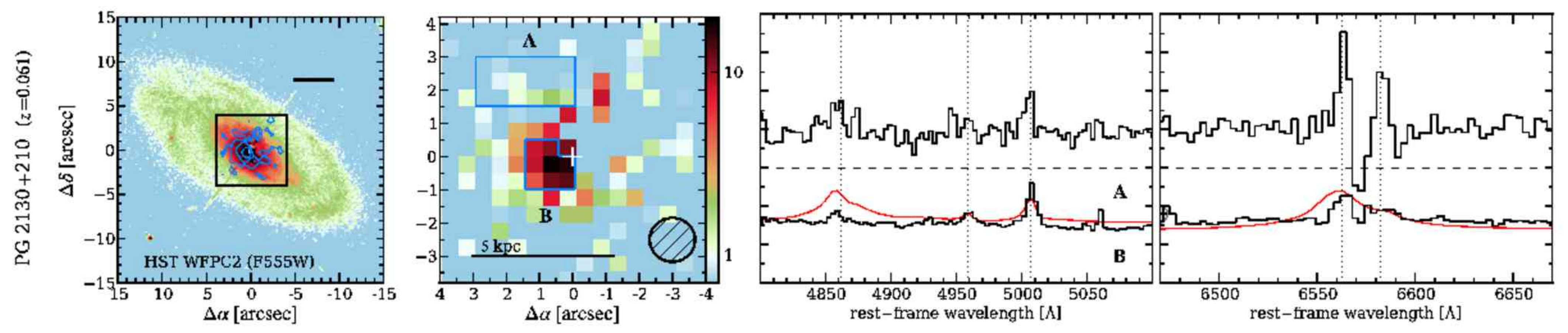}
\includegraphics[width=0.99\textwidth,clip]{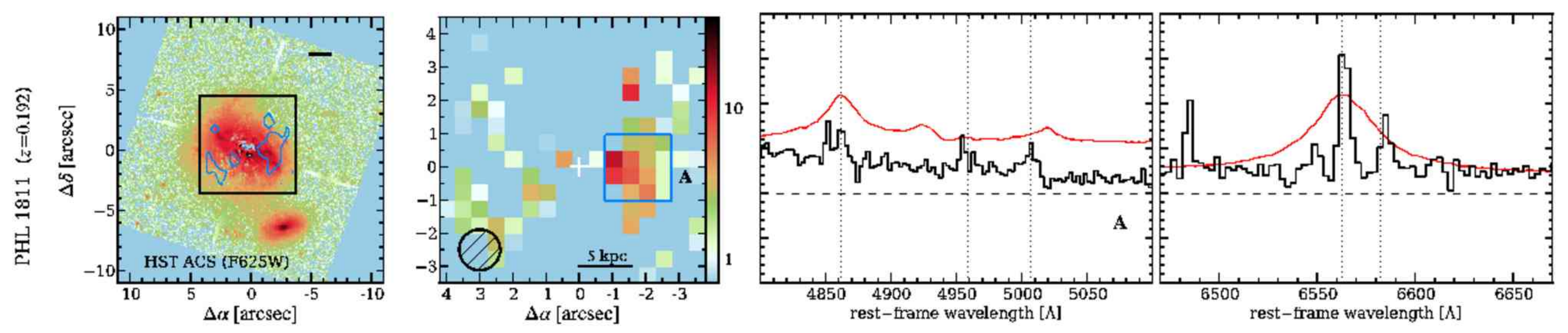}
\includegraphics[width=0.99\textwidth,clip]{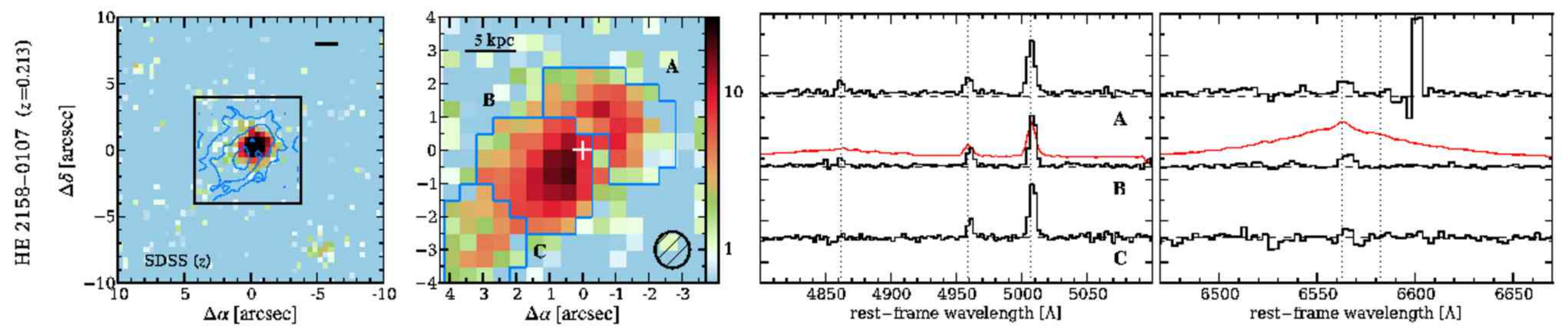}

\addtocounter{figure}{-1}
\caption{Continued.}
\end{figure*}

\begin{figure*}
 \centering
\includegraphics[width=0.99\textwidth,clip]{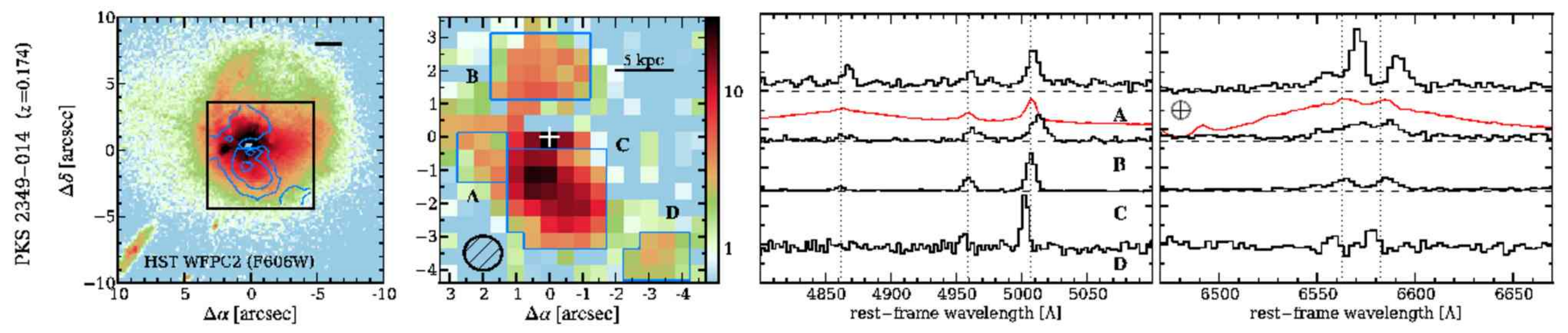}
\includegraphics[width=0.99\textwidth,clip]{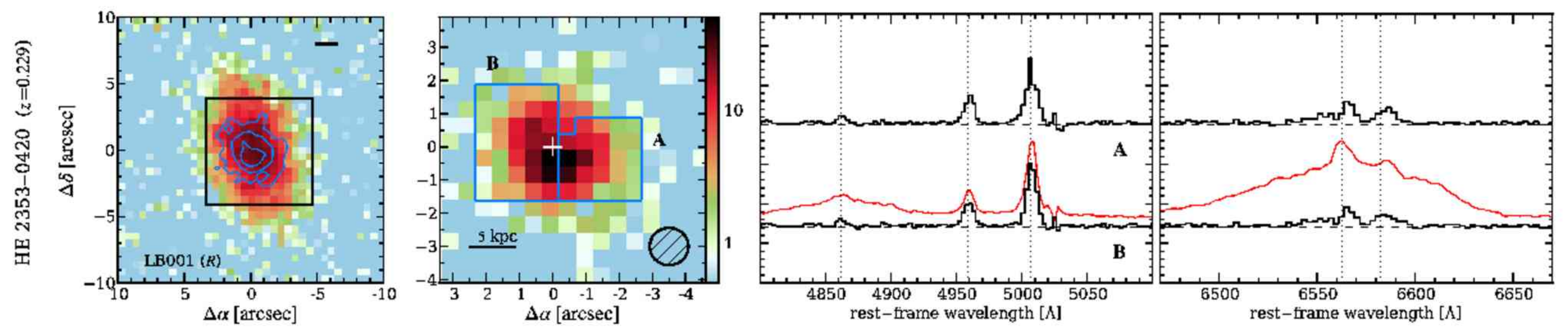}
\includegraphics[width=0.99\textwidth,clip]{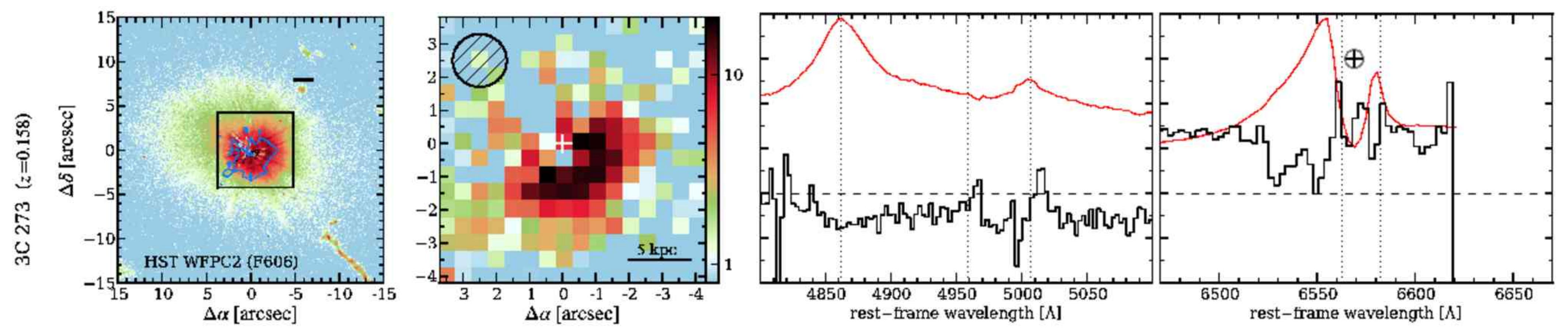}
\addtocounter{figure}{-1}
\caption[test]{Continued.}
\end{figure*}

\begin{figure*}
 \centering
\includegraphics[width=0.46\textwidth,clip]{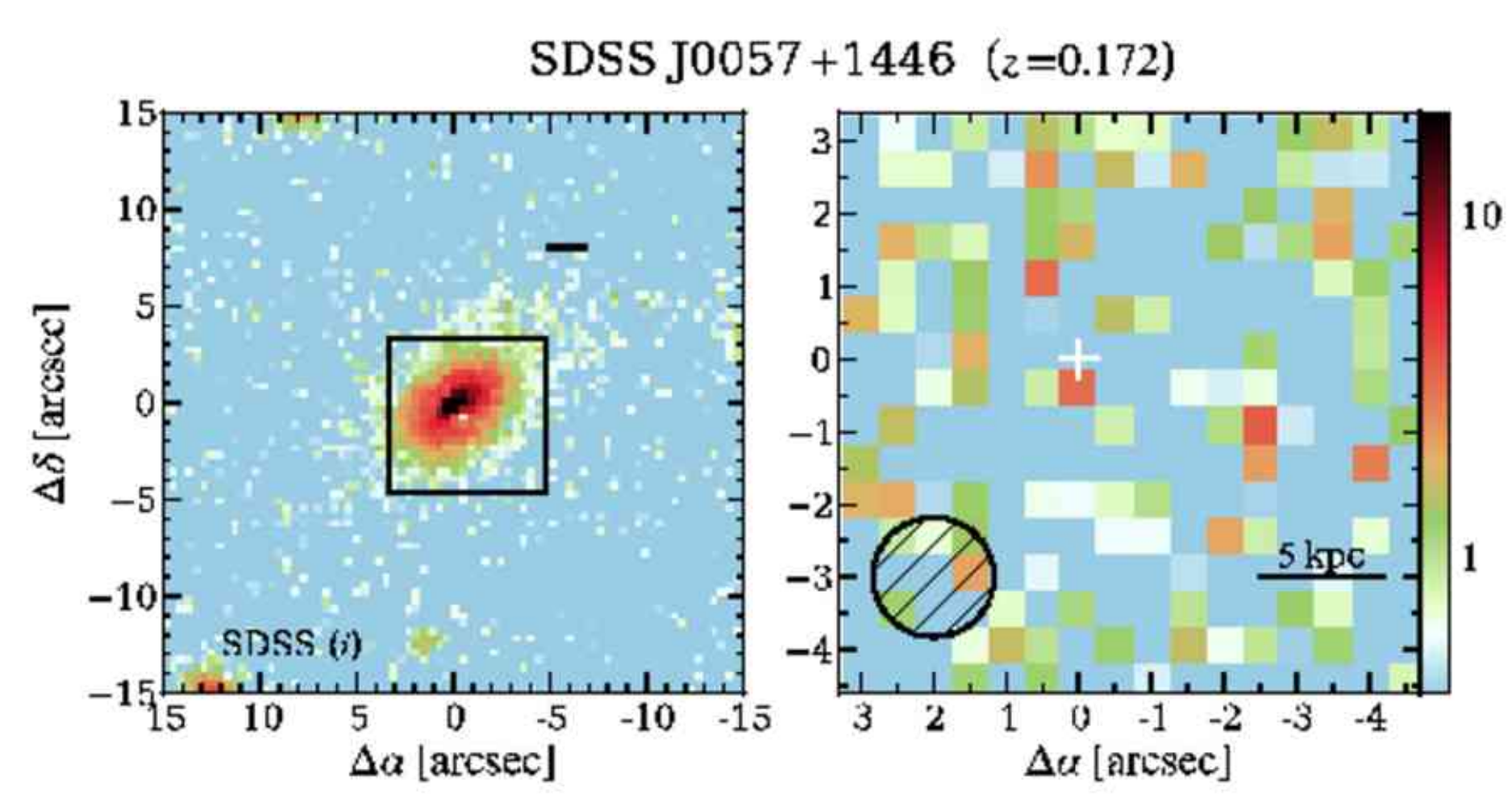}\hspace*{1cm}
\includegraphics[width=0.23\textwidth,clip]{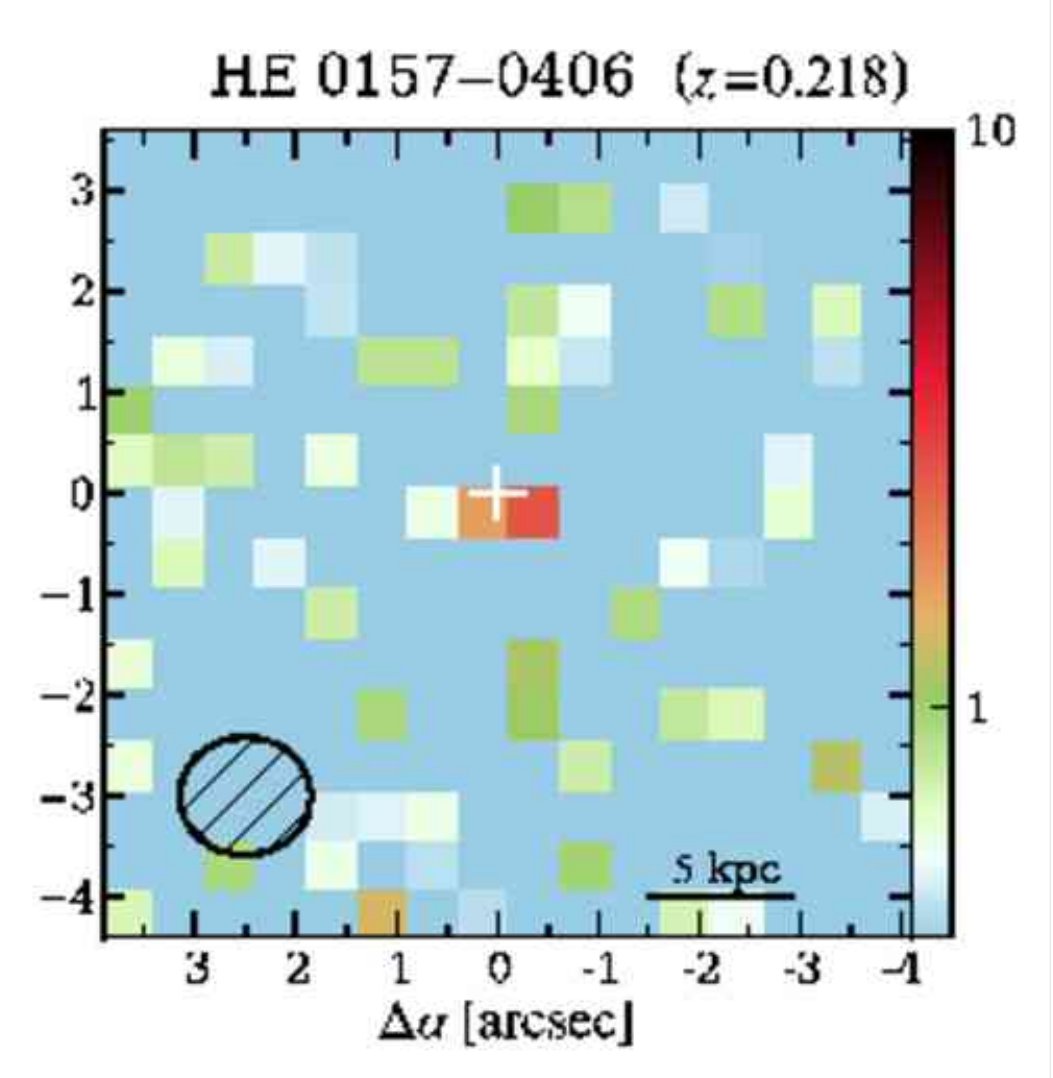}
\includegraphics[width=0.46\textwidth,clip]{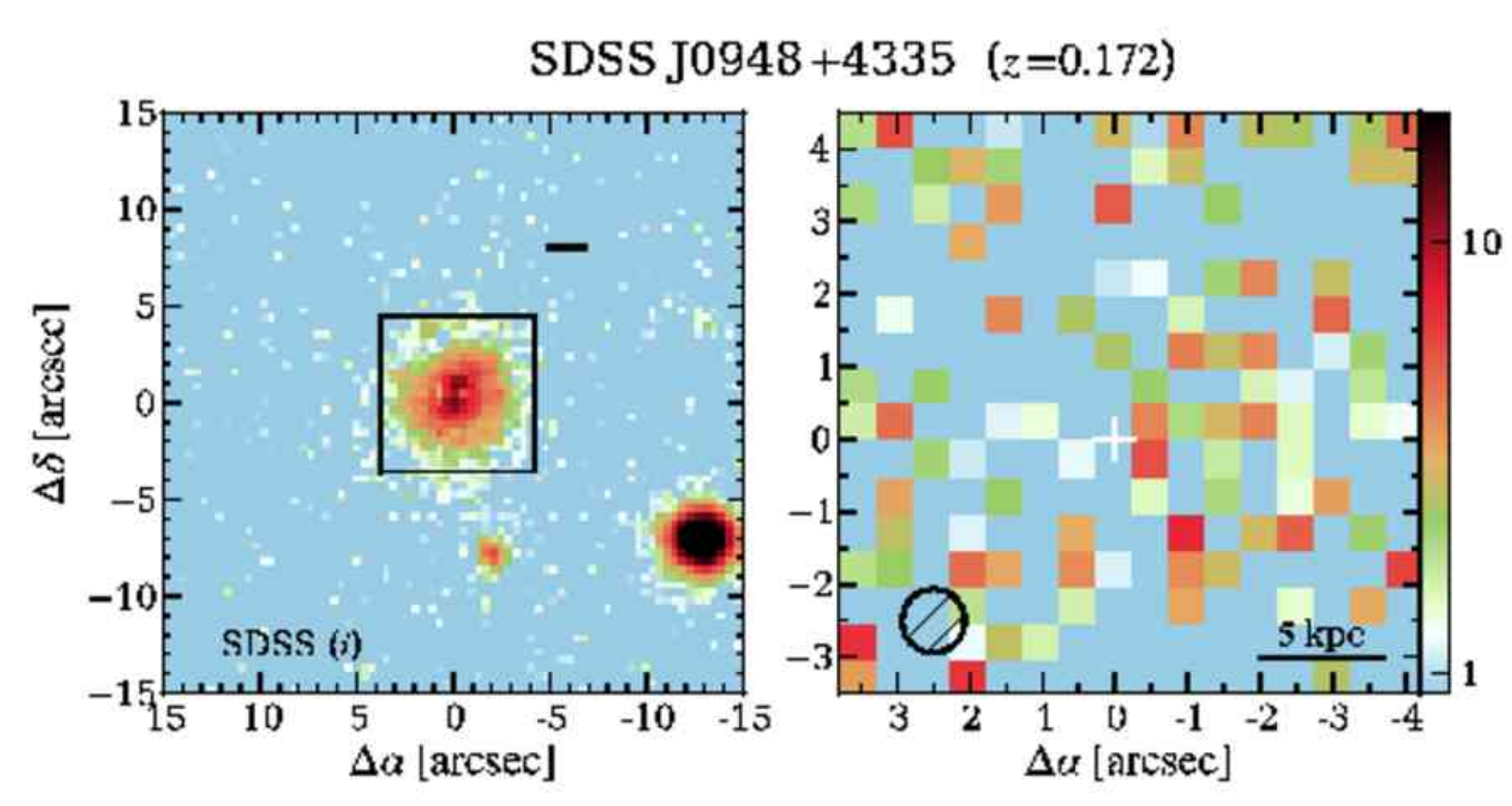}
\includegraphics[width=0.46\textwidth,clip]{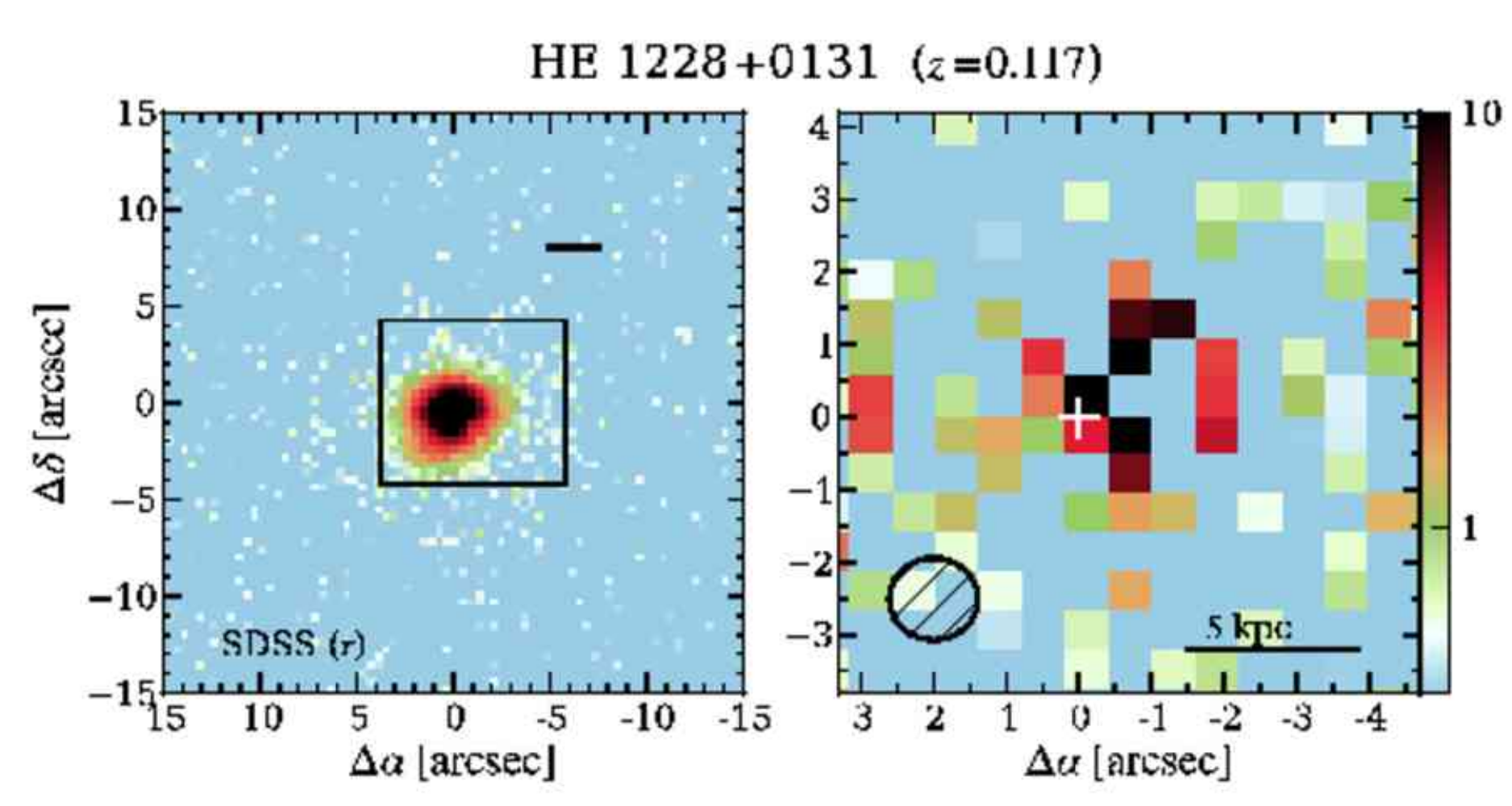}
\includegraphics[width=0.46\textwidth,clip]{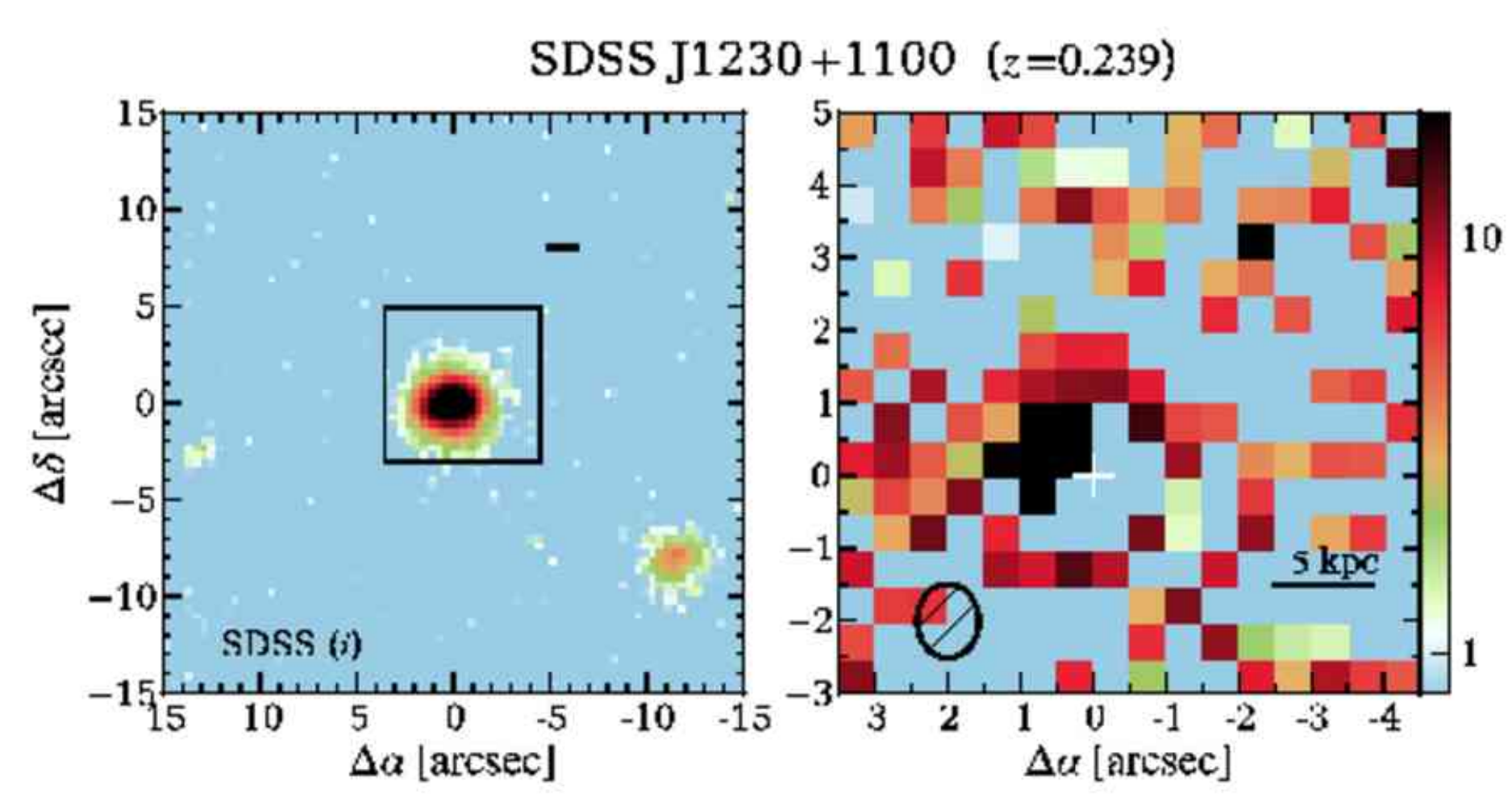}
\includegraphics[width=0.46\textwidth,clip]{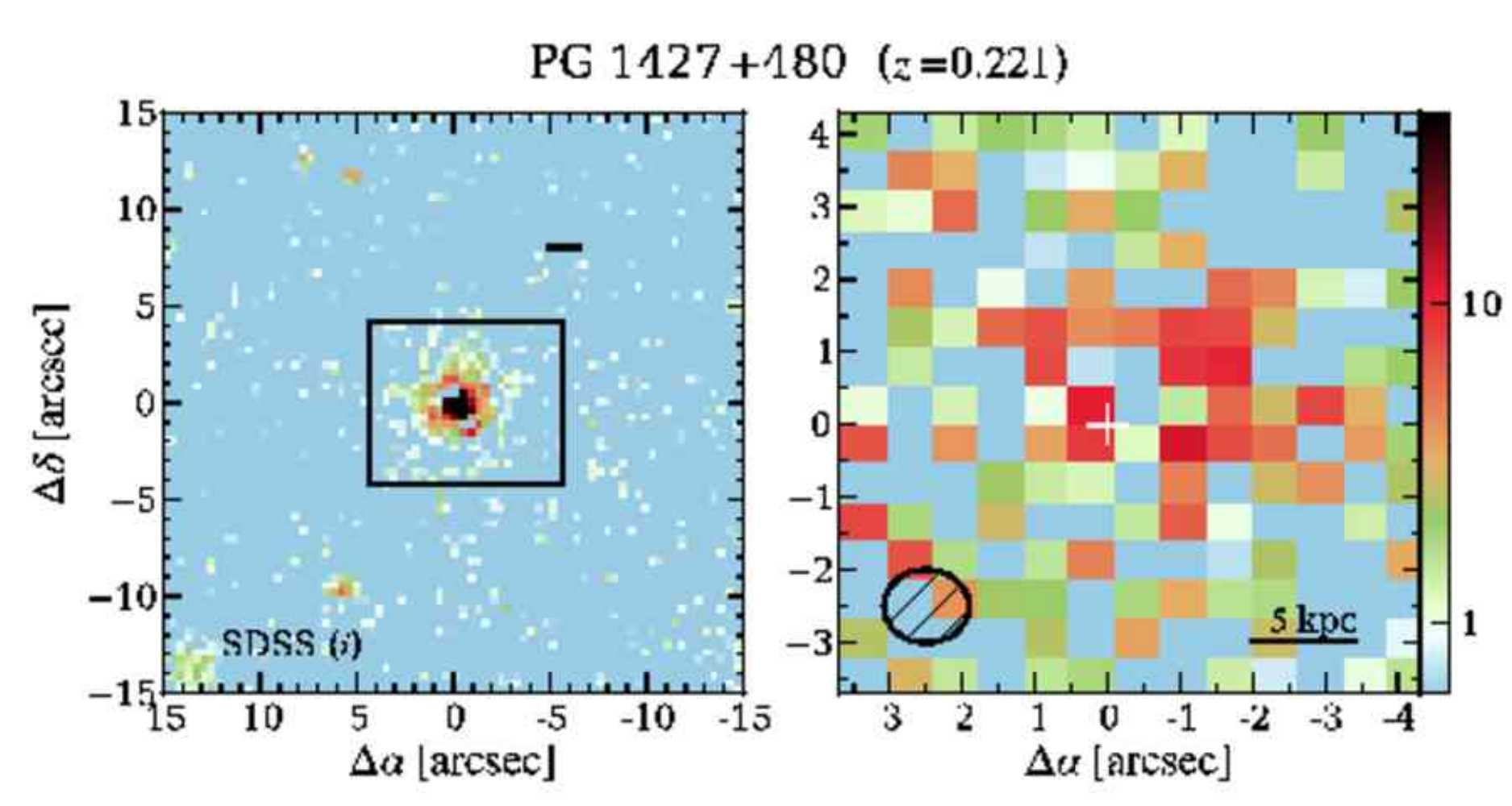}
\includegraphics[width=0.46\textwidth,clip]{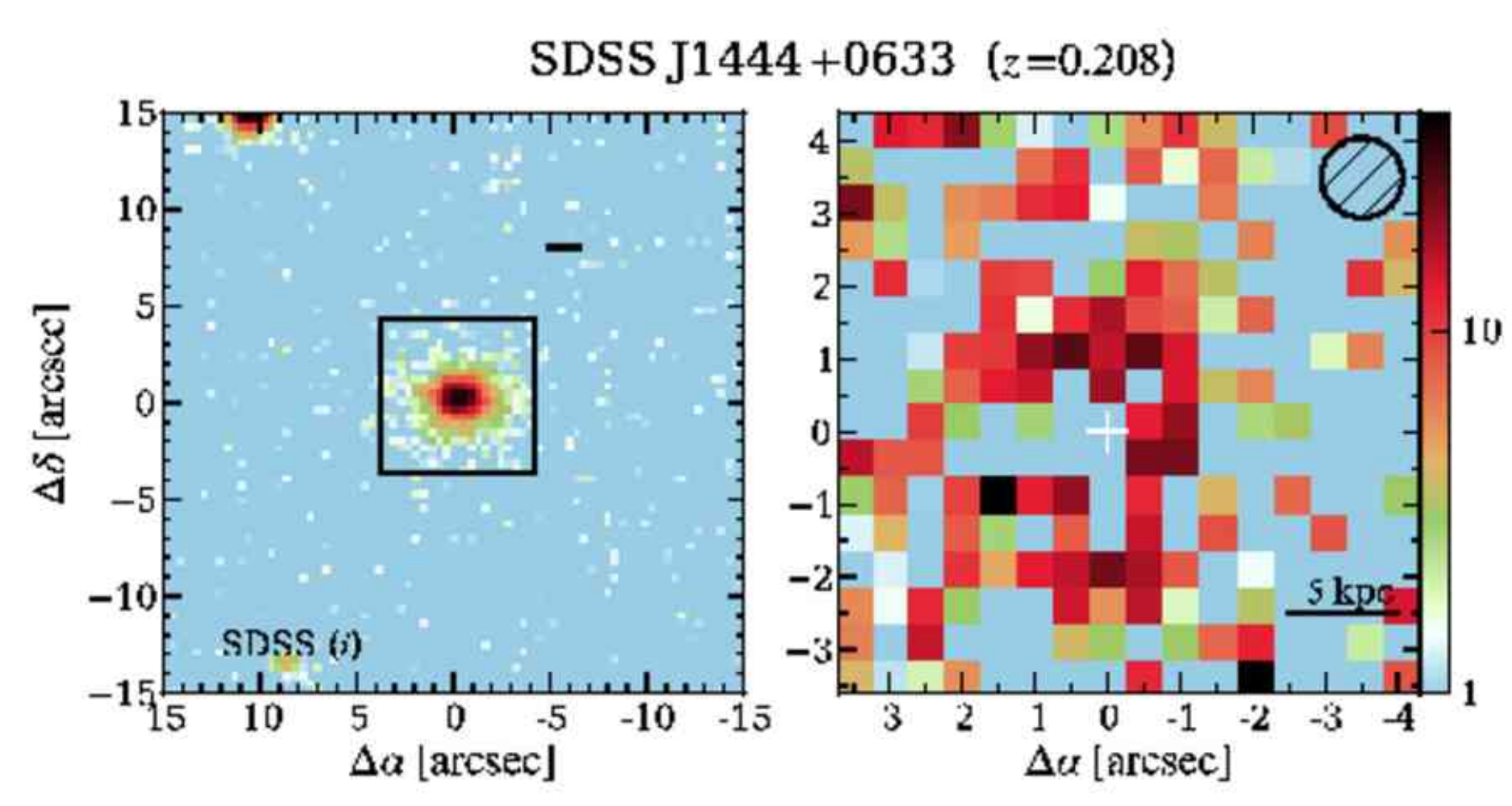}
\includegraphics[width=0.23\textwidth,clip]{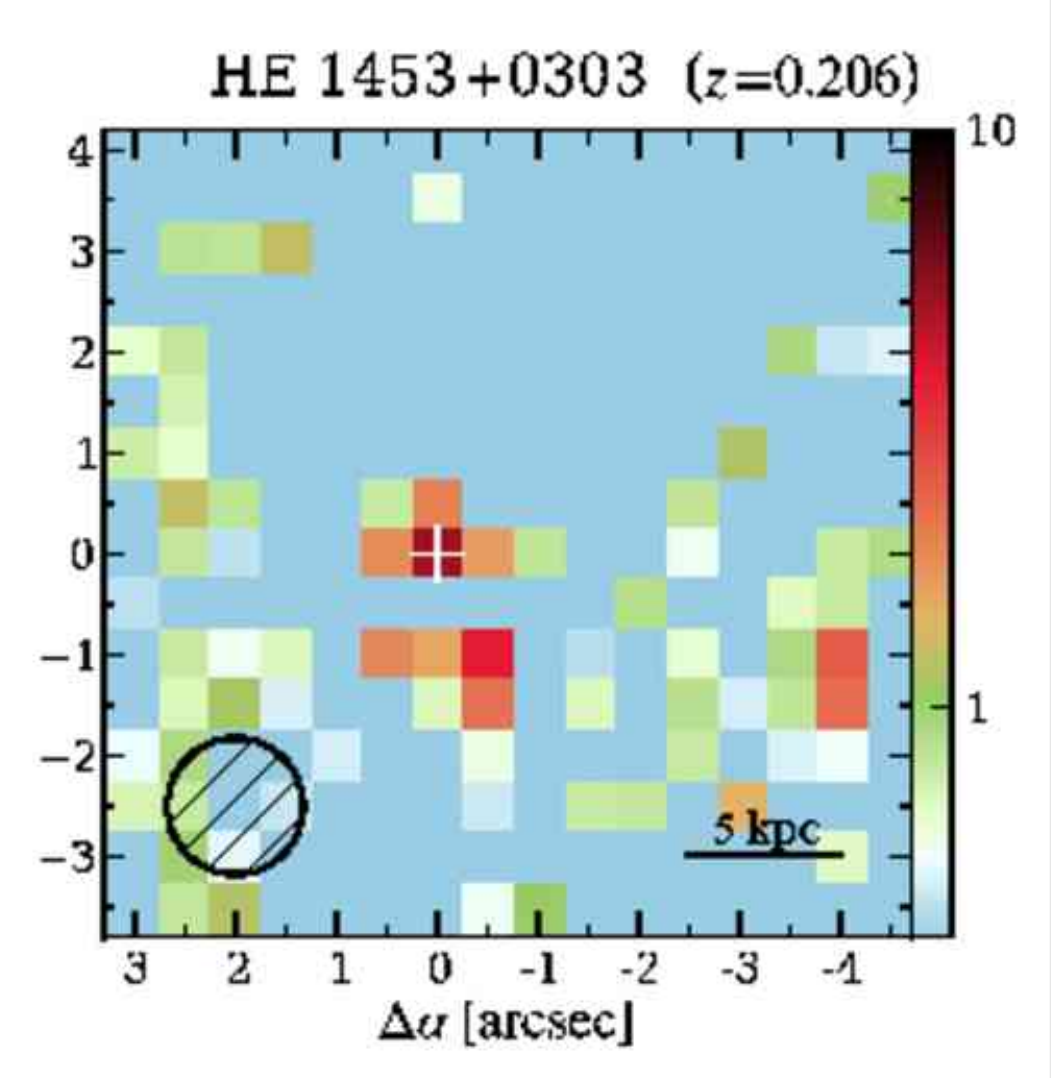}
\includegraphics[width=0.23\textwidth,clip]{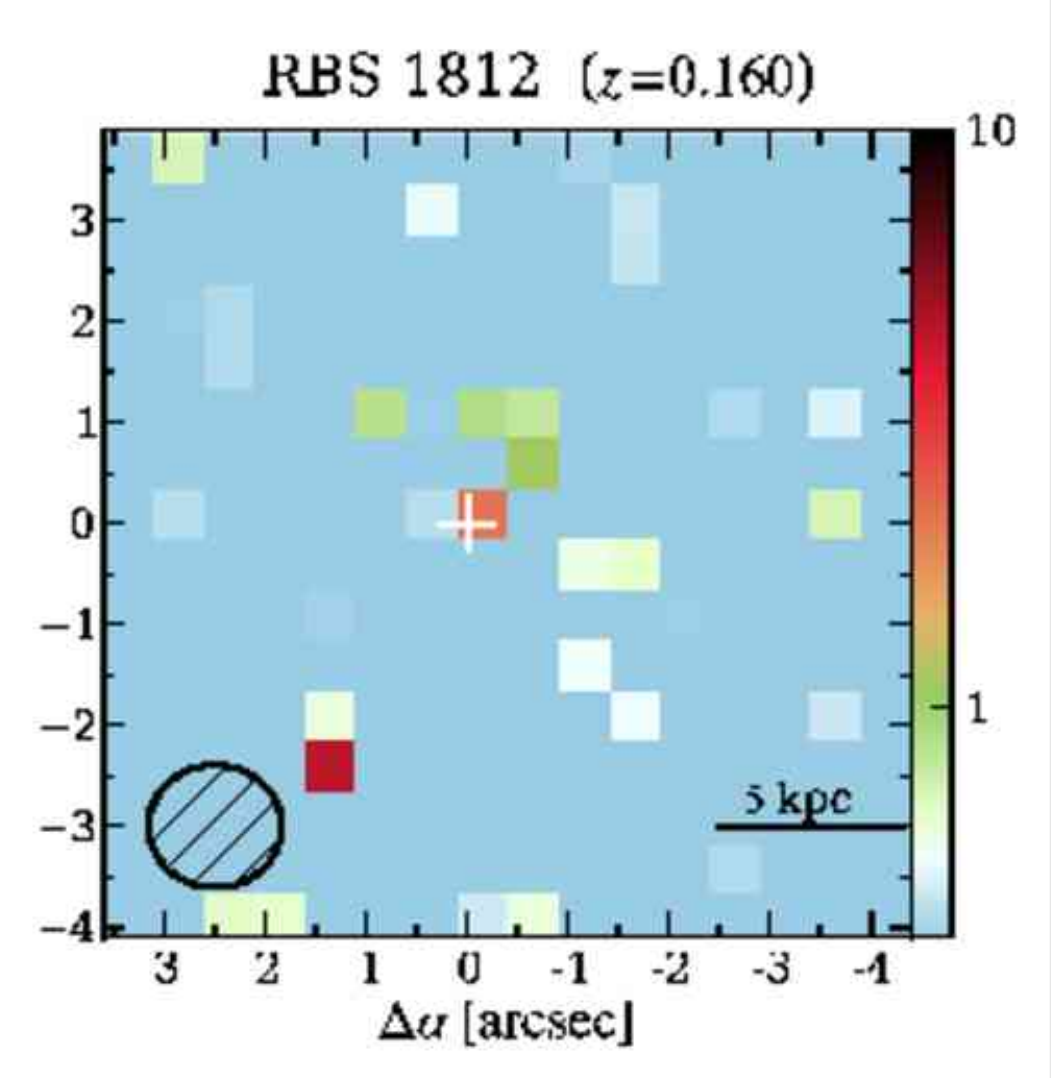}
\includegraphics[width=0.46\textwidth,clip]{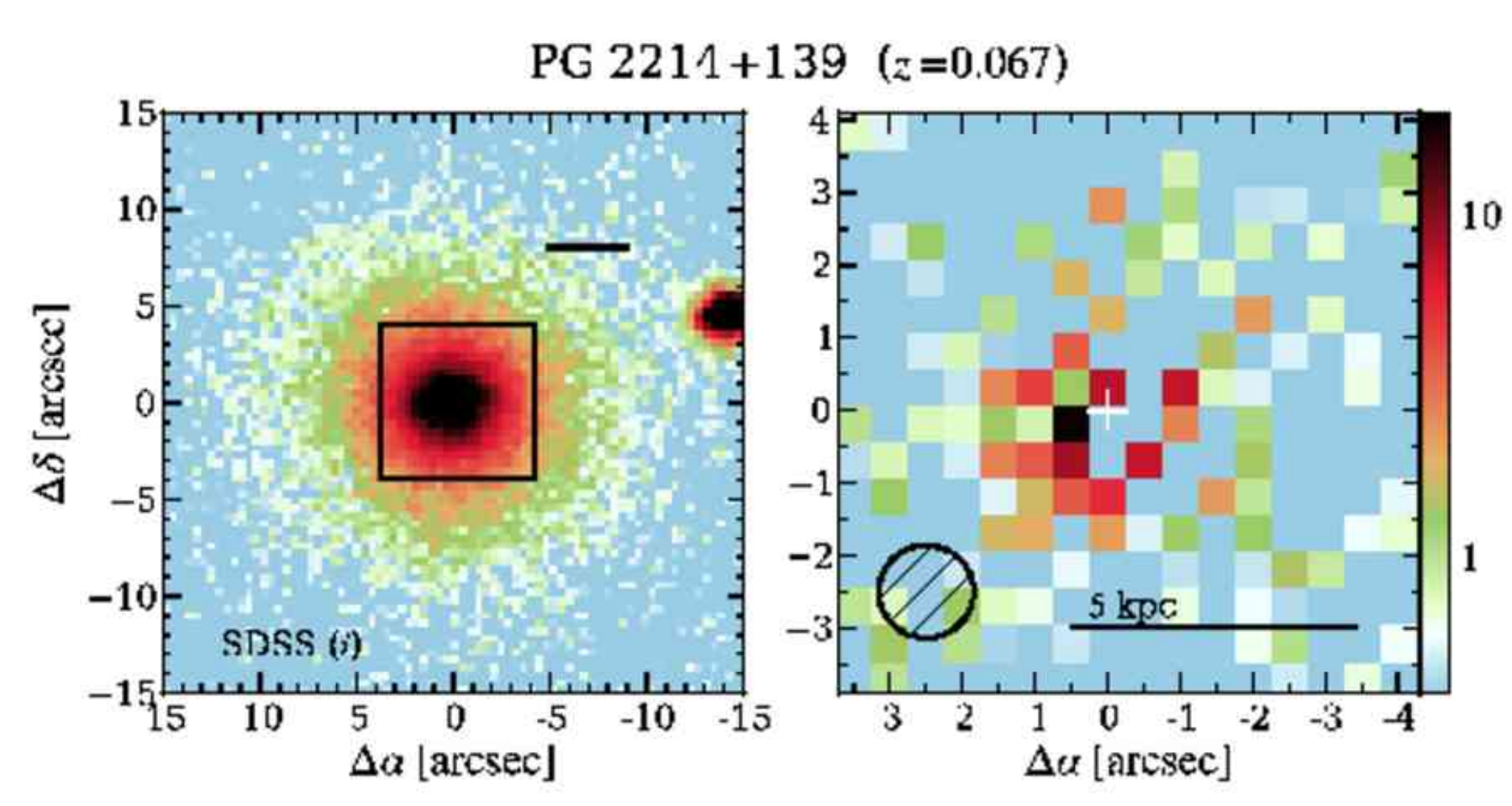}
\includegraphics[width=0.46\textwidth,clip]{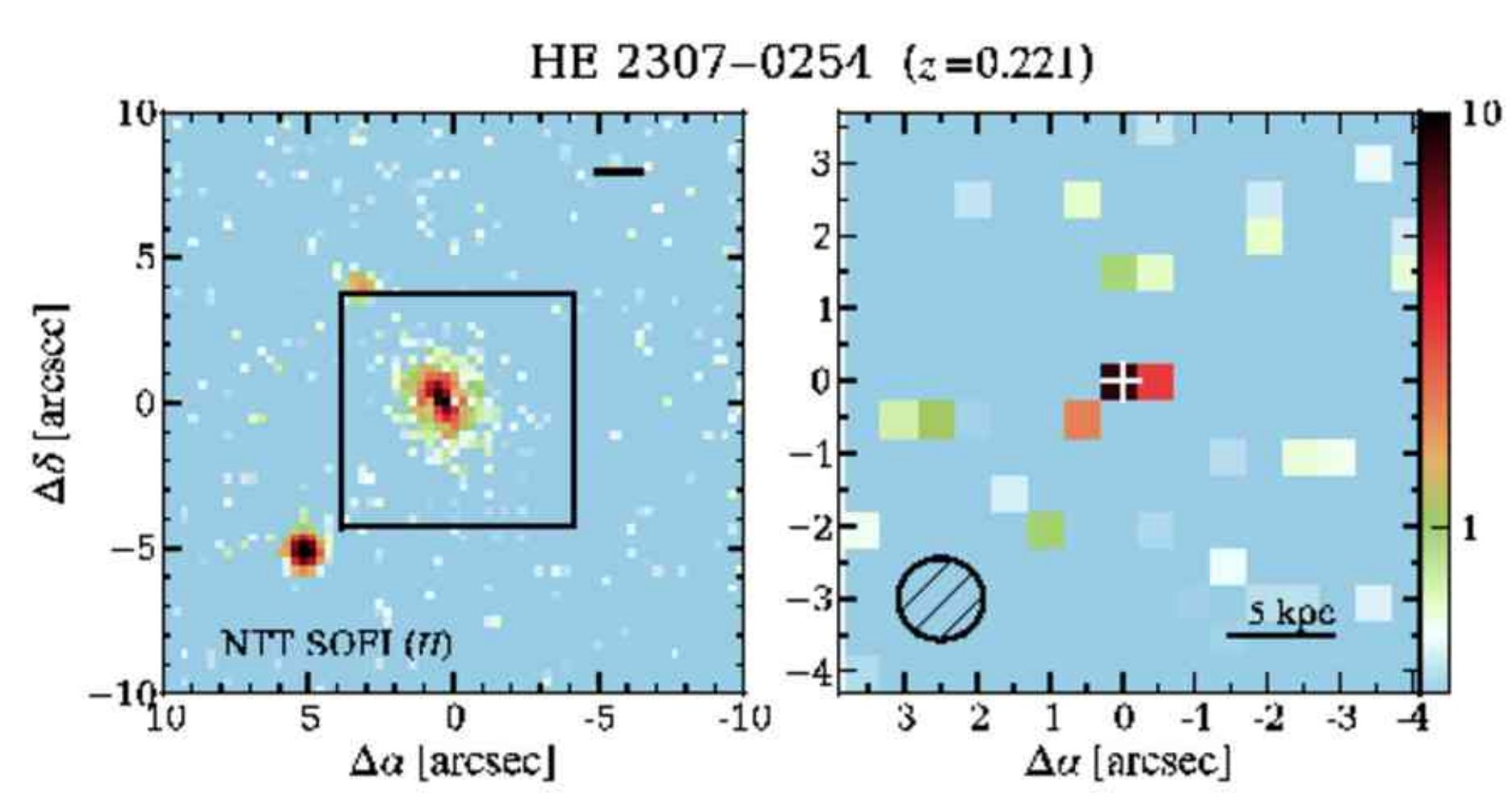}
\caption{Overview of the individual objects without detected extended emission. For each object the nucleus-subtracted broad-band images is presented, if available, in the left panel with the PMAS FoV indicated by the black rectangle. The black scale bar indicates a physical size of 5\,kpc at the redshift of the object. The nucleus-subtracted \mbox{[\ion{O}{iii}]}\ narrow-band images is presented in the right panels with the a white cross indicating the QSO position and the black ellipse indicates again the FWHM of the PSF. }
\label{fig_pmas:EELR_undet}
\end{figure*}

\begin{figure*}
\centering
\includegraphics[clip,width=\textwidth]{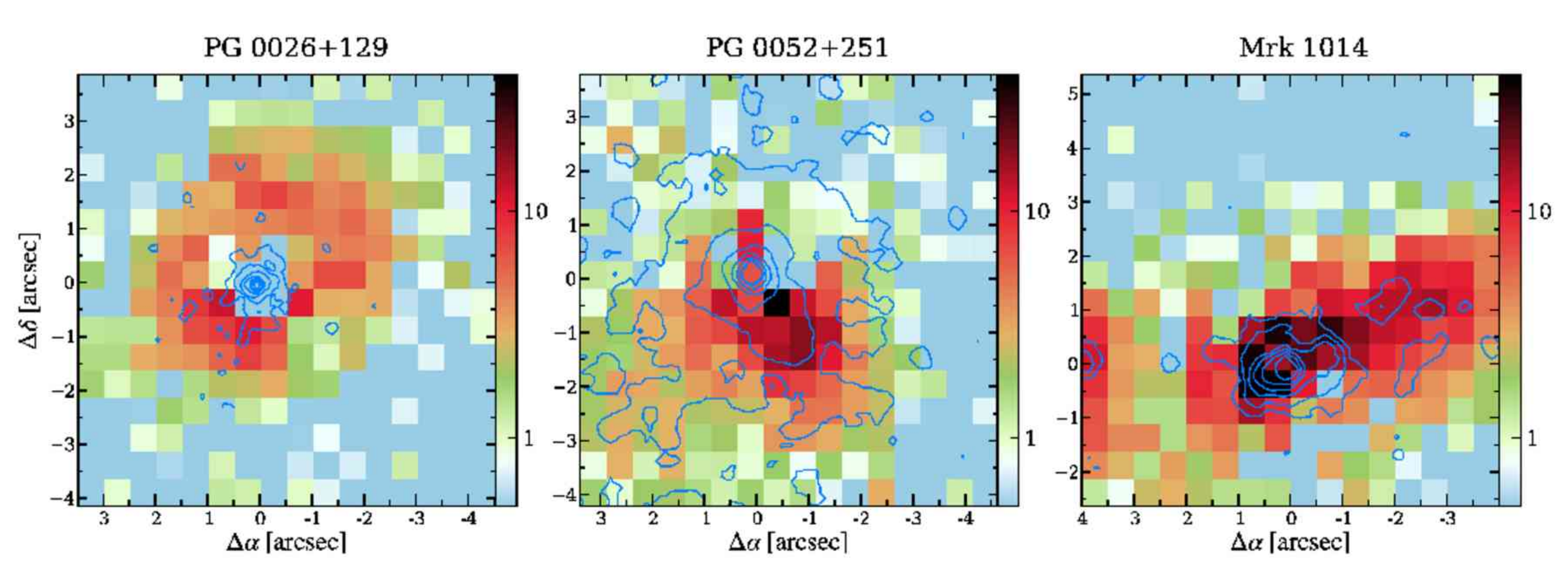}
\caption{Direct comparison of PMAS and HST \mbox{[\ion{O}{iii}]}\ narrow-band images. PMAS narrow-band images of PG~0026$+$129 (left panel), PG~0052$+$251 (middle panel) and Mrk~1014 (right panel) are shown in units of $10^{-16}\,\mathrm{erg}\,\mathrm{s}^{-1}\,\mathrm{cm}^{-2}\,\mathrm{arcsec}^{-2}$. The corresponding HST narrow-band images of \citet{Bennert:2002} are overplotted as blue contours also in a logarithmic scaling.}
\label{fig_pmas:HST_comp}
\end{figure*}

In Fig.~\ref{fig_pmas:EELR_det} we present the PMAS narrow-band images of all objects for which we detected extended ionised gas in comparison to the nucleus-subtracted broad-band host images, when available. The ionised gas  generally does not extend beyond the size of the host galaxies at the surface brightness limits of our observations. The only exception is HE~2158$-$0107 for which the ionised gas region is much more extended than the host galaxy and reaches even beyond the PMAS FoV with a projected size of $>$22\,kpc. A detailed discussion and interpretation of this interesting QSO is presented in \citet{Husemann:2011} and beyond the scope of this paper.

As the \mbox{H$\beta$}, \mbox{H$\alpha$}, and \mbox{[\ion{N}{ii}]}\ lines are rather weak in individual spaxels we decided to co-add several spaxels to increase their S/N. Based on the morphology of the extended emission we defined up to 4 characteristic spatial regions, avoiding spaxels with obvious signs of artefacts or QSO residuals that could affect the emission line measurements. The region boundaries are  shown in Fig.~\ref{fig_pmas:EELR_det} together with the corresponding co-added spectra in the rest-frame wavelength intervals 4800--5100\AA\ and 6450--6750\AA, if covered by our observation. The \mbox{H$\beta$}\ and \mbox{[\ion{O}{iii}]}\ emission lines were clearly detected in most of these objects. Essentially no contamination of  the broad \mbox{H$\beta$}\ emission  was left in these extended spectra, which demonstrates the robustness of the deblending process at least down to 0\farcs5 distance from the QSO position. The \mbox{H$\alpha$}\ and \mbox{[\ion{N}{ii}]}\ lines are often redshifted into one of the atmospheric absorption bands, which prevent a detection of these lines in those cases.

To quantitatively measure the observed emission line properties, we modelled each line in the spectra with a Gaussian profile taking into account the theoretically fixed line ratios of the \mbox{[\ion{O}{iii}]}\,$\lambda\lambda4960,5007$ and \mbox{[\ion{N}{ii}]}\,$\lambda\lambda6548,6583$ doublets. Furthermore,  we assumed that the lines are kinematically coupled having the same redshift and line width. This may not be true in all cases, because distinct ionised gas clouds with different kinematics and ionisation states can be aligned along our line-of-sight. The seeing may enhance this apparent confusion and in a few cases the redshift of the extended Balmer and \mbox{[\ion{O}{iii}]}\ lines do not agree in some regions of the host galaxy. Measured line fluxes, line widths and the rest-frame velocity offset with respect to the redshift of the narrow QSO \mbox{[\ion{O}{iii}]}\ emission line are reported in Table~\ref{tab_pmas:line_ratios}. Statistical errors on the flux measurements and $3\sigma$ upper limits for undetected emission lines were obtained based on the noise in the adjacent continuum.

On two EELR objects we have to comment in more detail:
The \mbox{H$\beta$}\ line flux for the region A and B of PG~1612$+$261 is unreliable due to a bad column on the PMAS CCD, but regions C and D remain unaffected. Off-nuclear long-slit spectra presented already by \citet{Boroson:1982b} showed that the \mbox{[\ion{O}{iii}]}/\mbox{H$\beta$}\ line ratio is $\sim$10 out to several arcseconds away from the nucleus. Thus, we assume that the gas is at least as highly ionised closer to the nucleus, although we were not able to reliably constrain the line ratios in that region. Furthermore, we find  that the \mbox{[\ion{O}{iii}]}\ emission line  in region A is significantly asymmetric requiring two independent Gaussian components to model the line profile. The QSO 3C~273 actually appears to be surrounded by some ionised gas, but it is by far the brightest object in our sample and was observed under poor seeing conditions. To check whether this feature may simply be an artefact of our data reduction or analysis we analysed the 3 individual exposures obtained for this target separately. The apparent structure could only be recovered in 1 of the 3 exposures implying that this may not be a real detection. Until observation with higher quality exists for 3C~273 we will treat it as a non-detection throughout the rest of this article.

No extended line emission was detected around the remaining 11 objects in the sample. We present their narrow-band images and available broad-band images in Fig.~\ref{fig_pmas:EELR_undet} for completeness.

\subsection{Comparison with HST narrow-band imaging}
\begin{table}\centering
\begin{footnotesize}
\caption{Surface brightness limits ($3\sigma$) for HST \mbox{[\ion{O}{iii}]}\ narrow-band images.}
\label{tbl_pmas:sfb_comparison}
 \begin{tabular}{lccc}\hline\hline\noalign{\smallskip}
Name &  HST & smoothed\tablefootmark{a} HST & PMAS  \\
     & \multicolumn{3}{c}{$\left[10^{-15}\,\mathrm{erg}\,\mathrm{s}^{-1}\,\mathrm{cm}^{-2}\,\mathrm{arcsec}^{-2}\right]$} \\\noalign{\smallskip}\hline\noalign{\smallskip}
PG 0026$+$129 & 10.73 & 3.79 & 0.14\\
PG 0052$+$251 & 2.77 & 0.80 & 0.18\\
Mrk 1014 & 2.74 & 0.91 & 0.21\\
\noalign{\smallskip}
\hline
\end{tabular}
\tablefoot{
\tablefoottext{a}{A median filter of $5\times5$ pixel box size was used.}
}

\end{footnotesize}
\end{table}
Three QSOs among our sample are within the sample of the HST narrow-band snapshot survey by \citet{Bennert:2002}.
This enables us to directly compare the narrow-band images obtained from both telescope (Fig.~\ref{fig_pmas:HST_comp}). We applied a median smoothing filter with a box size of 5$\times$5 pixels to the HST images in order to catch lower surface brightness features. The $3\sigma$ surface brightness limits for the HST and PMAS narrow-band images are listed in Table~\ref{tbl_pmas:sfb_comparison}. The estimated surface brightness limits support our visual impression from the comparison of the narrow-band images that the PMAS images are at least a factor of 4 deeper than the HST images even after smoothing.

PG~0026$+$129 was observed with the Planetary Camera (PC) that has a sampling of $0\farcs05\,\mathrm{pixel}^{-1}$ instead of $0\farcs1\,\mathrm{pixel}^{-1}$ compared to Wide Field Channel (WFC). Thus, the PMAS image for this particular object is roughly 20 times deeper than the one of HST, which was most likely the reason why the extended \mbox{[\ion{O}{iii}]}\ emission is completely missed by HST. The lowest surface brightness filaments around the nucleus in the HST image  may tentatively correspond to the high surface brightness structures in our PMAS images. It is obvious that \citet{Bennert:2002} underestimated the size of the extended \mbox{[\ion{O}{iii}]}\ emission for this object at least by factor of 3. 

The HST images of PG~0052$+$251 and Mrk~1014 obtained with the WFC are deeper so that high surface brightness structures of extended \mbox{[\ion{O}{iii}]}\ emission could be detected. Nevertheless, our ground-bases PMAS narrow-band images reach a lot deeper, in particular in the case of Mrk~1014 where diffuse emission is present between the high surface brightness structures. Similar IFU observation of Mrk~1014 obtained by \citet{Fu:2008} display the same \mbox{[\ion{O}{iii}]}\ light distribution as our PMAS observation suggesting that the structures are real. The low surface brightness regions south-east of PG~0052$+$251 was missed by \citet{Bennert:2002} and became visible only after smoothing the HST data. The maximum \mbox{[\ion{O}{iii}]}\ size was thus underestimated by a factor of $\sim$2 in these two cases. We return to the issue of NLR sizes in section~\ref{sect_pmas:ENLR_prop}.

For PG~0052$+$215 we note further that the contours of the smoothed HST images do not agree with the \mbox{[\ion{O}{iii}]}\ light distribution in our PMAS images north of the nucleus. We carefully checked that this is not introduced by our data reduction or data analysis. A possible origin for the rather symmetric diffuse light in the HST image around the nucleus might be continuum emission of the host. \citet{Bennert:2002} subtracted only the contribution of the QSO using a PSF-star image, because the S/N in the much shallower continuum image of the host was insufficient.  The unavailability of a continuum image also hampered the analysis of the narrow-band HST image of PKS~1545$+$210 (3C~323.1) as part of large HST snapshot program to study extended gas around radio galaxies \citep{Privon:2008}.

We conclude that HST is of course much better in terms of spatial resolution than any optical ground-based observations, but the narrow-band images strongly suffer from its low sensitivity which has already been noted by several authors \citep[e.g.][]{Bennert:2005,Privon:2008}. Additionally, narrow-band images require off-band images which are often not sufficiently deep or absent to subtract the stellar continuum contribution from the on-band images.

\onltab{
\begin{table*}
	\centering
	\begin{footnotesize}
	\caption{Kinematics and line fluxes for specific regions of the EELR.}
	\label{tab_pmas:line_ratios}
	\begin{tabular}{llccccccc}\hline\hline\noalign{\smallskip}
Object\tablefootmark{a} & &  $\Delta v_\mathrm{r}$\tablefootmark{b} & $\sigma_\mathrm{line}$\tablefootmark{c} & H$\beta$ & [\ion{O}{iii}]\,$\lambda$5007 & H$\alpha$  & [\ion{N}{ii}]\,$\lambda$6585 & Classification \\\cline{5-8}\noalign{\smallskip}
 &     &  $\left[\mathrm{km}\,\mathrm{s}^{-1}\right]$ & $\left[\mathrm{km}\,\mathrm{s}^{-1}\right]$ & \multicolumn{4}{c}{$\left[10^{-16}\,\mathrm{erg}\,\mathrm{s}^{-1}\,\mathrm{cm}^{-2}\right]$} & \\\noalign{\smallskip}\hline\noalign{\smallskip}
PG 0026$+$129 & A & $48\pm34$ & $158\pm14$ & $2.4\pm0.9$ & $25.1\pm1.0$ & ... & ... & AGN \\
 & B & $136\pm32$ & $153\pm9$ & $2.5\pm0.4$ & $19.8\pm0.5$ & ... & ... & AGN \\
\noalign{\smallskip}
PG 0050$+$124 & A & $-312\pm27$ & $83\pm24$ & ... & ... & $9.3\pm1.4$ & $7.9\pm0.3$ & \ion{H}{ii} \\
 & B & $-475\pm28$ & $18\pm23$ & ... & ... & $6.9\pm2.8$ & $7.7\pm0.4$ & \ion{H}{ii} \\
 & C & $-215\pm36$ & $80\pm30$ & ... & ... & $9.8\pm1.9$ & $5.8\pm0.7$ & \ion{H}{ii} \\
\noalign{\smallskip}
PG 0052$+$251 & A & $-127\pm36$ & $110\pm31$ & $3.0\pm0.5$ & $9.4\pm0.2$ & $9.5\pm0.4$ & $<2.5$ & \ion{H}{ii} \\
 & B & $-232\pm32$ & $82\pm27$ & $6.3\pm1.1$ & $38.5\pm0.2$ & $22.1\pm0.8$ & $<3.4$ & \ion{H}{ii}? \\
 & C & $-264\pm41$ & $65\pm28$ & $5.5\pm1.2$ & $7.0\pm0.4$ & $14.1\pm1.1$ & $<5.3$ & AGN \\
\noalign{\smallskip}
RBS 219 & A & $-95\pm40$ & $76\pm22$ & ... & ... & $12.1\pm1.2$ & $<2.5$ & \ion{H}{ii} \\
 & B & $17\pm39$ & $0\pm25$ & ... & ... & $5.8\pm1.7$ & $<4.6$ & \ion{H}{ii} \\
 & C & $-105\pm47$ & $0\pm39$ & ... & ... & $4.7\pm2.9$ & $<8.3$ & \ion{H}{ii} \\
\noalign{\smallskip}
SDSS J0155$-$0857 & A & $134\pm40$ & $73\pm12$ & $0.4\pm0.3$ & $3.5\pm0.3$ & ... & ... & AGN \\
 & B & $-7\pm42$ & $129\pm13$ & $1.7\pm0.8$ & $7.0\pm0.6$ & ... & ... & AGN \\
\noalign{\smallskip}
Mrk 1014 & A & $-100\pm28$ & $155\pm22$ & $9.1\pm1.8$ & $53.9\pm0.5$ & $26.1\pm5.1$\tablefootmark{d} & $14.6\pm0.7$ & AGN \\
 & B & $67\pm32$ & $229\pm17$ & $14.0\pm3.5$ & $40.1\pm0.9$ & $40.2\pm10.0$\tablefootmark{d} & $14.2\pm0.8$ & AGN \\
 & C & $-222\pm31$ & $113\pm28$ & $5.0\pm1.0$ & $23.0\pm0.5$ & $14.2\pm2.7$\tablefootmark{d} & $6.8\pm1.0$ & AGN \\
\noalign{\smallskip}
SDSS J0846$+$4426 & A & $-88\pm46$ & $132\pm3$ & $18.2\pm2.2$ & $198.2\pm2.3$ & ... & ... & AGN \\
 & B & $-229\pm46$ & $143\pm3$ & $14.2\pm1.4$ & $146.0\pm1.5$ & ... & ... & AGN \\
 & C & $-154\pm50$ & $130\pm8$ & $<5.4$ & $58.6\pm1.8$ & ... & ... & AGN \\
\noalign{\smallskip}
SDSS J1131$+$2623 & A & $-44\pm69$ & $117\pm3$ & $1.9\pm0.9$ & $32.6\pm0.6$ & ... & ... & AGN \\
 & B & $-32\pm68$ & $87\pm4$ & $1.6\pm0.3$ & $25.7\pm0.4$ & ... & ... & AGN \\
 & C & $-424\pm72$ & $54\pm8$ & $<1.3$ & $8.6\pm0.4$ & ... & ... & AGN \\
 & D & $131\pm76$ & $105\pm10$ & $<2.4$ & $13.9\pm0.8$ & ... & ... & AGN \\
\noalign{\smallskip}
PKS 1217+023 & A & $45\pm35$ & $75\pm27$ & $2.5\pm0.6$ & $9.3\pm0.3$ & ... & ... & AGN \\
 & B & $-61\pm37$ & $80\pm27$ & $3.4\pm1.1$ & $19.4\pm0.7$ & $6.1\pm0.8$ & $2.8\pm0.5$ & AGN \\
\noalign{\smallskip}
SDSS J1230$+$6621 & A & $299\pm24$ & $114\pm6$ & $4.0\pm0.9$ & $35.3\pm0.9$ & ... & ... & AGN \\
 & B & $67\pm23$ & $80\pm5$ & $2.9\pm0.4$ & $19.5\pm0.5$ & ... & ... & AGN \\
\noalign{\smallskip}
PKS 1545$+$210 & A & $309\pm47$ & $93\pm29$ & $4.5\pm2.4$ & $36.9\pm1.8$ & ... & ... & AGN \\
 & B & $-134\pm64$ & $212\pm51$ & $6.0\pm4.1$ & $35.5\pm4.7$ & ... & ... & AGN \\
\noalign{\smallskip}
PG 1612$+$261 & A & $70\pm29$ & $77\pm5$ & $<18.9$ & $111.0\pm4.7$ & ... & ... & AGN \\
 & B & $54\pm28$ & $91\pm3$ & $<8.2$ & $289.1\pm2.0$ & ... & ... & AGN \\
 & C & $248\pm28$ & $82\pm4$ & $18.2\pm3.7$ & $199.8\pm1.2$ & ... & ... & AGN \\
 & D & $-92\pm33$ & $105\pm8$ & $<4.8$ & $29.3\pm1.2$ & ... & ... & AGN \\
\noalign{\smallskip}
SDSS J1655$+$2146 & A & $-85\pm26$ & $161\pm2$ & $1.3\pm0.8$ & $127.9\pm0.9$ & ... & ... & AGN \\
 & B & $-85\pm27$ & $183\pm3$ & $20.4\pm2.6$ & $319.3\pm3.0$ & ... & ... & AGN \\
 & C & $-78\pm31$ & $197\pm7$ & $8.1\pm1.0$ & $44.3\pm1.2$ & ... & ... & AGN \\
 & D & $-128\pm30$ & $94\pm5$ & $<4.1$ & $43.5\pm1.4$ & ... & ... & AGN \\
\noalign{\smallskip}
PG 1700$+$518 & A & $-889\pm67$ & $158\pm39$ & $2.1\pm0.6$ & $9.2\pm0.8$ & ... & ... & AGN \\
\noalign{\smallskip}
PG 2130$+$099 & A & $-99\pm68$ & $180\pm105$ & $5.9\pm3.0$ & $3.5\pm0.7$ & $7.7\pm1.6$ & $3.8\pm0.4$ & \ion{H}{ii} \\
 & B & $-262\pm36$ & $176\pm32$ & $10.1\pm2.8$ & $22.1\pm1.3$ & $21.6\pm1.1$ & $9.5\pm0.3$ & AGN? \\
\noalign{\smallskip}
PHL 1811 & A & $49\pm27$ & $33\pm15$ & $3.4\pm0.4$ & $3.4\pm0.4$ & $14.6\pm1.0$ & $5.7\pm0.5$ & \ion{H}{ii} \\
\noalign{\smallskip}
HE 2158$-$0107 & A & $42\pm44$ & $69\pm26$ & $8.1\pm1.4$ & $39.6\pm1.5$ & $11.5\pm1.6$ & $<3.8$ & AGN \\
 & B & $-50\pm39$ & $39\pm23$ & $11.2\pm1.8$ & $108.5\pm1.9$ & $28.4\pm3.6$ & $<8.2$ & AGN \\
 & C & $-50\pm47$ & $0\pm30$ & $<3.6$ & $18.7\pm1.2$ & ... & ... & AGN \\
\noalign{\smallskip}
PKS 2349$-$014 & A & $-33\pm49$ & $172\pm24$ & $4.2\pm0.8$ & $6.9\pm0.3$ & $10.3\pm0.7$ & $5.7\pm0.4$ & \ion{H}{ii} \\
 & B & $-269\pm57$ & $299\pm21$ & $5.6\pm2.1$ & $20.8\pm1.1$ & $11.7\pm2.2$ & $13.2\pm1.2$ & AGN \\
 & C & $97\pm39$ & $146\pm22$ & $12.2\pm1.2$ & $96.6\pm0.9$ & $34.0\pm2.2$ & $36.8\pm1.6$ & AGN \\
 & D & $418\pm47$ & $0\pm26$ & $1.3\pm0.1$ & $5.8\pm0.3$ & ... & ... & AGN \\
\noalign{\smallskip}
HE 2353$-$0420 & A & $85\pm51$ & $225\pm17$ & $7.6\pm1.5$ & $77.9\pm2.0$ & $22.5\pm1.6$ & $18.0\pm1.0$ & AGN \\
 & B & $66\pm46$ & $189\pm18$ & $6.0\pm1.3$ & $116.5\pm1.8$ & $32.5\pm2.2$ & $26.6\pm1.6$ & AGN \\
\noalign{\smallskip}
\noalign{\smallskip}
\hline
\end{tabular}
\tablefoot{
\tablefoottext{a}{Name of QSOs followed with the alphabetically ordered character of the specific co-added spectra. }
\tablefoottext{b}{Radial velocity offset with respect to the rest-frame of the nuclear \mbox{[\ion{O}{iii}]}\ line. Negative values corresponds to a blue shift.}
\tablefoottext{c}{Emission line dispersion after the instrumental dispersion was subtracted in quadrature. A zero dispersion means that the line was unresolved.}
\tablefoottext{d}{The \mbox{H$\alpha$}\ line was strongly affected by telluric absorption and its flux was estimates from the \mbox{H$\beta$}\ line assuming zero dust extinction (\mbox{H$\alpha$}/\mbox{H$\beta$}$=$2.87).}
}

	\end{footnotesize}
\end{table*}
}
\subsection{Ionisation sources of the extended ionised gas}\label{pmas_sect:BPT_diagnostic}
Narrow-band imaging observations of large AGN samples have been the main tool for detecting and characterising EELRs \citep[e.g][]{Stockton:1987,Baum:1988,Mulchaey:1996,Bennert:2002,Schmitt:2003a,Schmitt:2003b,Privon:2008}. Most of these projects focused on a single emission line, which does not allow to constrain the origin of the gas ionisation. Three different ionisation mechanisms for the ISM on galactic scales need to be distinguished: 1. Photoionisation by the hard UV radiation field. 2. Shock ionisation due to winds or radio jets. 3. Photoionisation due to the ionising UV photons of young and massive OB stars in star-forming region, i.e. classical \ion{H}{ii} regions. 
\begin{figure}
\resizebox{\hsize}{!}{\includegraphics[clip]{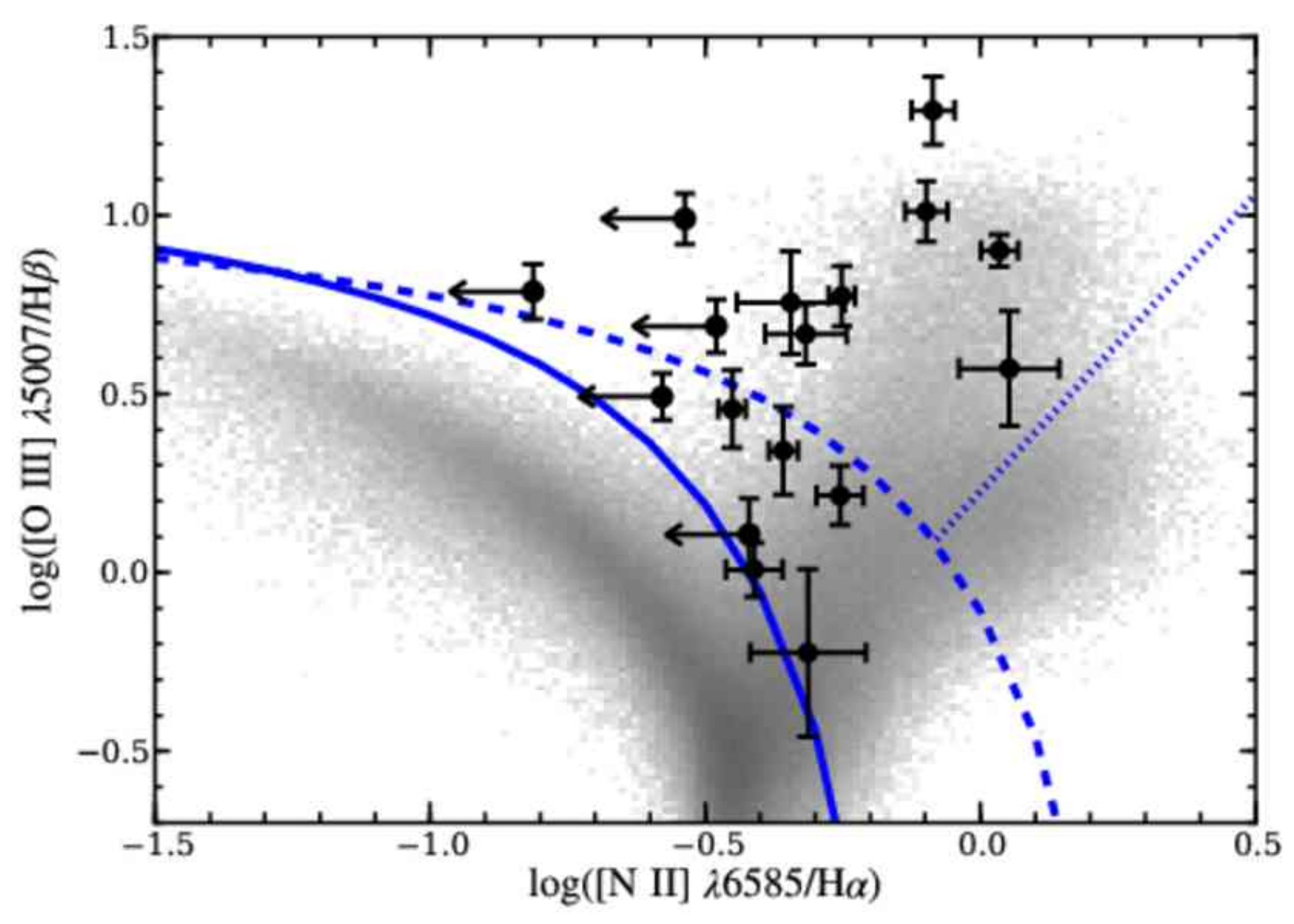}}
\caption{The standard BPT $\mbox{[\ion{O}{iii}]} \lambda 5007/\mbox{H$\beta$}$ vs. $\mbox{[\ion{N}{ii}]} \lambda6583/\mbox{H$\alpha$}$ diagram. The 2D histogram of the line ratios for more than 800\,000 emission-line galaxies taken from the SDSS DR7 value-added MPA/JHU galaxy catalogue are shown in the grey (logarithmic) scale image. All EELR regions of our sample for which both line ratios could be measured, even if one ratio is just a limit, are plotted as black symbols. Demarcation curves by \citet{Kewley:2001}, \citet{Kauffmann:2003} and \citet{Stasinska:2008} are drawn as solid, dashed and dotted blue lines, respectively.}
\label{fig_pmas:BPT_diagram} 
\end{figure}
It is possible to distinguish between different excitation mechanisms with the aid of diagnostic emission-line ratio diagrams \citep[e.g.][]{Baldwin:1981,Veilleux:1987,Villar-Martin:1997,Allen:1998}. The most commonly used diagnostic diagram in the optical is the \mbox{[\ion{O}{iii}]}\ $\lambda5007$/\mbox{H$\beta$}\ vs. \mbox{[\ion{N}{ii}]}\ $\lambda6583$/\mbox{H$\alpha$}\ diagram \citep{Veilleux:1987}, also known as the BPT diagram \citep{Baldwin:1981}. Those lines are usually the strongest in the optical spectrum and their ratios are most insensitive to reddening. Unfortunately, AGN photoionisation and shock-ionisation lead to very similar line ratios in the BPT diagram and cannot be reliable distinguished just from those lines. Anyway, both of these ionisation mechanisms are likely related to the AGN and we consider all corresponding line ratios being AGN ionised. AGN ionised  and \mbox{\ion{H}{ii}}\ regions populate two distinct branches  on this BPT diagram as shown in Fig.~\ref{fig_pmas:BPT_diagram}.  A theoretical limit for the emission-line ratios of \mbox{\ion{H}{ii}}\ regions that can be reached during a maximum starburst event was derived by \citet{Kewley:2001} as
\begin{equation}
 \log(\mbox{[\ion{O}{iii}]}/\mbox{H$\beta$} ) = 0.61/(\log(\mbox{[\ion{N}{ii}]}/\mbox{H$\alpha$})-0.47) + 1.19\ .
\end{equation}
Based on a large sample of SDSS emission-line galaxies, \citet{Kauffmann:2003} found that the AGN branch in the BPT diagram separates from \ion{H}{ii} branch well below the Kewley et al. demarcation curve and empirically defined  a new line that tightly encloses the \mbox{\ion{H}{ii}}\ branch being parametrised as
\begin{equation}
 \log(\mbox{[\ion{O}{iii}]}/\mbox{H$\beta$} ) = 0.61/(\log(\mbox{[\ion{N}{ii}]}/\mbox{H$\alpha$})-0.05) + 1.3\ .
\end{equation}
The region in between the two demarcation lines is often referred to as a composite region \citep{Kauffmann:2003,Kewley:2006} where the emission-line gas could be partially ionised by both mechanisms. However, this classification approach may be  oversimplified as pointed out by \citet{CidFernandes:2010}, because even emission-line ratios in the AGN area above the Kewley et al. curve may have some hidden contribution from \mbox{\ion{H}{ii}}\ region. Conversely, a region below that curve could be entirely powered by a massive starburst without requiring an AGN contribution. Furthermore, the AGN branch is composed of Seyfert type emission lines and Low Ionisation Nuclear Emission Regions \citep[LINERs,][]{Heckman:1980} with lower \mbox{[\ion{O}{iii}]}/\mbox{H$\beta$}\ at a given \mbox{[\ion{N}{ii}]}/\mbox{H$\alpha$}\ ratio. \citet{Stasinska:2008} defined a demarcation line between the two classes at a  angle of $59\degr$, counter clock-wise from a line of constant \mbox{[\ion{O}{iii}]}/\mbox{H$\beta$}, going through the point ($\log\mbox{[\ion{N}{ii}]}/\mbox{H$\alpha$}=-0.43,\log\mbox{[\ion{O}{iii}]}/\mbox{H$\beta$}=-0.49$). As can be seen from Fig.~\ref{fig_pmas:BPT_diagram}, LINER-type line ratios are not found for any extended ionised gas region in our sample.

We created a BPT diagram (Fig.~\ref{fig_pmas:BPT_diagram}) for all the co-added spectra in which  we could measure both line ratios, \mbox{[\ion{O}{iii}]}\ $\lambda5007$/\mbox{H$\beta$}\ \emph{and} \mbox{[\ion{N}{ii}]}\ $\lambda6583$/\mbox{H$\alpha$}. In the case that either \mbox{H$\beta$}\ or \mbox{[\ion{N}{ii}]}\,$\lambda6538$ was below the detection limit we used the $3\sigma$ upper limit to constrain at least an upper/lower limit on the corresponding line ratio. Most of the emission-line ratios are on the AGN side above the Kewley et al. line for which we assume that AGN ionisation is the \emph{dominant} process to ionise the gas. Region C of PG~0052+251, region A of PG~2130+099 and region A of PHL~1811 are almost directly on the \mbox{\ion{H}{ii}}\ branch and represent regions of ongoing star formation. In case of PHL~1811 the star forming region is coincident with spiral-like structure detected in the high-resolution HST image that was reported by \citet{Jenkins:2005}. Region A of PKS~2349$-$014 correspond to a physical companion located 4\arcsec\ West of the nucleus and its emission-line ratios would place it in the composite region of the BPT diagram. However, the H$\beta$ and H$\alpha$ lines are redshifted by $\sim$200\,$\mathrm{km}\,\mathrm{s}^{-1}$ with respect to the \mbox{[\ion{O}{iii}]}\ lines. This can be explained most likely due to an seeing induced superposition of emission lines corresponding to an \mbox{\ion{H}{ii}}\ region and AGN ionised gas and we classify this particular emission-line region as solely being due to star formation. These knots of star formation are likely to found in interacting/merging systems, besides a large AGN ionised region, as reported by \citet{Villar-Martin:2011a} from observations of type 2 QSO at redshift 0.3-0.6.

For the majority of regions we are only able to infer a single emission line ratio, either \mbox{[\ion{O}{iii}]}\ $\lambda5007$/\mbox{H$\beta$}\ or \mbox{[\ion{N}{ii}]}\ $\lambda6583$/\mbox{H$\alpha$}, so that the ionisation mechanism can generally not be determined unambiguously. We assumed that all regions with only an H$\alpha$ detection are characteristic for \mbox{\ion{H}{ii}}\ regions. If these regions were instead ionised by AGN, the \mbox{[\ion{O}{iii}]}\ line would be at least as bright as the H$\alpha$ line assuming a theoretical Balmer decrement of  H$\alpha$/H$\beta \sim 2.87$ \citep{Osterbrock:2006} and a \mbox{[\ion{O}{iii}]}/\mbox{H$\beta$}\ line ratio of $>$3 as a threshold for AGN ionised gas. Dust extinction along our line-of-sight within the host galaxy can dilute the \mbox{[\ion{O}{iii}]}\ and \mbox{H$\beta$}\ emission further, but a high dust extinction is often found around \mbox{\ion{H}{ii}}\ regions and is directly proportional to the current star formation rate (SFR) \citep[e.g.][]{Hopkins:2001,Sullivan:2001}. Thus 
AGN ionisation would be an unlikely scenario in those cases. 

On the other hand, it is clear from the two branches in the BPT diagram that the \mbox{[\ion{O}{iii}]}\ $\lambda5007$/\mbox{H$\beta$}\ line ratio alone is not sufficient to constrain the ionisation mechanism. An emission-line ratio of $\log(\mbox{[\ion{O}{iii}]}\ \lambda5007/\mbox{H$\beta$})>0.9$ can still be securely assigned to AGN ionisation which is the case for 7 objects. \citet{Ho:1997a} defined a threshold value of $\mbox{[\ion{O}{iii}]}\ \lambda5007/\mbox{H$\beta$}>3$ above which AGN ionisation is likely to be the dominant process. This can be physically explained by the fact that the star forming branch in the BPT diagram represents actually a sequence of increasing gas-phase metallicity from high to low \mbox{[\ion{O}{iii}]}/\mbox{H$\beta$}\ line ratios \citep[e.g.][]{Dopita:2000,Pettini:2004}. A high $\mbox{[\ion{O}{iii}]}\ \lambda5007/\mbox{H$\beta$}$ line ratio can reasonably excluded as a signature for star formation for  luminous AGN by considering the mass-metallicity relation of galaxies \citep[e.g.][]{Tremonti:2004}.  Luminous AGN reside almost exclusively in massive galaxies above $M_*>10^{10}\mathrm{M}_{\sun}$ \citep{Kauffmann:2003,Heckman:2004}, corresponding to a high gas-phase metallicity ($Z>\mathrm{Z}_{\sun}$).  Thus, we will rely on the Ho et al. criterion to classify the ionisation of our EELRs, but we caution that the argument as outlined above might not be correct for all EELRs as the gas-phase metallicity in galaxies typically decreases with distance from the galaxy centre \citep[e.g.][]{Henry:1999}.

From the detected regions of extended ionised gas we classify 13 of 19 to be entirely ionised by the AGN, 3 of 19 to be powered by \mbox{\ion{H}{ii}}\ regions only, and 3 of 19 exhibit signatures for both mechanisms dominating in spatially separated regions.  Interestingly, all the \mbox{\ion{H}{ii}}\ regions are located in disc-dominated galaxies or in the strongly interacting system PKS~2349$-$014. \citet{Letawe:2007} reported from their on-axis longslit spectroscopy of luminous QSOs that the ISM was in most cases ionised either by young stars or shocks. They found a completely AGN-ionised ISM only in the case of bulge-dominated galaxies, while their disc-dominated system or morphological disturbed galaxies show signatures of \mbox{\ion{H}{ii}}\ regions in almost every case. Since inactive disc-dominated galaxies are known to have young stellar population and continue to from star in their disc, it may not be surprising that this is true also for QSO host galaxies. 

\subsection{Dependence on nuclear properties}
Compared to our earlier work presented in Hu08, we increased our sample by 11 objects and improved the QSO-host deblending algorithm. Our main result in Hu08 was that the presence of an EELR is closely correlated with certain spectral properties of the QSO itself, most prominently the EW of \ion{Fe}{ii} and the FWHM of the broad H$\beta$ line. Both are the main drivers for the Eigenvector 1  parameter space \citep{Boroson:1992} describing most of the variance among rest-frame optical QSO spectra. While this result is still valid, some of the new EELR detections require additional notes. 

We classified I~Zw~1, RBS~219, PG~1700$+$518, PG~2130$+$099, and PHL~1811 as EELR non-detections in Hu08 for which we now find evidence for an EELR. All of them have a narrower H$\beta$ width than the proposed threshold at 4000\,$\mathrm{km}\,\mathrm{s}^{-1}$ and significant \ion{Fe}{ii} emission in their spectra. The main point about the EELRs around I~Zw~1, RBS~219 and PHL~1811 is that all the extended emission is ionised solely by star formation rather than AGN radiation. Thus, we refine our statement in Hu08 such that only the \emph{presence of extended AGN-ionised gas on kpc scales} is correlated with the nuclear properties as described above. Our improved deblending technique allows us also to detect a small ENLR close to the nucleus of PG~1700$+$518 and PG~2130$+$099. Similar to Mrk~1014, which we noted as an outlier in Hu08, these objects are known to host a triple radio source. Radio jets may therefore be an additional ingredient altering the general trends, which we will explore more closely in the next sections.

The additional 11 SDSS QSOs were specifically selected from their nuclear properties as candidates to show luminous ENLRs. This could be confirmed in most cases. Of the 4 EELR non-detections, 3 were observed through clouds (cf. Table~\ref{tbl:pmas_obslog})  which substantially reduced our chance to detect low surface brightness features. Thus, our statistics are not significantly increased compared to Hu08 and we focus on a detailed study of the ENLR\footnote{Of course an ENLR is also an EELR, but we want to emphasize the AGN-dominated ionisation source with a different naming following the literature. Thus, an ENLR may either be embedded in a larger EELR structure, or representing the entire EELR.}. The detected ENLRs in the remaining 7 new objects fit well into the correlations of Hu08, reinforcing the results reported in that paper.

\begin{figure*}
\includegraphics[width=\textwidth]{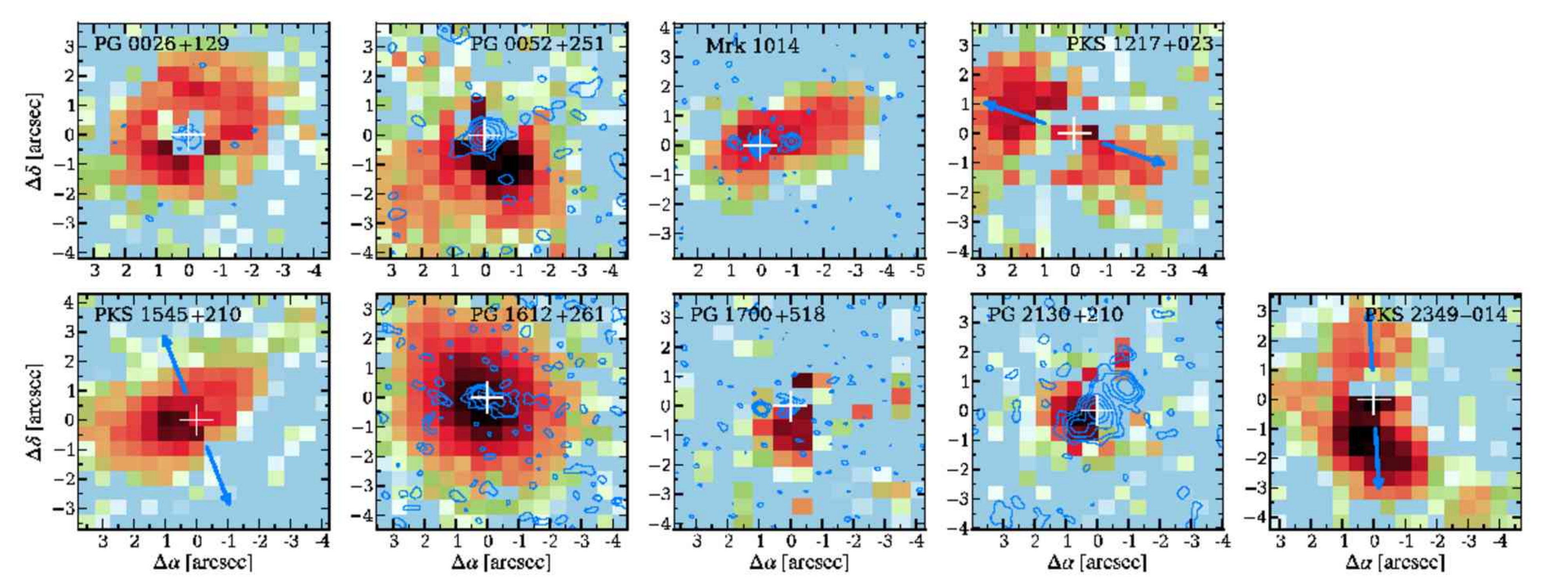}
\caption{The nucleus-subtracted ENLR light distribution of the \mbox{[\ion{O}{iii}]}\ line in comparison to the resolved radio emission. For the RQQs the high spatial resolution VLA radio images are overplotted as blue contours, while the primary radio axis of the radio lobes in RLQs is indicated by the blue arrows. The white cross marks the position of the QSO on which the radio emission is centred.}
\label{fig_pmas:EELR_radio_comp}
\end{figure*}
\section{Extended narrow-line regions around QSOs}\label{sect_pmas:ENLR_prop}
\subsection{The ENLR morphology}
\begin{table}
\centering
\begin{footnotesize}
\caption[test]{Luminosity and morphological parameters of the ENLR.}
\label{tab_pmas:EELR_flux_lum}
\begin{tabular}{lcccc}\hline\hline\noalign{\smallskip}
Name  & $\log L_\mathrm{ENLR}([\ion{O}{iii}])$  & Morph. & $e$ & PA  \\
      &  [$\mathrm{erg}\,\mathrm{s}^{-1}$]                 &         &    &  [\degr]     \\\hline\noalign{\smallskip}
\multicolumn{5}{c}{detected ENLRs}\\\hline\noalign{\smallskip}
PG 0026$+$129 & $41.80\pm 0.14$ &     round & $0.30$ & $-57$\\
PG 0052$+$251 & $41.75\pm 0.13$ & one-sided & ... & $34$  \\
SDSS J0155$-$0857 & $41.43\pm 0.19$ &     round & $0.39$ & $5$\\
Mrk 1014  & $42.22\pm 0.14$ & biconical & $0.60$ & $-65$\\
SDSS J0846$+$4426 & $43.08\pm 0.09$ &     round & $0.09$ & $83$\\
SDSS J1131$+$2623 & $42.11\pm 0.14$ & biconical & $0.64$ & $18$\\
PKS 1217+023 & $41.82\pm 0.16$ & biconical & $0.44$ & $62$\\
SDSS J1230$+$6621 & $41.90\pm 0.15$ &     round & $0.32$ & $5$\\
PKS 1545$+$210 & $42.35\pm 0.13$ & biconical & $0.45$ & $-49$\\
PG 1612$+$261  & $42.80\pm 0.12$ &     round & $0.20$ & $39$\\
SDSS J1655$+$2146 & $43.03\pm 0.09$ &     round & $0.19$ & $43$\\
PG 1700$+$518 & $41.81\pm 0.18$ & one-sided & ... & $-27$  \\
PG 2130$+$099 & $40.90\pm 0.32$ & one-sided & ... & $-28$  \\
HE 2158$-$0107  & $42.32\pm 0.13$ & biconical & $0.46$ & $-46$\\
PKS 2349$-$014 & $42.11\pm 0.11$ & biconical & $0.18$ & $6$\\
HE 2353$-$0420 & $42.47\pm 0.12$ &     round & $0.15$ & $57$\\
\noalign{\smallskip}
\hline
\multicolumn{5}{c}{undetected ENLRs}\\\hline\noalign{\smallskip}
I~Zw~1 & $<41.07$ & ... & ... & ...  \\
SDSS~J0057$+$1446 & $<41.90$ & ... & ... & ...  \\
RBS 219 & $<41.57$ & ... & ... & ...  \\
HE~0157$-$0406 & $<41.84$ & ... & ... & ...  \\
SDSS~J0948$+$4335 & $<42.66$ & ... & ... & ...  \\
3C~273 & $<42.10$ & ... & ... & ...  \\
HE~1228$+$0131 & $<41.51$ & ... & ... & ...  \\
SDSS~J1230$+$1100 & $<42.86$ & ... & ... & ...  \\
PG~1427$+$480 & $<42.42$ & ... & ... & ...  \\
SDSS~J1444$+$0633 & $<42.97$ & ... & ... & ...  \\
HE~1453$-$0303 & $<41.77$ & ... & ... & ...  \\
PHL 1811 & $<41.72$ & ... & ... & ...  \\
RBS 1812 & $<41.51$ & ... & ... & ...  \\
HE~2307$-$0254 & $<41.63$ & ... & ... & ...  \\
PG~2214$+$139 & $<41.03$ & ... & ... & ...  \\
\hline
\end{tabular}

\end{footnotesize}
\end{table}
As a first step, we classified the ENLR morphologically by eye into three categories: 'one-sided' (3 objects), 'biconical' (6 objects) and 'round' (7 objects). Assuming elliptical isophotes, we also measured the ellipticity $(1-b/a)$ of the ENLR from the light distribution at a level of $2\sigma$ above the background noise in the reconstructed [\ion{O}{iii}] narrow-band images. Furthermore, we estimated the position angle (PA) of the major ENLR axis by searching for the maximum flux within bipolar cones of 20\degr\ opening angle. All these parameters are summarised in Table~\ref{tab_pmas:EELR_flux_lum}. The ellipticities of the ENLRs agree with our visual impression of round and biconical morphologies with a transition value at $e\simeq0.4$. An exception is PKS~2349$-$014 because its ENLR is quite lopsided and the southern high surface brightness region dominates the whole ENLR structure.

We find a large fraction  of 37.5\% biconical ENLRs which even increases to 56\% when the one-sided ENLRs are also assumed to be intrinsically biconical. The biconical fraction is in agreement with the one determined by \citet{Schmitt:2003b} for type 1 Seyfert galaxies. However, it is expected that the opening angle of AGN ionisation cones increases with the luminosity of the AGN in the ``receding torus'' model \citep{Lawrence:1991}. Thus, biconical/elongated ENLRs should be less frequent than in lower luminosity Seyfert galaxies. This may indicate that the ENLR morphology is significantly affected by secondary properties, such as the gas density distribution with respect to the ionisation cones \citep{Mulchaey:1996b}, or by the presence and direction of a radio jet. 

A close alignment between the EELR and radio axis particularly around high-redshift RLQs and radio galaxies has been well established \citep{McCarthy:1987,McCarthy:1995}, which is often recognised also for lower luminosity Seyfert galaxies \citep{Wilson:1994,Capetti:1996,Falcke:1998,Nagar:1999,Schmitt:2003b} on much smaller physical scales. In Fig.~\ref{fig_pmas:EELR_radio_comp} we directly compare the ENLR light distribution against the radio morphology  for those QSO with available interferometric radio data. Among the 7 objects for which the PA of the radio jet is known, only PKS~1545$+$210 is unaligned with respect to the radio axis, while the axes are aligned within $\pm$20\degr\ (rms) for the majority of QSOs. This supports the view that radio jets even with relatively low power ($<3\times10^{24}\,\mathrm{W}\,\mathrm{Hz}^{-1}$ at 1.4GHz) can have a significant influence on the morphology and characteristics of the ENLR.

\subsection{ENLR size measurements}\label{sect_pmas:ENLR_sizes}
Previous ENLR surveys used various definitions for the ENLR size. Ground-based narrow-band images of 47 Seyfert galaxies were analysed by \citet{Mulchaey:1996}, who measured isophotal ENLR sizes at a limiting surface brightness of $2\times10^{-16}\,\mathrm{erg}\,\mathrm{s}^{-1}\,\mathrm{cm}^{-2}\,\mathrm{arcsec}^{-2}$. Circular annuli were used by \citet{Bennert:2002} to infer the ENLR radii at which the \mbox{[\ion{O}{iii}]}\ surface brightness dropped below the background surface brightness limit. Finally, \citet{Schmitt:2003a} estimated the maximum projected sizes of the \mbox{[\ion{O}{iii}]}\ light distribution by eye from their HST narrow-band images.  Thus, different ENLR size definitions  are not directly comparable with each other, and some of them depend critically on the depth of the data.

We measured three different ENLR sizes from the \mbox{[\ion{O}{iii}]}\ light distribution at $2\sigma$ above the background noise: 1) The $r_{95}$ radius defined as the circular aperture containing 95\%  of the ENLR \mbox{[\ion{O}{iii}]}\ flux, 2) the effective radius $r_\mathrm{e}$, defined as the luminosity-weighted mean of projected pixel distances to the centre, and 3) the maximum isophotal radius $r_\mathrm{iso}$ out to a limiting \mbox{[\ion{O}{iii}]}\ surface brightness of $2\times10^{-16}\,\mathrm{erg}\,\mathrm{s}^{-1}\,\mathrm{cm}^{-2}\,\mathrm{arcsec}^{-2}$. The measured ENLR sizes are reported in Table~\ref{tab_pmas:EELR_size} with errors based on Monte Carlo simulations of the QSO-host deblending process. We generated 200 different realisations of each object datacube consistent with the variance in each pixel, which we similarly analysed to infer 1$\sigma$ errors as the rms of the parameters. Because of the low spatial sampling of our reconstructed PMAS narrow-band images, we further added 0\farcs5 to the uncertainty of $r_{95}$, and $r_\mathrm{iso}$ as well as 0\farcs25 to the uncertainty of $r_\mathrm{e}$.  The seeing certainly increases the apparent size of the ENLR, but its strength strongly depends on the intrinsic emission line distribution and also on the size definition. We emphasise that any correction for the seeing would be a small effect given the large apparent ENLR sizes of several arcseconds, so that we did not perform any seeing correction. The ENLR apparently reaches beyond the PMAS FoV in three cases, HE~2158$-$0107, Mrk~1014, and PG~1612$+$261, for which we set lower limits to their ENLR sizes. For all other objects the surface brightness in the ENLR is sufficiently low at the edge of the PMAS FoV so that no significant extension of the ENLR is expected beyond the FoV.

\begin{table}
\centering
\begin{footnotesize}
\caption{Measured sizes of the ENLR.}
\label{tab_pmas:EELR_size}
\begin{tabular}{lcccccc}\hline\hline\noalign{\smallskip}
Name  & $r_{95}$ & $r_\mathrm{e}$ & $r_{\mathrm{iso}}$ \\
      &  [\arcsec] & [\arcsec] & [\arcsec]  \\\hline\noalign{\smallskip}
PG 0026$+$129 & $3.4\pm 0.7$ & $1.9\pm 0.3$ & $3.1\pm 0.8$ \\
PG 0052$+$251 & $3.4\pm 0.6$ & $1.7\pm 0.4$ & $4.1\pm 0.8$ \\
SDSS J0155$-$0857 & $2.0\pm 1.3$ & $1.0\pm 0.4$ & $2.0\pm 1.5$ \\
Mrk 1014 & $>3.9$ & $>1.9$ & $>4.3$ \\
SDSS J0846$+$4426 & $2.6\pm 1.1$ & $1.3\pm 0.4$ & $3.1\pm 1.4$ \\
SDSS J1131$+$2623 & $3.3\pm 0.8$ & $1.4\pm 0.4$ & $4.5\pm 1.0$ \\
PKS 1217+023 & $3.3\pm 0.6$ & $2.1\pm 0.4$ & $3.2\pm 0.7$ \\
SDSS J1230$+$6621 & $1.6\pm 0.9$ & $0.8\pm 0.4$ & $2.0\pm 0.8$ \\
PKS 1545$+$210 & $3.4\pm 0.7$ & $1.8\pm 0.4$ & $3.2\pm 0.7$ \\
PG 1612$+$261 & $>2.7$ & $>1.3$ & $>3.3$ \\
SDSS J1655$+$2146 & $2.3\pm 0.6$ & $1.0\pm 0.3$ & $3.1\pm 0.7$ \\
PG 1700$+$518 & $3.6\pm 1.0$ & $1.8\pm 0.4$ & $2.2\pm 1.8$ \\
PG 2130$+$099 & $2.5\pm 1.1$ & $1.3\pm 0.5$ & $1.8\pm 1.4$ \\
HE 2158$-$0107 & $>3.4$ & $>1.9$ & $>4.8$ \\
PKS 2349$-$014 & $3.1\pm 0.6$ & $1.8\pm 0.3$ & $5.1\pm 1.2$ \\
HE 2353$-$0420 & $2.5\pm 0.6$ & $1.1\pm 0.3$ & $2.3\pm 0.6$ \\
\noalign{\smallskip}
\hline
\end{tabular}

\end{footnotesize}
\end{table}

Because of the high frequency of elongated ENLR structures, we tested by how much longslit spectroscopic observations would underestimate the ENLR size if not properly aligned along the ENLR major axis. We extracted synthetic longslit surface brightness distributions from the narrow-band images along the major and minor axis of the ENLR. We find that ENLR sizes are underestimated on average by a factor of 2 when the slit is oriented along the major axis of the ENLR. For individual objects even a factor of 6 can be reached. Thus, the choice of the longslit position angle has a stronger impact on the inferred ENLR size than the 30\% reported by \citet{Greene:2011} based on their multiple longslit observations of type 2 QSOs. One possible explanation for this discrepancy could be the confusion of extended and unresolved \mbox{[\ion{O}{iii}]}\ emission of the nucleus, enhanced by the seeing. We can deblend these two component in our QSO using the broad emission lines, but those are not visible in the spectra of type 2 QSOs by definition.

\subsection{ENLR size-luminosity relation}

\begin{figure*}
 \includegraphics[clip,width=0.91\textwidth]{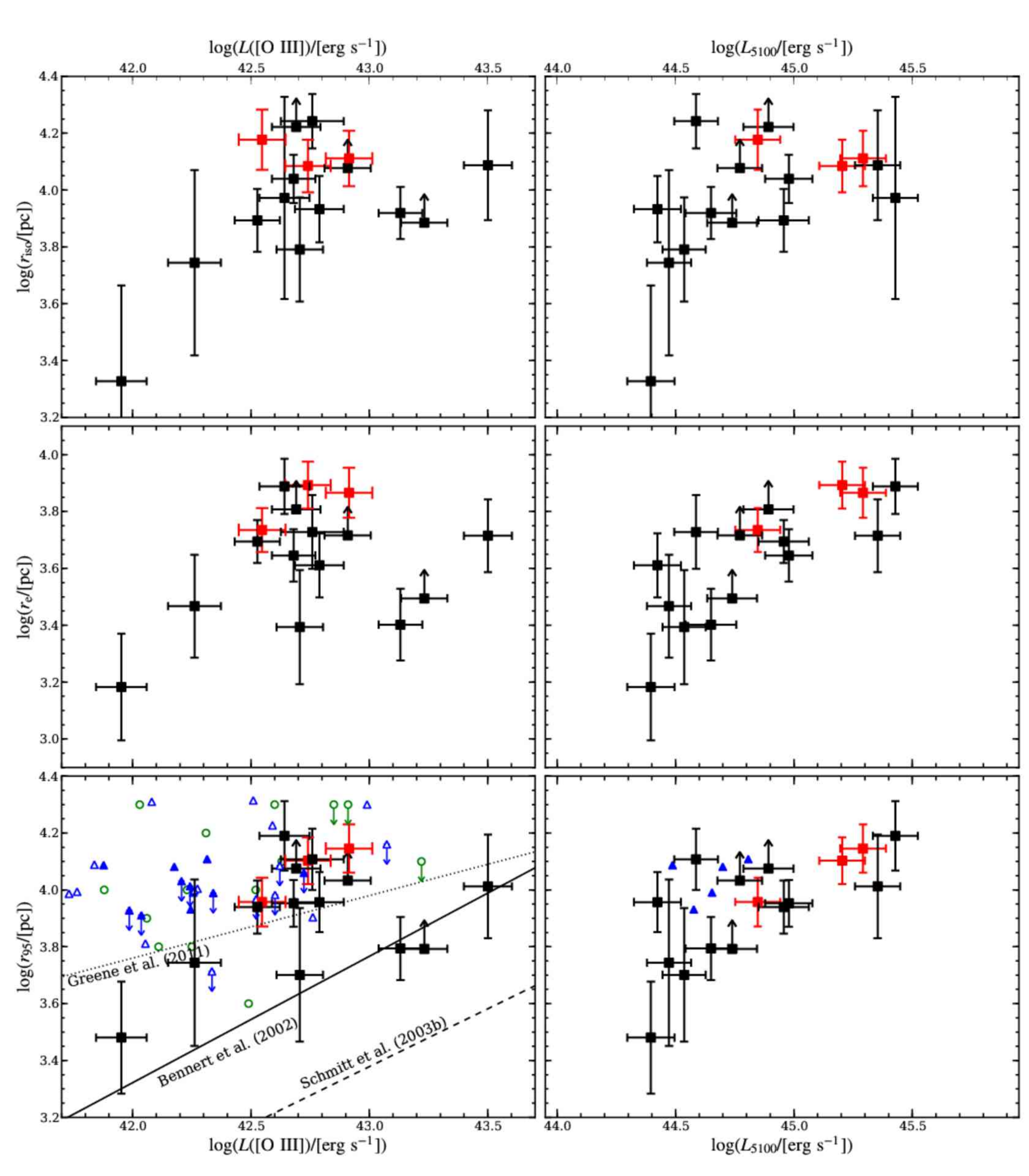}
  \caption{ENLR sizes as a function of total \mbox{[\ion{O}{iii}]}\ luminosity (left panels) and the QSO continuum luminosity at 5100\AA\ (right panels). The used size definitions are isophotal radius $r_\mathrm{iso}$ (upper panels), effective radius $r_\mathrm{e}$ (middle panels), and the radius containing 95\%\ of the ENLR flux $r_{95}$ (lower panels). The solid, dashed and dotted lines in the lower left panel correspond to the ENLR size-luminosity relations of \citet{Bennert:2002}, \citet{Schmitt:2003b}, and \citet{Greene:2011}, respectively. The black filled symbols represent RQQs whereas the red filled symbols highlight the few RLQs in our sample. The green opened circles are measurement for luminous type 2 QSOs as reported by \citet{Greene:2011} and the blue triangles are measurements for double-peaked AGN (open symbols denote type 2 and filled type 1 AGN) provided by \citet{Fu:2012}.}
  \label{fig_pmas:ENLR_size_lum}
\end{figure*}

 In Fig.~\ref{fig_pmas:ENLR_size_lum} we compare the projected physical ENLR size  against the \emph{total} \mbox{[\ion{O}{iii}]}\ luminosity (ENLR+QSO)  (left panels) and the bolometric luminosity $L_\mathrm{bol}$ of the QSO (right panels). The ENLR size appears to be almost uncorrelated with the \mbox{[\ion{O}{iii}]}\ luminosity given the low Spearman rank correlation coefficient and a high probability for no correlation ($\rho=0.27,P=0.3$, for $r_\mathrm{95}$). We find that  the correlation coefficient is significantly  higher for the QSO continuum luminosity $L_{5100}$ ($\rho=0.65,P=6\times10^{-3}$, for $r_\mathrm{95}$).  
The correlation between the ENLR size and the QSO continuum luminosity has not been investigated so far, because all previous studies focused either on type 2 AGN or on low-luminosity Seyfert nuclei which often exhibit a substantial contamination by the host galaxy continuum. \citet{Bennert:2002} compared the ENLR size also with the broad \mbox{H$\beta$}\ luminosity of the QSOs  which is well correlated with the QSO continuum luminosity \citep[e.g.][]{Greene:2005,Schulze:2009,Punsly:2011}. Their data also imply a higher correlation coefficient for the ENLR size as a function of \mbox{H$\beta$}\ luminosity which however they did not notice in the paper. The fact that ENLR size is much better correlated with the AGN continuum luminosity than with the normally used \mbox{[\ion{O}{iii}]}\ luminosity shows that the \mbox{[\ion{O}{iii}]}\ luminosity is a less reliable proxy for the amount of ionising photons compared to the continuum luminosity, although the \mbox{[\ion{O}{iii}]}\ luminosity has been commonly found to scale well with the QSO luminosity \citep[e.g.][]{Zakamska:2003,Heckman:2004}.

Because reliable ENLR characterisation are scarce for lower luminosity Seyfert 1 galaxies, we cannot study the ENLR size-luminosity relation in detail in this paper and restrict ourselves to average quantities. The mean size of the detected ENLRs for our type 1 QSOs is $r_\mathrm{e}=5\,\mathrm{kpc}\pm2\,\mathrm{kpc}$ and $r_\mathrm{95}=10\,\mathrm{kpc}\pm3\,\mathrm{kpc}$ with a similar value for $r_\mathrm{iso}$ at a median \mbox{[\ion{O}{iii}]}\ luminosity of $\log( L([\ion{O}{iii}])/[\mathrm{erg}\mathrm{s}^{-1}])= 42.7\pm0.15$. Just a few objects seem to follow the scaling relation of \citet{Bennert:2002}, while the majority are a factor of 2 larger than the extrapolation of their relation to the corresponding \mbox{[\ion{O}{iii}]}\ luminosities. The discrepancy is even larger for the scaling relation of \citet{Schmitt:2003b} for their Seyfert 1 sub-sample. This is may be unsurprising given the fact that the limiting surface brightness of the HST narrow-band images is at least a factor of 10 higher.  The samples of \citet{Greene:2011} and \citet{Fu:2012} have \mbox{[\ion{O}{iii}]}\ luminosities and limiting surface brightnesses ($\sim$1$\times10^{-16}\,\mathrm{erg}\,\mathrm{s}^{-1}\,\mathrm{cm}^{-2}\,\mathrm{arcsec}^{-2}$) comparable to our sample of type 1 QSOs. Their reported ENLR sizes are of the order of 10\,kpc, which matches with the results from our sample,  although their ENLR sizes seem a bit larger at lower \mbox{[\ion{O}{iii}]}\ luminosities (cf. Fig.~\ref{fig_pmas:ENLR_size_lum}). However, when considering the continuum luminosities for the few type 1 QSOs within the sample of \citeauthor{Fu:2012} this  discrepancy is largely reduced. 

The combined data of all studies strongly suggests that the ENLR scaling relation cannot be as different for type 1 and type 2 AGN as implied  by the results of \citet{Schmitt:2003b}. If we extrapolate their relation to an \mbox{[\ion{O}{iii}]}\ luminosity $\log( L([\ion{O}{iii}])/[\mathrm{erg}\mathrm{s}^{-1}])= 42.5$, it would imply an ENLR size at least a factor of 2 lower for type 2 QSOs than for type 1 QSOs.  Additionally, it is obvious that the zero-point of the ENLR size-luminosity relation based on HST data is too low because of the surface brightness sensitivity as discussed above. The slope of the relation is a different issue and may be a physically more interesting parameter, but cannot be inferred from our small range in luminosities. \citet{Greene:2011} determined a shallow ENLR size-luminosity relation only because ENLR sizes from Seyfert 2 galaxies were taken into account. Those were estimated by extrapolation of the NLR surface brightness profiles to large radii based on longslit spectroscopy \citep{Fraquelli:2003}. Whether this is a robust procedure to obtain reliable and unbiased (E)NLR sizes for low-luminosity AGN still needs to be confirmed by direct observations.

\citet{Netzer:2004} emphasised that the ENLR size-luminosity relation should not be a universal law. The ENLR size of luminous QSOs at high redshift, implied by the current relation, would exceed those of the host galaxies by more than an order of magnitude. Our QSOs are just at the limit where the ISM of the entire host galaxies is ionised. so that we do not yet see a break in the ENLR size-luminosity relation. On the other hand, we were unable to detect any ENLR for about half of our QSO sample, at similar QSO continuum luminosities. Thus, it is not clear whether a universal ENLR size-luminosity relation does actually exist. A few of our ENLR non-detections are associated with compact and apparently undisturbed host galaxies, such as PG~1427$+$480 (cf. Fig~\ref{fig_pmas:EELR_det}), which might have a comparably small ENLR size that cannot be resolved by our observations. It is possible that the host galaxy size imposes a natural limit for the ENLR because of the sharp decline in gas density. Alternatively, galaxy interactions/mergers might \emph{increase} the amount of available gas illuminated by the QSO in contrast to objects with non-detected ENLRs as suggested by \citet{Matsuoka:2012}, which would enhance the visibility of the ENLR and possibly also its size. We already emphasized in Hu08 that the detection or non-detection of an ENLR correlates with certain spectral properties of the QSO that drive the variance along the Eigenvector 1  parameter space \citep{Boroson:1992}. It is puzzling how the EW \ion{Fe}{ii} in the nuclear spectrum can be causally connected with properties of the ENLR on kpc scales. We doubt that this can be ascribed solely by galaxy interactions;  fundamental physical properties of the nucleus that are yet to be identified probably play an important role.

\subsection{ENLR luminosities}

\begin{figure}
\resizebox{\hsize}{!}{\includegraphics[clip]{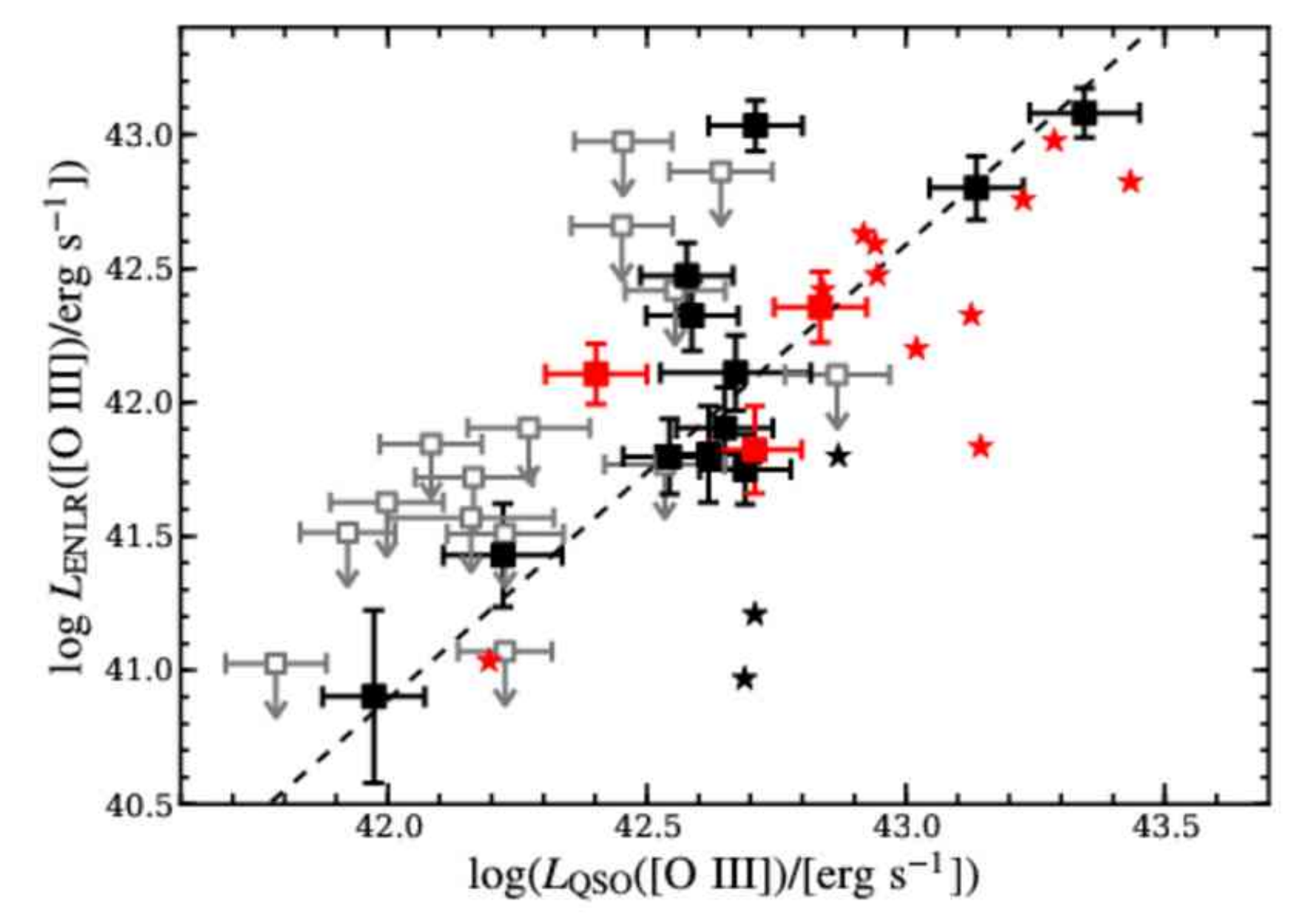}}
\caption{\mbox{[\ion{O}{iii}]}\ luminosity of the ENLR as a function of the unresolved QSO \mbox{[\ion{O}{iii}]}\ luminosity. Squared symbols indicate our own PMAS measurements and star symbols refer to the literature data of \citet{Stockton:1987}. Radio-quiet and radio-loud objects are marked as black and red symbols, respectively. The open grey symbol represent upper limits and the dashed line is a linear fit to our PMAS data (squared symbols) only.}
\label{fig_pmas:EELR_lum_O3_lum}
\end{figure}

In addition to the \mbox{[\ion{O}{iii}]}\ luminosities of the resolved ENLR we also obtained robust upper limits for $L_\mathrm{ENLR}(\mbox{[\ion{O}{iii}]})$ in the undetected cases.  We added ENLRs with the observed shapes to the original datacubes until the input luminosity reached the detection limit. We consider an ENLR to be detected at $3\sigma$ significance when the fluxes in 9 adjacent spaxel were above the $2\sigma$ background noise level in 95\%\ of 200 different Monte Carlo realisations. Here we implicitly assume that the ENLR luminosity is the limiting factor to detect an ENLR and not its size. The \mbox{[\ion{O}{iii}]}\ luminosity of the ENLRs ($L_\mathrm{ENLR}(\mbox{[\ion{O}{iii}]})$) is reported for all QSOs in Table~\ref{tab_pmas:EELR_flux_lum} and compared with the unresolved \mbox{[\ion{O}{iii}]}\ luminosity of the nucleus ($L_\mathrm{QSO}(\mbox{[\ion{O}{iii}]})$) in Fig.~\ref{fig_pmas:EELR_lum_O3_lum}.
Our measurements indicate a correlation of the two different luminosities that is in agreement with the derived upper limits. With the Astronomy Survival Analysis  package  \citep[ASURV,][]{LaValley:1992}  we determine a generalised Spearman rank correlation coefficients of $\rho=0.76$, taking the censored data into account. The probability for a non-correlation is $P<10^{-5}$ for the combined data set, which covers more than 2 orders of magnitude in \mbox{[\ion{O}{iii}]}\ luminosity. This possibly suggests that the nebulae are the extensions of the unresolved NLRs in terms of the covering factor for the ionising radiation of the AGN.
A linear regression analy\-sis with the Buckley-James method as part of ASURV  yields a relation of 
\begin{equation}
 \log(L_{\mathrm{ENLR}}([\ion{O}{iii}])) = (1.7\pm0.3) \log(L_{\mathrm{QSO}}([\ion{O}{iii}])) - (30.4\pm2.2)
\end{equation}

A similar trend was already reported from the \mbox{[\ion{O}{iii}]}\ narrow-band survey of low-redshift QSOs by \citet{Stockton:1987} and the narrow-band survey of local Seyfert galaxies carried out by \citet{Mulchaey:1996}. The detected EELRs in the \citet{Stockton:1987} sample are mostly associated with RLQs and appear to follow a similar scaling relation as implied by our data at high \mbox{[\ion{O}{iii}]}\ luminosities. That their measurements are \emph{systematically} lower than our relation can be easily explained by the fact that they excluded a central aperture with a radius of $\sim$3\arcsec\ to avoid contamination by the QSO light.

\begin{figure}
\resizebox{\hsize}{!}{\includegraphics[clip]{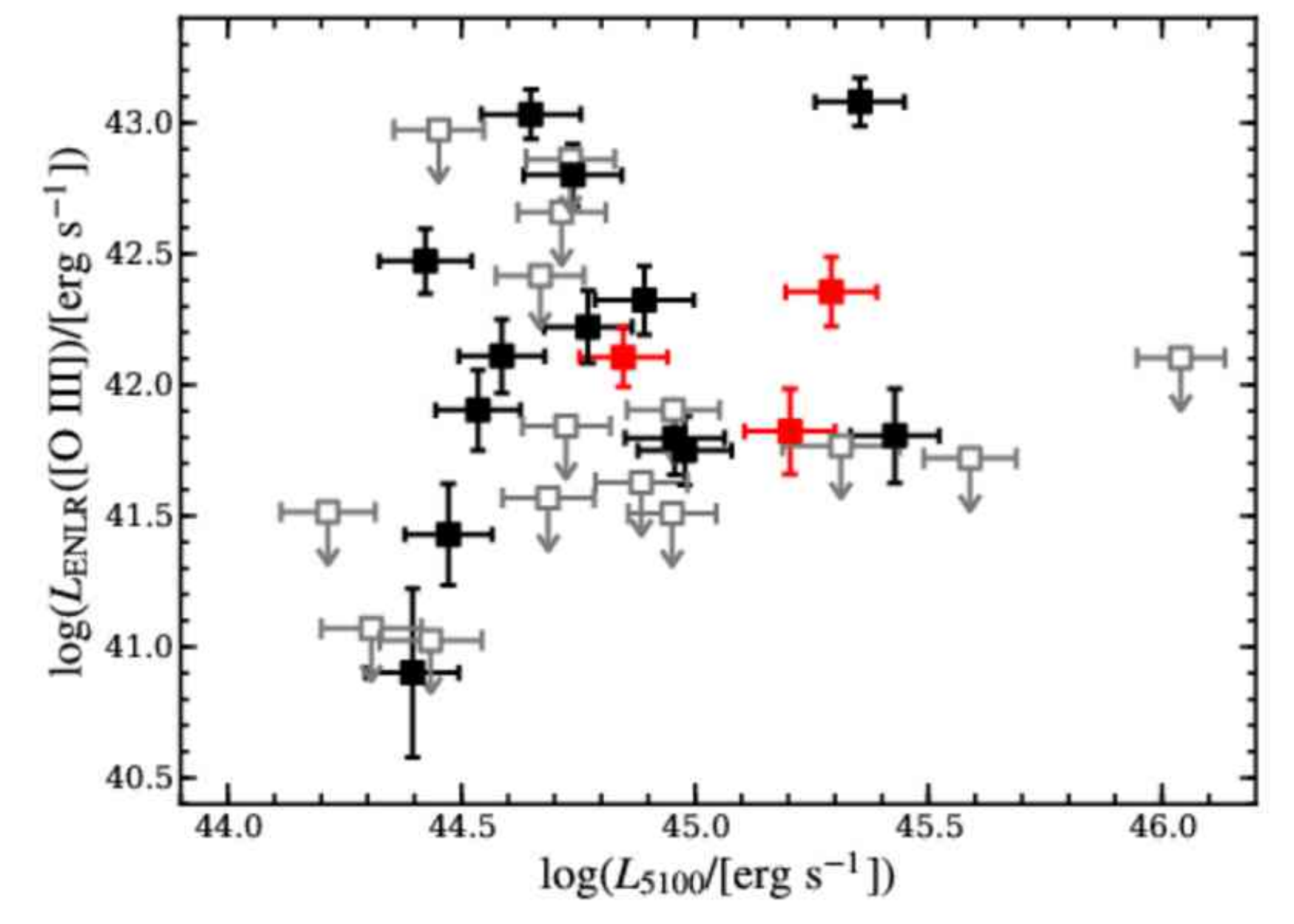}}
\caption{\mbox{[\ion{O}{iii}]}\ luminosity of the ENLR as a function of QSO continuum luminosity at 5100\AA. Symbols are as defined in Fig.~\ref{fig_pmas:EELR_lum_O3_lum}}
\label{fig_pmas:EELR_lum_bol_lum}
\end{figure}

If the \mbox{[\ion{O}{iii}]}\ luminosity was in general an indicator for the amount of ionising photons emitted by the AGN, we would also expect a significant correlation of the ENLR luminosity with the QSO continuum luminosity $L_{5100}$. We already reported in Hu08 that this is not the case and present the corresponding dia\-gram now for our extended sample in Fig.~\ref{fig_pmas:EELR_lum_bol_lum}. The generalised Spearman rank correlation coefficient computed with ASURV is $\rho=0.108$ with a probability of $P=0.55$ to be completely uncorrelated. This is rather surprising because of the strong correlation between the ENLR size and the continuum luminosity as shown in Fig.~\ref{fig_pmas:ENLR_size_lum}. 

\begin{figure}
\resizebox{\hsize}{!}{\includegraphics[clip]{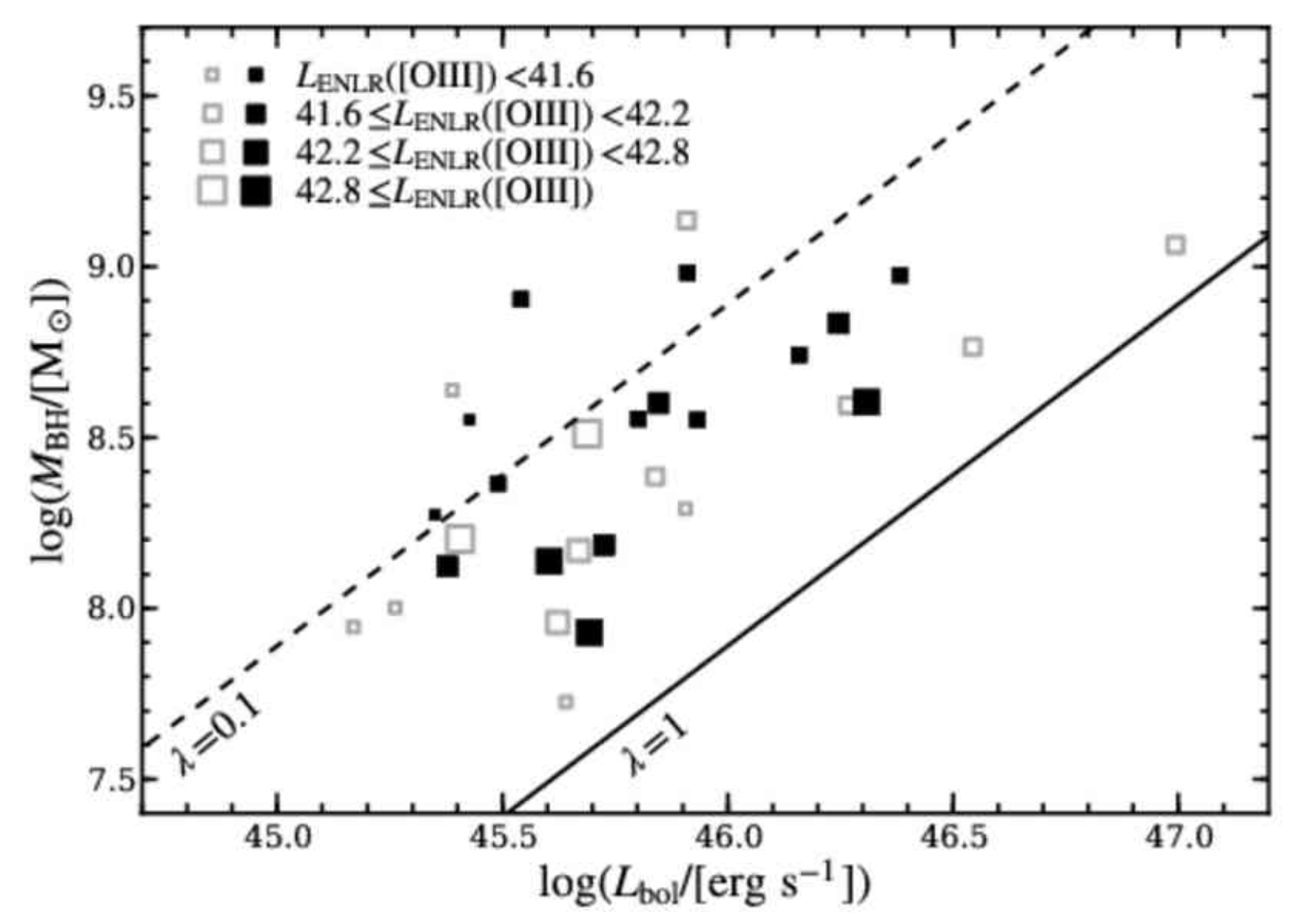}}
\caption{Distribution of QSOs with detected and non-detected ENLRs in the $M_\mathrm{BH}$-$L_\mathrm{bol}$ plane. The ENLR luminosity or its upper limit is encoded in the symbol size. QSO with detected ENLR are indicated by filled symbols and non-detections by grey open symbols, respectively. Solid and dashed lines correspond to constant Eddington ratios of $\lambda=1$ and $\lambda=0.1$, respectively.}
\label{fig_pmas:EELR_Mbh_QSO}
\end{figure}

To investigate the cause of this lack of correlation, \citet{Matsuoka:2012} compared the AGN continuum luminosity against the Eddington ratio using a large compilation of available QSO data including the one presented in this article. They quote to see a clear anti-correlation between the Eddington ratio and the ENLR detection rate and argue that this is likely caused by the availability of gas given the signs of interaction found in these systems \citep[e.g.][]{Husemann:2010,Villar-Martin:2011a}. We show the distribution of our QSOs with ENLR detections and non-detection in the $M_\mathrm{BH}$-$L_\mathrm{bol}$ plane in Fig.~\ref{fig_pmas:EELR_Mbh_QSO}. Many QSOs with a non-detected ENLR appear to populate a similar region than the ones with a detected ENLR. A two dimensional Kolmogorov-Smirnov (K-S) test \citep{Peacock:1983,Fasano:1987} confirms a high probability of $P=0.74$ that both samples (ENLR detections and non-detections) are drawn from the same parent population, which questions the trend proposed by \citet{Matsuoka:2012}. What we can confirm is that we see a deficit of ENLR for QSOs close to Eddington ratio ($\log \lambda>-0.5$), but the number of QSOs in our sample is too small in that regime to make a robust statements.

Nevertheless, the ENLR luminosity apparently cannot be determined solely by the amount of ionising photons, so that it seems indeed required to invoke the mass and distribution of gas in the host galaxy as additional parameters. Another aspect is the increasing evidence supporting the clumpy torus model \citep{Nenkova:2008,Mor:2009,Alonso-Herrero:2011}. In this model it is just a matter of probability if our line-of-sight is intercept by  dusty gas clouds of sufficient column density to obscure the nucleus. Thus, the intrinsic fraction of radiation that is able to escape into the host galaxy can have a large range even for apparent type 1 QSOs. It is therefore not surprising that the covering factor has been shown to be the dominant parameter that controls the [\ion{O}{iii}] luminosity in QSO spectra \citep{Baskin:2005}. Interestingly, it was recently reported by \citet{Xu:2012} that also the gas density of the ISM correlates with the nuclear properties in the sense that objects with narrow H$\beta$ lines and stronger \mbox{\ion{Fe}{ii}}\ emission have a systematically denser ISM. This could imply that our non-detected ENLRs may be explained simply by the low efficiency of ionising photons to ionise the dense ISM clumps due to self-shielding, It may also even explain the presence of star formation, as high gas density clumps are more favourable to have cool molecular gas cores.

\subsection{Dependence on radio properties}\label{sect:radio_prop}

\begin{figure}
\resizebox{\hsize}{!}{\includegraphics[clip]{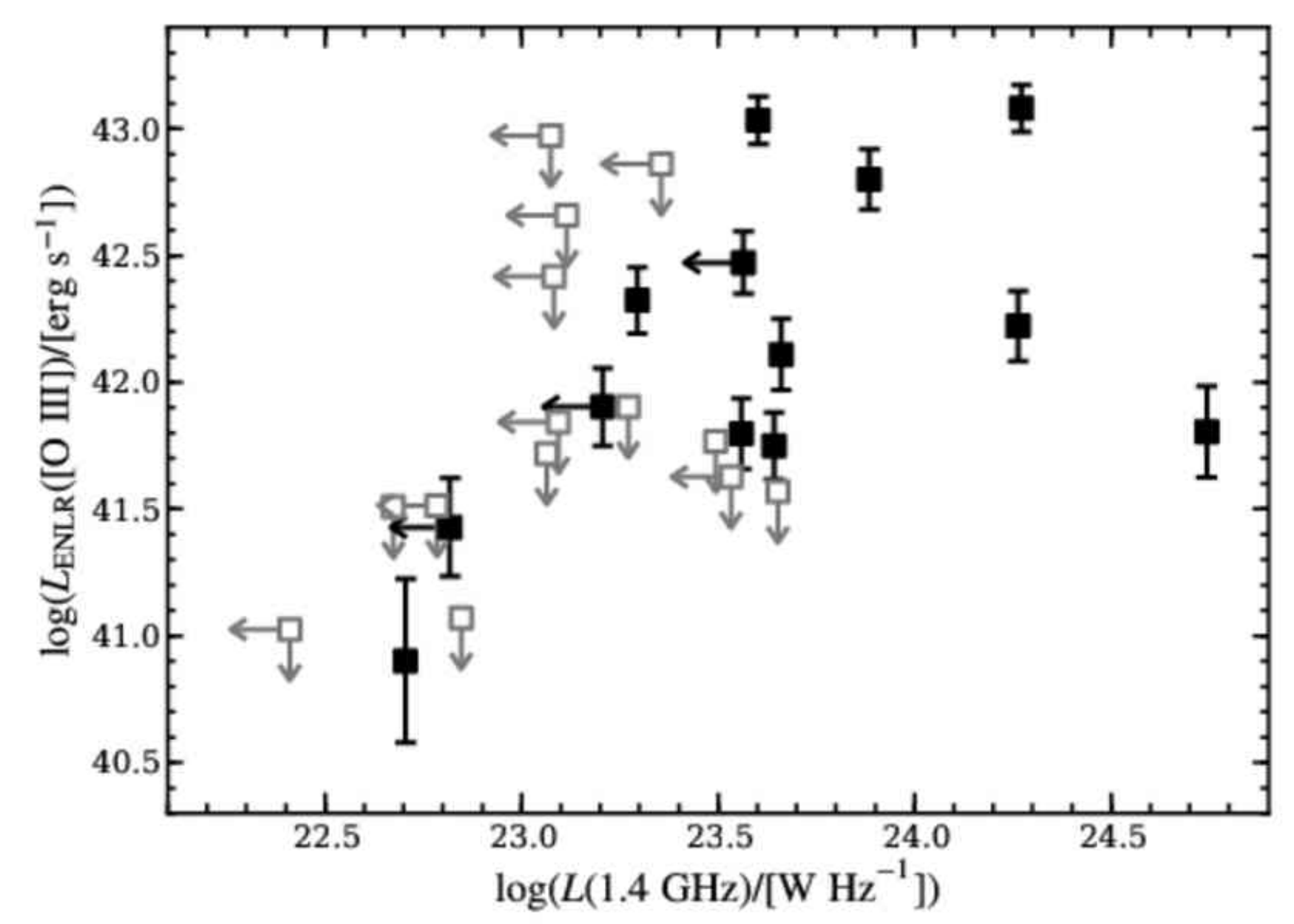}}
\caption{\mbox{[\ion{O}{iii}]}\ luminosity of the ENLR against radio luminosity at 1.4\,GHz. Black solid symbols refer to objects with detected and grey open symbols to objects with undetected ENLR (upper limits).}
\label{fig_pmas:EELR_lum_radio_pow}
\end{figure}

Radio observations have shown that the nuclear \mbox{[\ion{O}{iii}]}\ luminosity is correlated with the continuum radio luminosity in radio galaxies and RLQs \citep[e.g.][]{Baum:1989b,Rawlings:1991,Xu:1999} as well as in RQQs \citep[e.g.][]{Miller:1993,Xu:1999,Ho:2001}. The radio (6\,cm) to optical (4400\,\AA) flux ratio, the $R$ parameter, has often been used to distinguish between RQQs ($R<1$) and RLQs ($R>10$), so that their reported \mbox{[\ion{O}{iii}]}\ scaling relations often appear significantly offset from each other in terms of radio luminosity.  Because the radio emission in RLQs is often dominated by their extended lobes, the two classes may follow a common relation only if the radio core power is considered \citep{Xu:1999}. Here we compare $L_\mathrm{ENLR}(\mbox{[\ion{O}{iii}]})$ with the total radio luminosity at 1.4\,GHz (Fig.~\ref{fig_pmas:EELR_lum_radio_pow}) excluding the few known RLQs in our sample. It is difficult to argue about a possible correlation between these two quantities because of the numerous upper limits in both axes, radio  and ENLR luminosity. The generalised Spearman rank correlation coefficient is $\rho=0.56$ with $P=0.003$ for the non-correlation case.  Nevertheless, it seems that the ENLR luminosity depends more strongly on the radio luminosity than on the QSO continuum luminosity, which we find is supported by three facts:
\begin{itemize}
 \item[1.]  An ENLR is detected around \emph{all} QSOs with $\log(L(1.4\,\mathrm{GHz})/[\mathrm{W}\,\mathrm{Hz}^{-1}])>23.6$ (8 objects).
 \item[2.]  Ten of thirteen RQQs with a detected ENLR (77\%)  have a radio detection, while only five of fourteen RQQs with a non-detected ENLR (40\%) are detected in the radio. 
 \item[3.] In terms of radio-loudness, we detected an ENLR only around 26\% of the RQQs with $R<1$, but around 87.5\% of the QSOs with $1<R<10$.
\end{itemize}

The ENLR luminosities of RLQs (not included in Fig.~\ref{fig_pmas:EELR_lum_radio_pow}) are of the order of their radio-quiet counterparts, although the integrated radio luminosity is more than two orders of magnitude higher. Whether this is an intrinsic property or caused by the limited FoV of our IFU observation is something we are not able to say at present. 

\citet{Sulentic:2000} found that the broad H$\beta$ line width of RLQs is almost always $>$4000$\,\mathrm{km}\,\mathrm{s}^{-1}$, and defined two different populations of QSOs. Population A with $\mathrm{FWHM}_{\mathrm{H}\beta}$\,$\leq$4000$\,\mathrm{km}\,\mathrm{s}^{-1}$ is entirely dominated by RQQs  and Population B  with $\mathrm{FWHM}_{\mathrm{H}\beta}$\,$>$4000$\,\mathrm{km}\,\mathrm{s}^{-1}$ contains all RLQs and a number of RQQs with otherwise similar characteristics. These two  populations represent opposite regimes along the famous Eigenvector 1 \citep{Boroson:1992}, which ultimately links the width of \mbox{H$\beta$}, the strength of the \mbox{\ion{Fe}{ii}}\ complexes and the relative strength of \mbox{[\ion{O}{iii}]}\ with respect to \mbox{H$\beta$}. The physical drivers for this sequence in parameter space are thought to be $M_\mathrm{BH}$ and $\lambda$ \citep{Boroson:2002,Marzini:2003b}. In Hu08 we emphasised that a luminous ENLR is preferentially found around QSOs with $\mathrm{FWHM}_{\mathrm{H}\beta}>4000\,\mathrm{km}\,\mathrm{s}^{-1}$ and weak \mbox{\ion{Fe}{ii}}\ emission that would correspond exclusively to Population A. Our extended sample now contains a few objects, such as the radio-intermediate QSOs PG~1612$+$261 and SDSS~1655$+$2146, which are surrounded by a luminous ENLR but have a $\mathrm{FWHM}_{\mathrm{H}\beta}<3000\,\mathrm{km}\,\mathrm{s}^{-1}$ and therefore do not correspond to Population A as one would have na\"ively expected. This is in agreement with the results of \citet{Zamfir:2008} who found that radio-intermediate QSOs also cover the full range in broad \mbox{H$\beta$}\ widths like RQQs and are not restricted to Population A like RLQs.

\begin{figure}
\resizebox{\hsize}{!}{\includegraphics[clip]{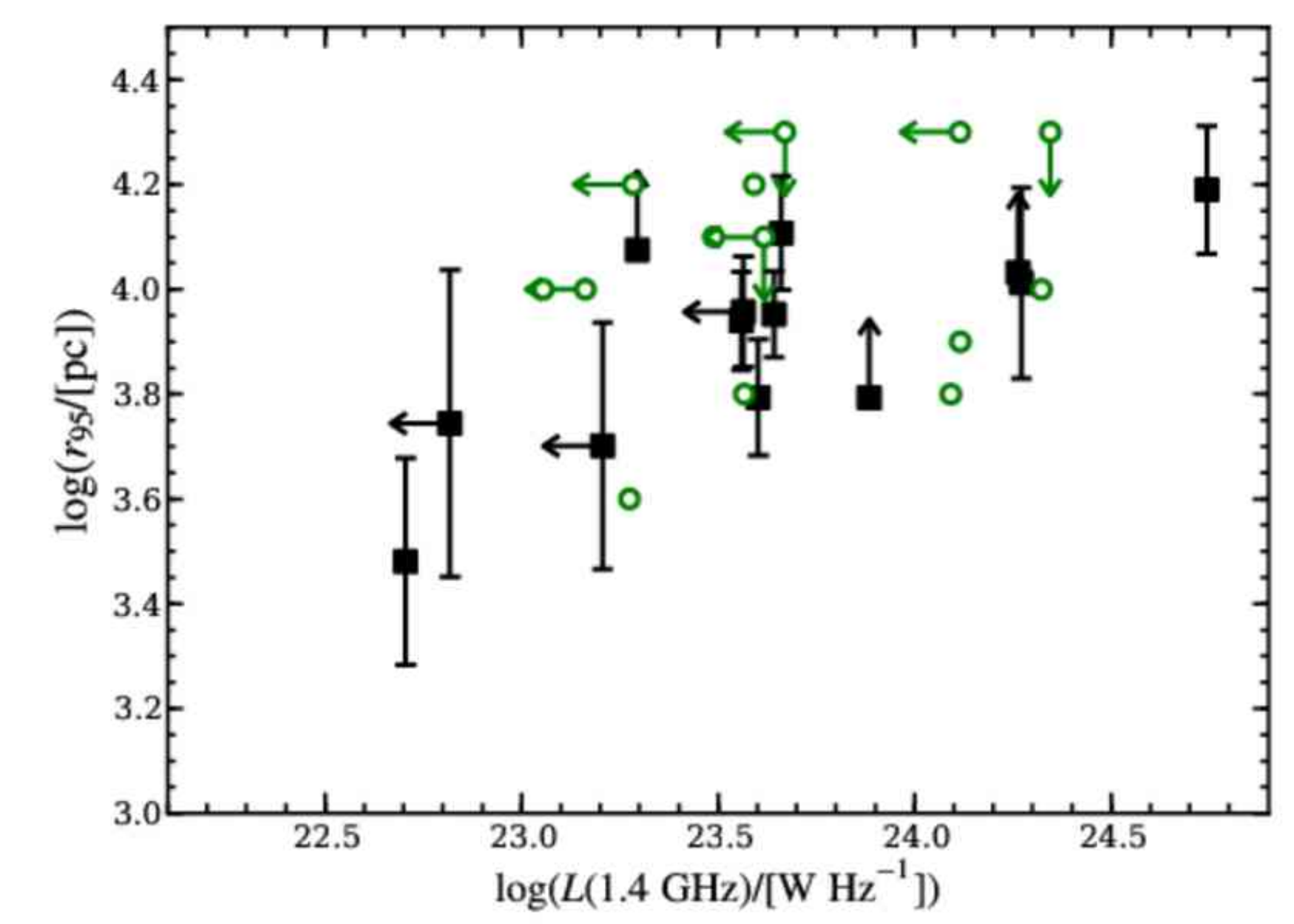}}
\caption{ENLR size as a function of radio luminosity at 1.4\,GHz. Measurements for our RQQs with detected ENLR are indicated by the black filled symbols and open green symbols denote measurements for type 2 QSOs as reported by \citet{Greene:2011}. }
\label{fig_pmas:EELR_size_radio_pow}
\end{figure} 

Furthermore, \citet{Leipski:2006b} reported a significant trend between the size of the radio jet in RQQs and the ENLR. Because the radio emission is resolved only for very few of our QSOs, we investigate instead whether the ENLR size depends on the radio luminosity (Fig.~\ref{fig_pmas:EELR_size_radio_pow}) including the recent measurements for type 2 QSOs by \citet{Greene:2011}. Only a weak trend between the two quantities seems to be present if  there is one at all. Given our tentative evidence that the radio luminosity is more strongly correlated with the ENLR luminosity than its size, we speculate that a radio jet increases the covering factor of the gas for the ionising photons of the nucleus, reaching out to larger distances. This would well match with the scenario proposed by \citet{Hopkins:2010}, who argued that a weak outflow in a hot diffuse medium can possibly increase the impact of the AGN on the host galaxy via a two-stage process. First, cold molecular gas clouds expand or are destroyed by the outflow induced pressure gradients and instabilities, which could be generated by a radio jet. Second, the molecular clouds are more efficiently illuminated by the QSO radiation, because of their increased cross-sections and covering factor, so that the AGN can dissipate more energy into the medium.

\section{Kinematics of the extended ionised gas}\label{sect_pmas:EELR_kin}
\begin{figure*}
\hspace*{2cm}\includegraphics[width=0.75\textwidth,clip]{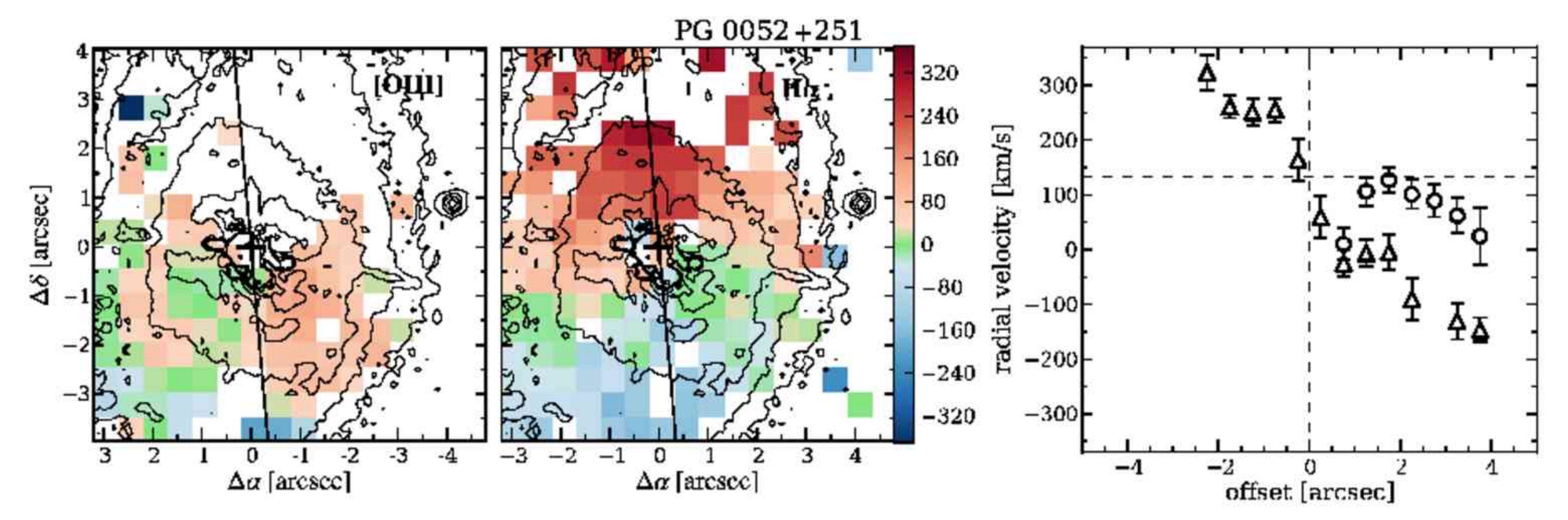}
\includegraphics[width=0.5\textwidth,clip]{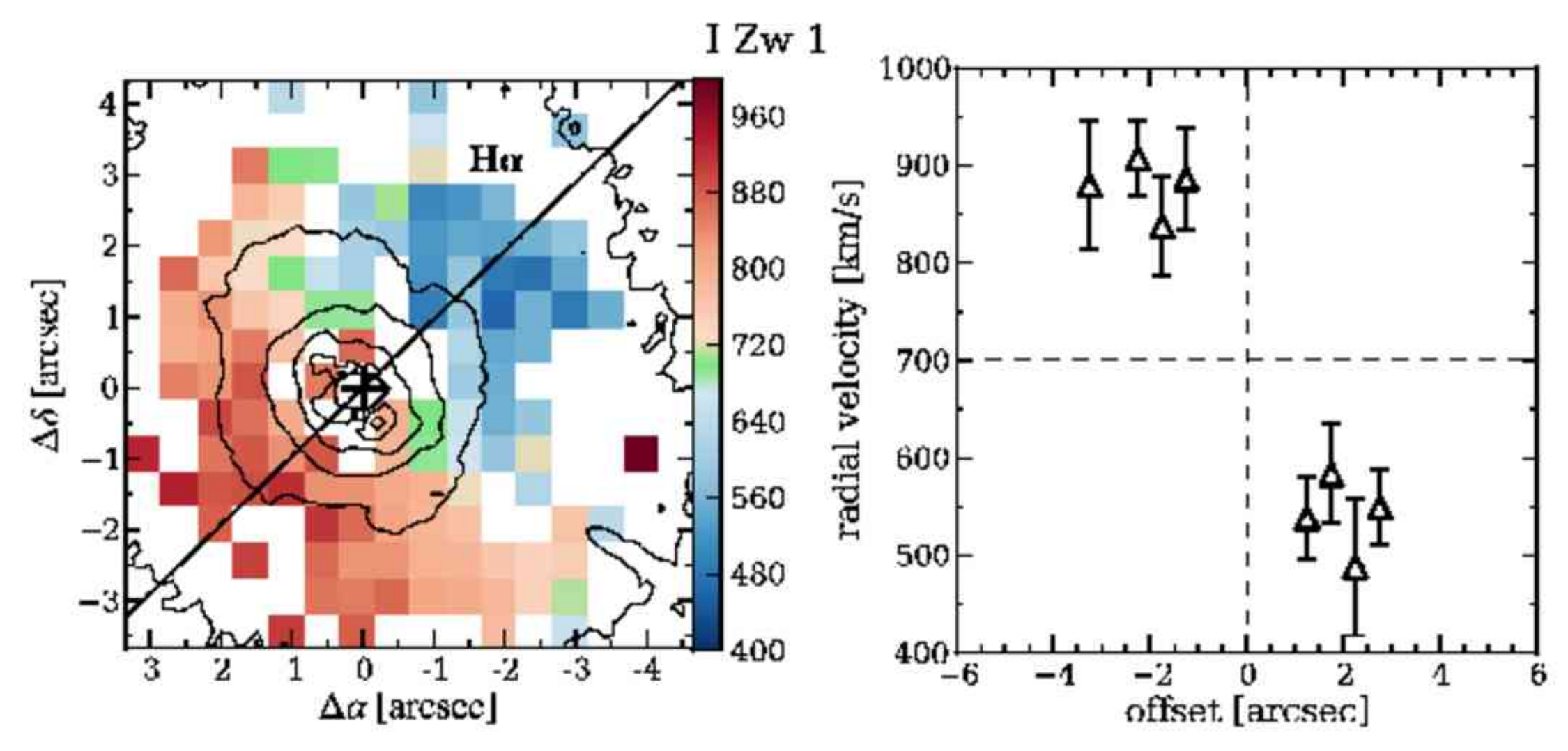}
\includegraphics[width=0.5\textwidth,clip]{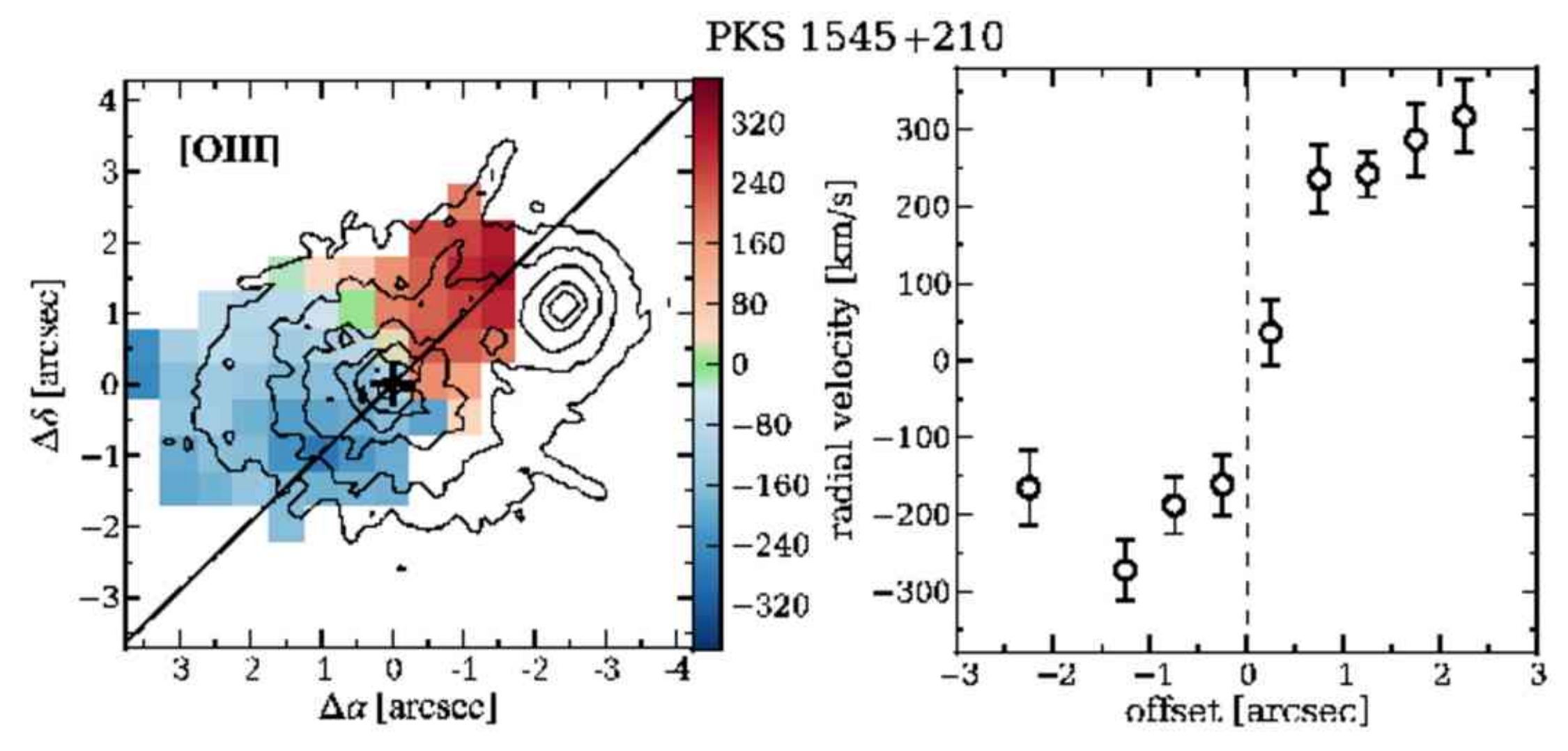}
\includegraphics[width=0.5\textwidth,clip]{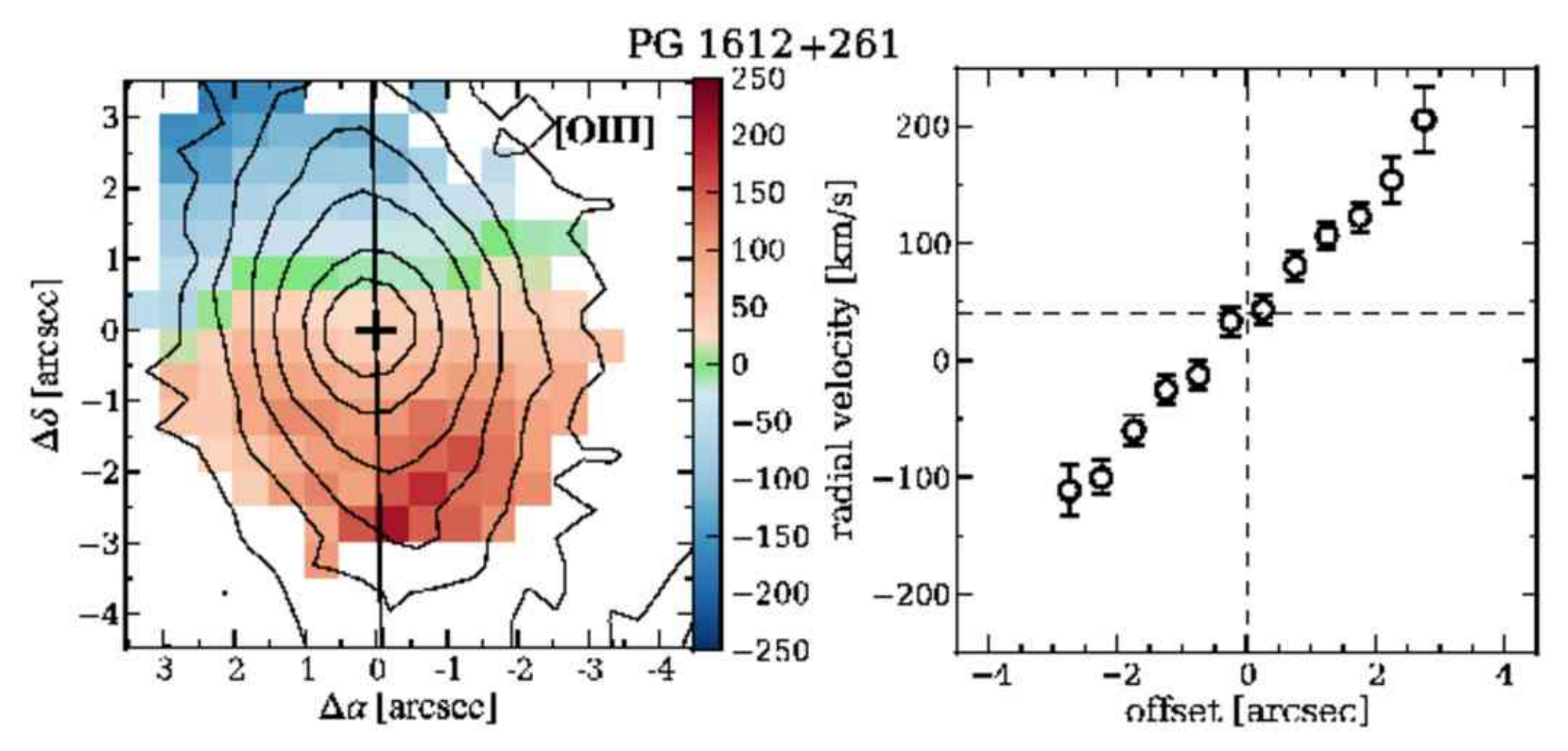}
\includegraphics[width=0.5\textwidth,clip]{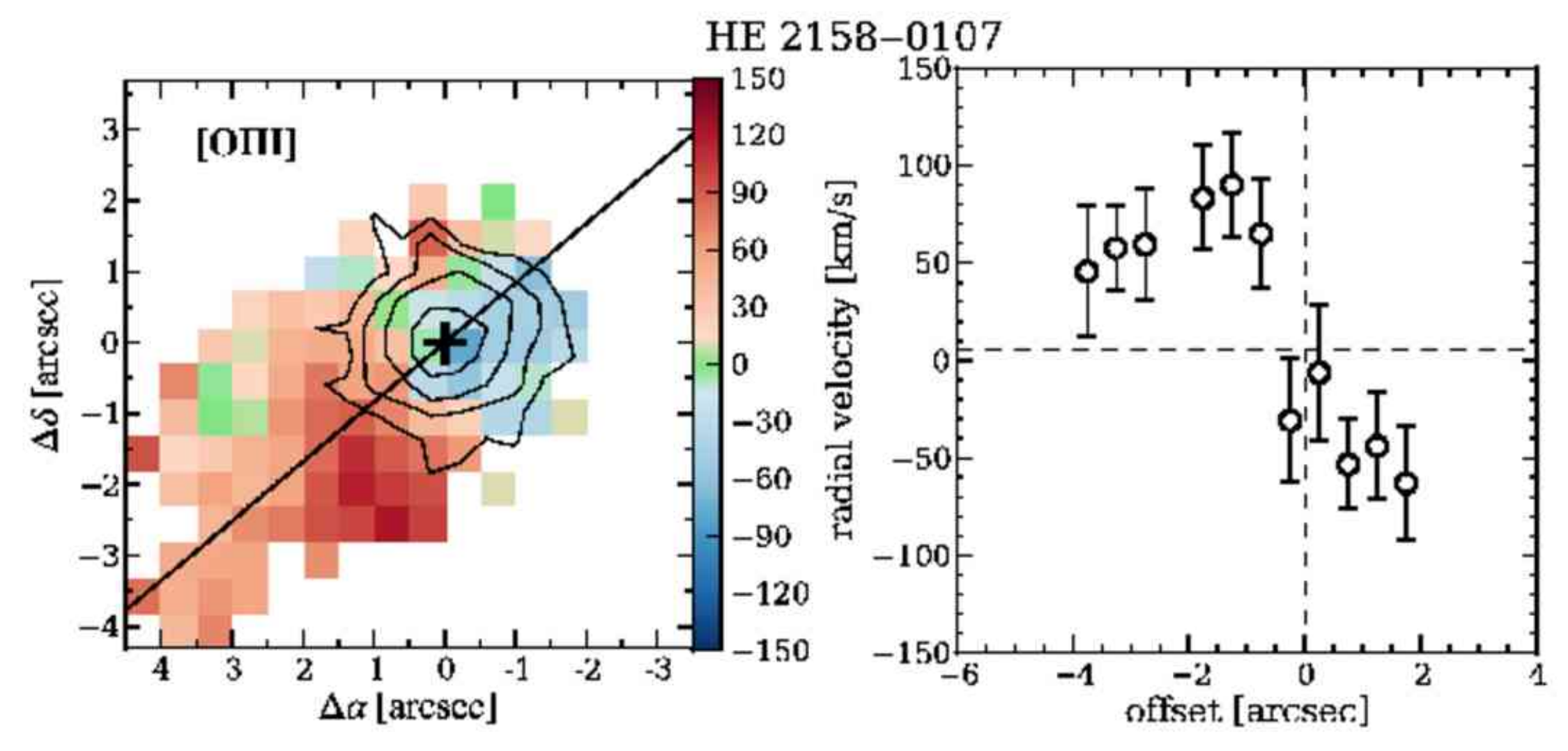}
\includegraphics[width=0.5\textwidth,clip]{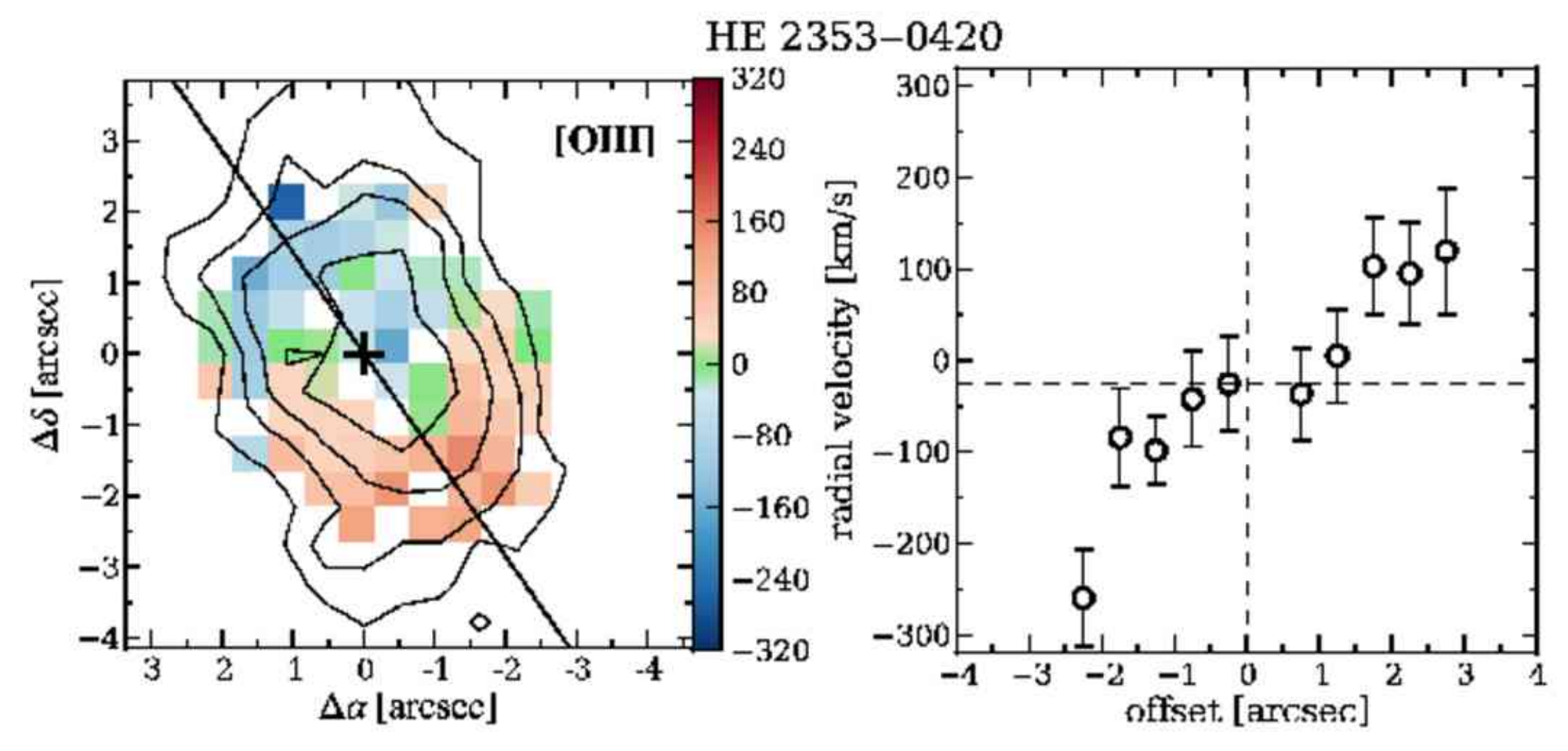}
\caption{2D velocity maps for objects with apparently large-scale rotational motion gas motion. The continuum light distribution of broad-band images are overplotted as thin solid black contours for comparison. A ``longslit'' velocity curves along the apparent major kinematic axis is extracted from the maps as described in the text. A thick solid black line indicates the orientation of the major kinematic axis. The radial velocity are estimated for each spaxel by modelling the corresponding emission line with a  Gaussian profile. The vertical dashed line highlights the position of the QSO along the synthetic slit and the horizontal dashed line indicates the measured velocity zero-point of the velocity curve.}
\label{fig_pmas:EELR_kin_rot}
\end{figure*}

\subsection{Construction of 2D velocity fields}
In order to study the kinematics of the EELR we used the brightest emission lines to infer the 2D radial velocity distribution of the ionised gas. We modelled the \mbox{[\ion{O}{iii}]}\ $\lambda\lambda4960,5007$ doublet  and/or the H$\alpha$ emission line adopting a simple Gaussian line profile in each spaxel of the nucleus-subtracted datacubes. The line widths of the doublet lines were coupled and their line ratio was fixed to the theoretical value. We also modelled each spectrum assuming a Gauss-Hermite line profile up to the fourth order following \citet{van-der-Marel:1993} to identify asymmetric line profiles in the EELR not caused by noise. Asymmetric lines are not frequently found in the EELR of our QSO sample (see co-added spectra in Fig.~\ref{fig_pmas:EELR_det}). In the following, we first present the 2D velocity fields for the bulk gas motion only. When line asymmetries are significant we replace the radial velocity of the single Gaussian fit with the mean radial velocity of the Gauss-Hermite result at that location. In addition to the 2D velocity maps, we constructed synthetic ``longslit'' curves along a certain PA directly from the maps in a geometrically manner.  We averaged the velocities of all spaxel that are covered by a hypothetical slit of 1\farcs3 width at an angular separation of 0\farcs5 along the slit axis.
The QSOs PG~1700$+$518, PHL~1811 and PG~2130$-$099 were excluded from this 2D analysis, as their EELR are too small in spatial extent. Still, it is worthwhile to note that we measured a redshifted radial velocity of $\sim$800$\,\mathrm{km}\,\mathrm{s}^{-1}$ for the EELR structure of PG~1700$+$518 (cf. Table~\ref{tab_pmas:line_ratios}). Furthermore, we needed to exclude RBS~219 and SDSS~J0155$-$0857, for which the low S/N of the extended emission lines did not allow a spatially resolved analysis.  

\begin{figure*}
\includegraphics[width=0.5\textwidth,clip]{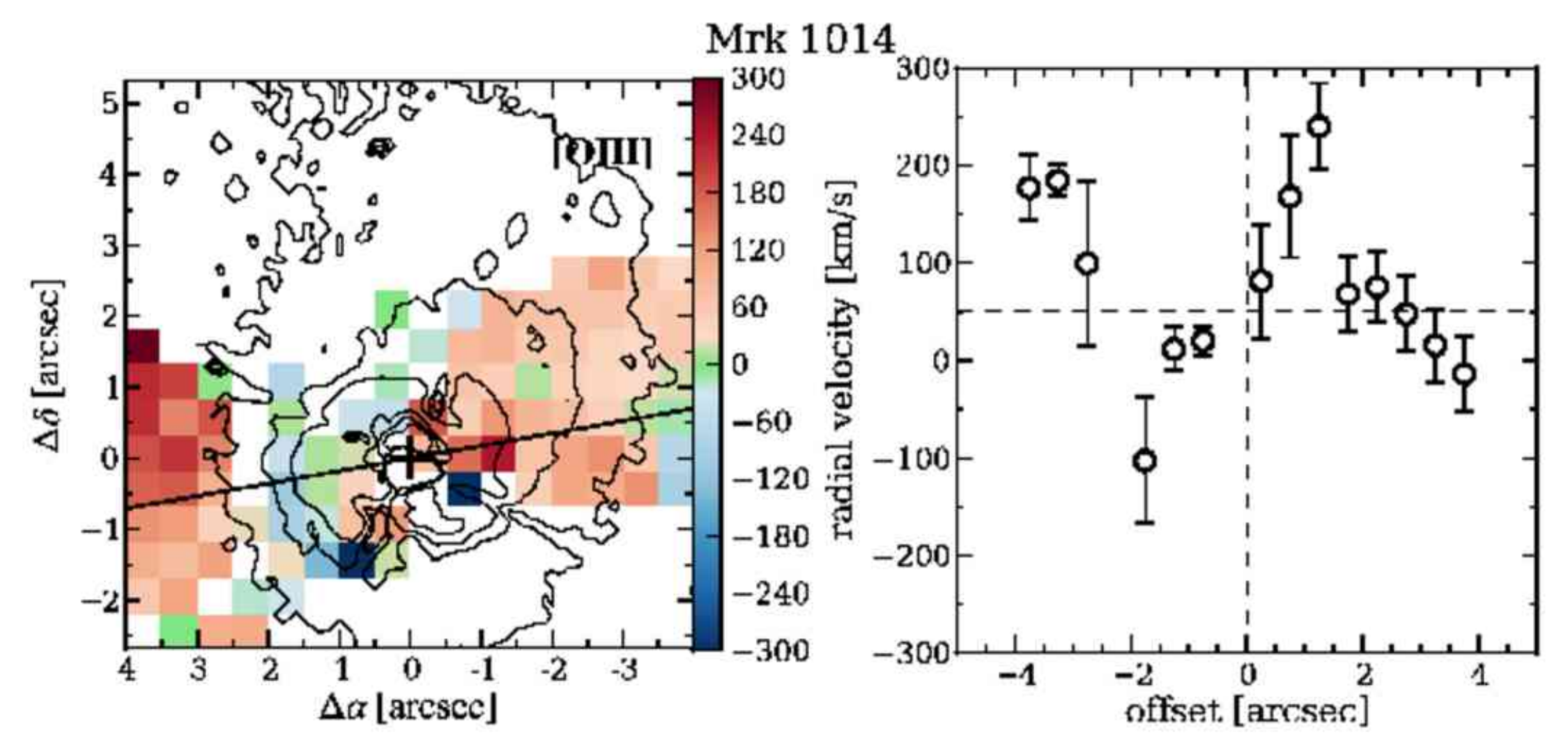}
\includegraphics[width=0.5\textwidth,clip]{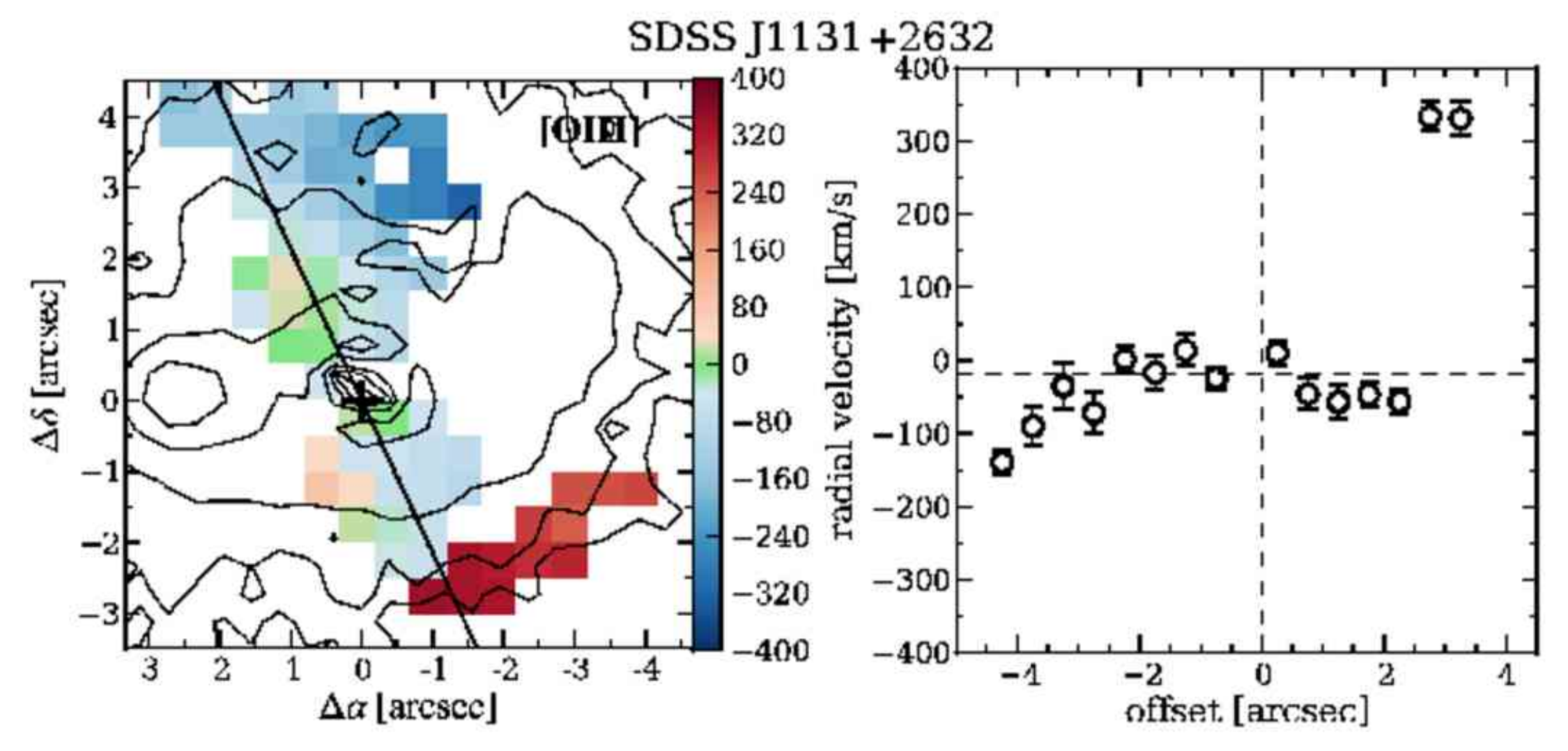}
\hspace*{2cm}\includegraphics[width=0.75\textwidth,clip]{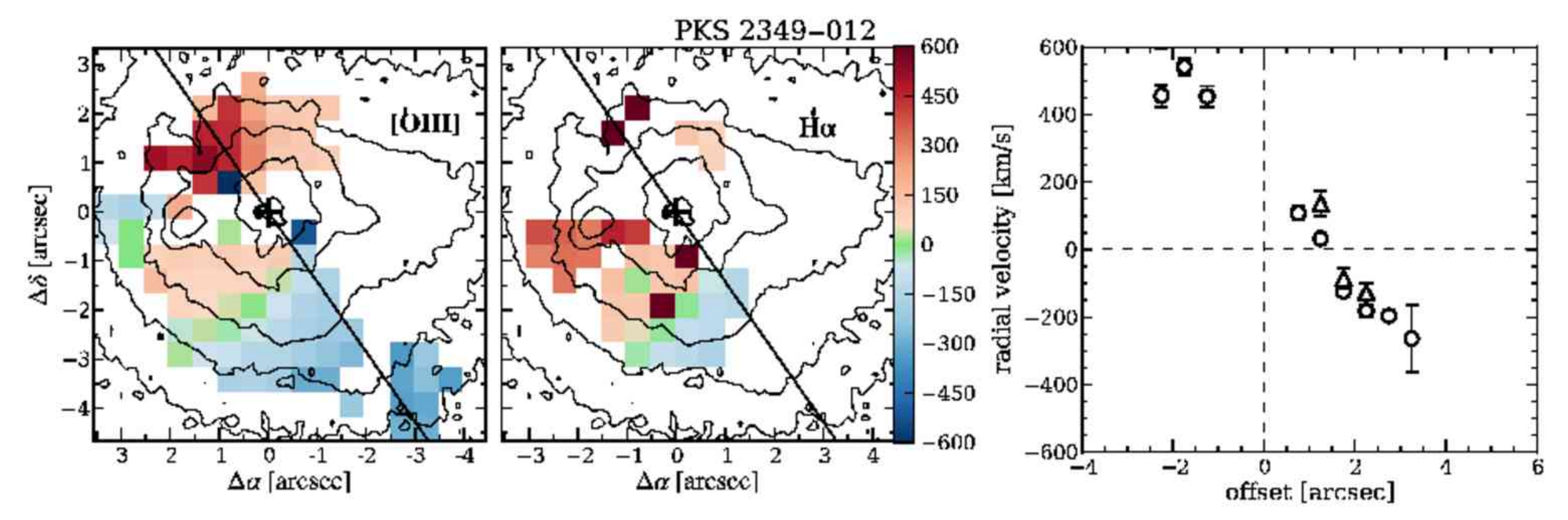}

\caption{Same as Fig.~\ref{fig_pmas:EELR_kin_rot} showing the velocity fields of the 3 QSO hosts morphologically classified as ongoing major mergers.}
\label{fig_pmas:EELR_kin_merger}
\end{figure*}

\begin{figure*}
\includegraphics[width=0.5\textwidth,clip]{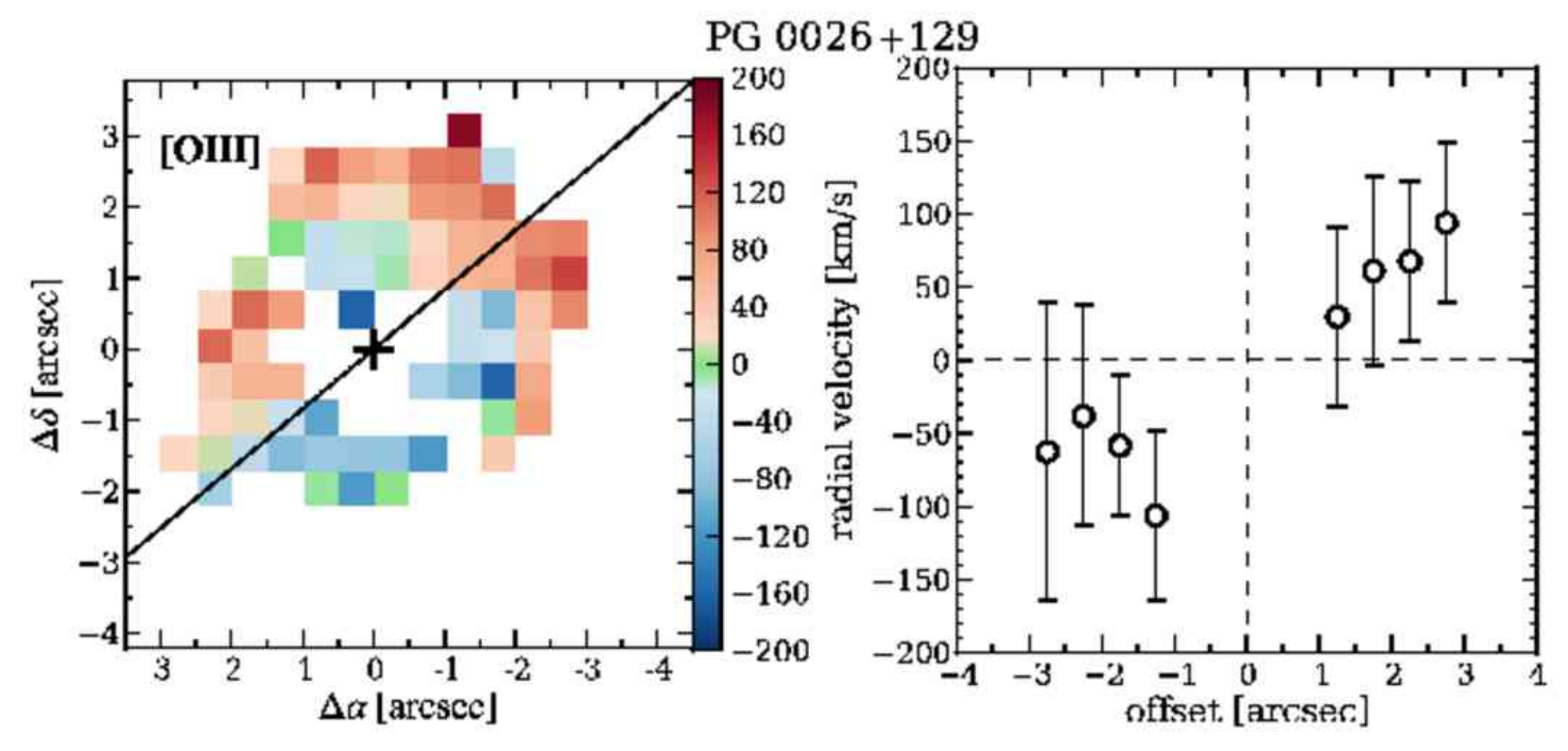}
\includegraphics[width=0.5\textwidth,clip]{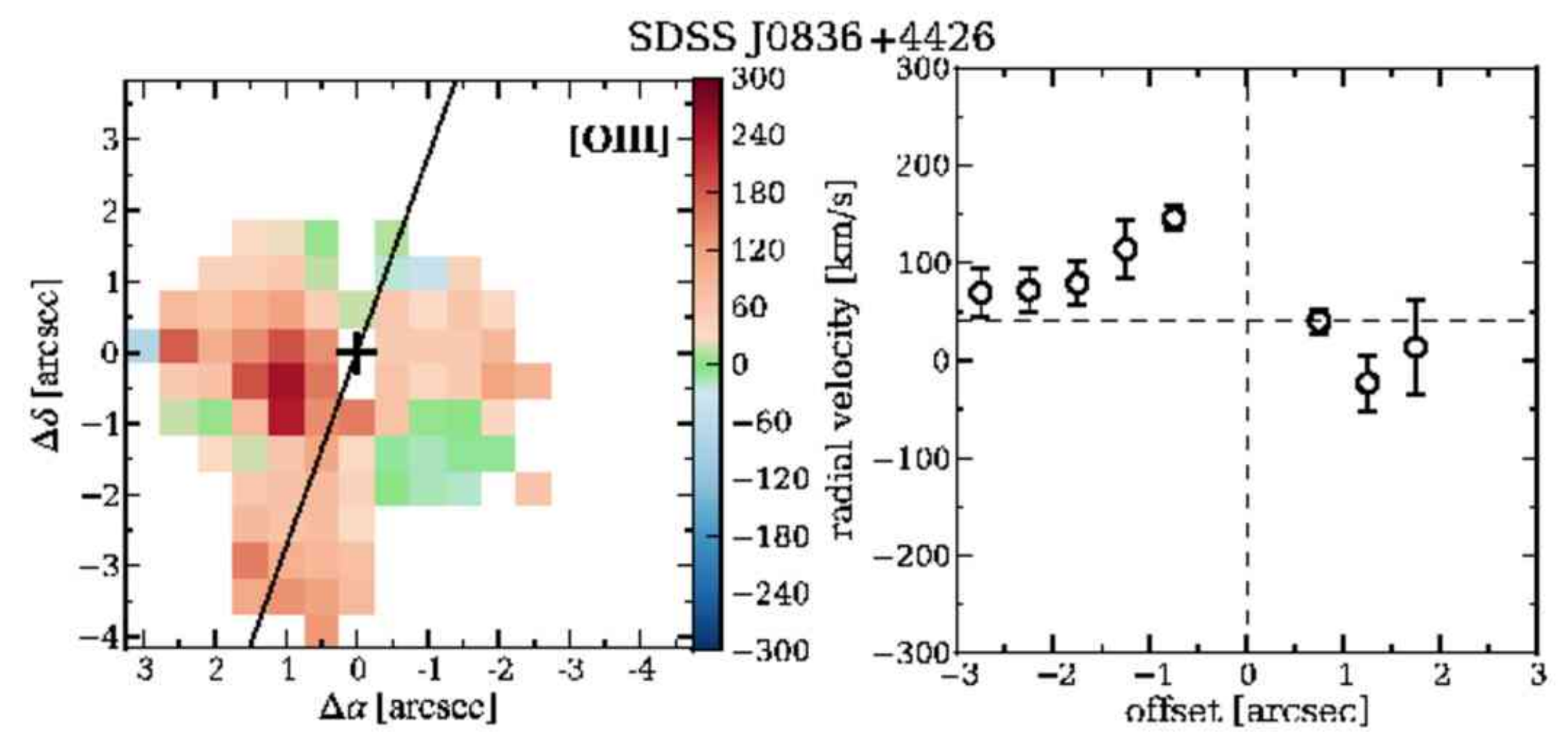}
\includegraphics[width=0.5\textwidth,clip]{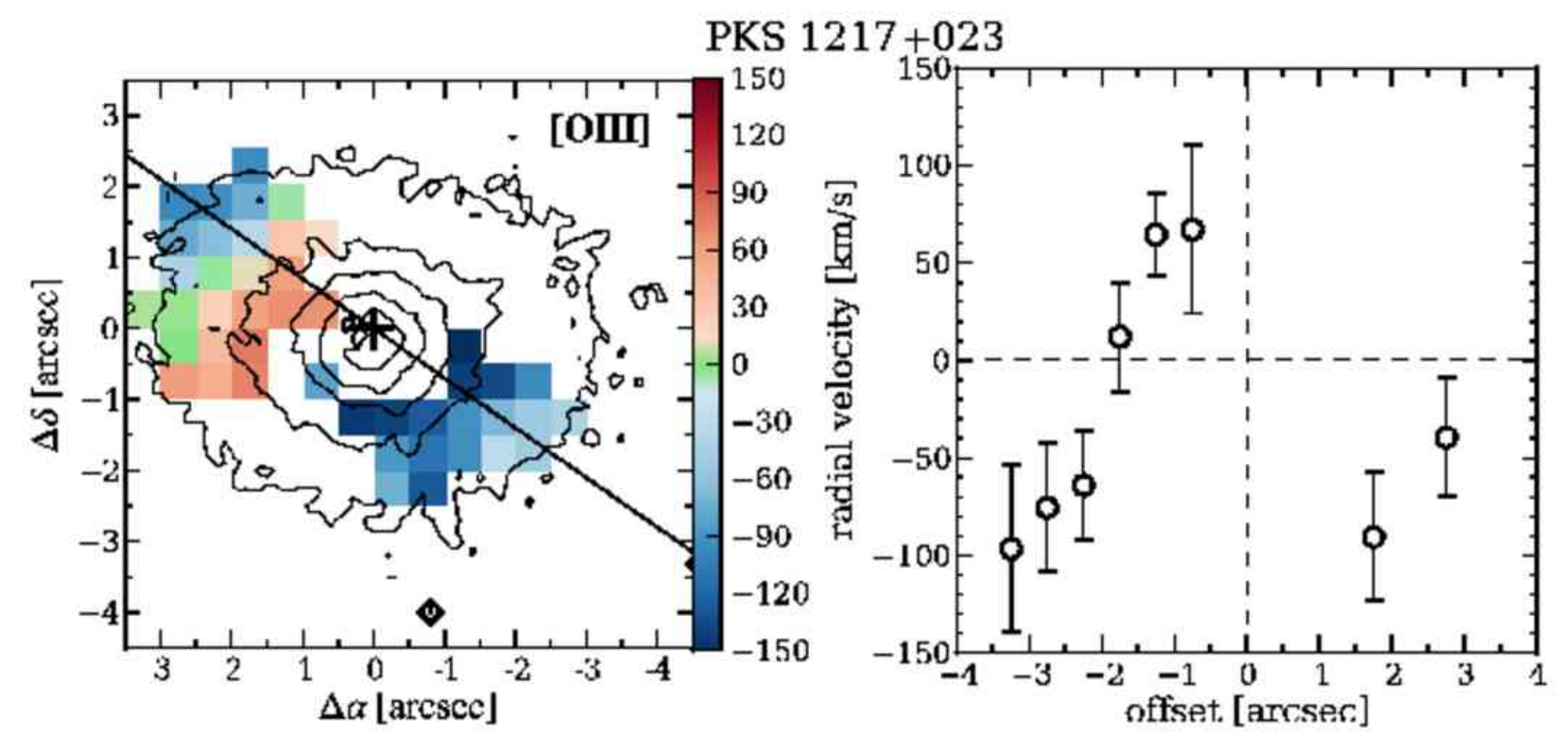}
\includegraphics[width=0.5\textwidth,clip]{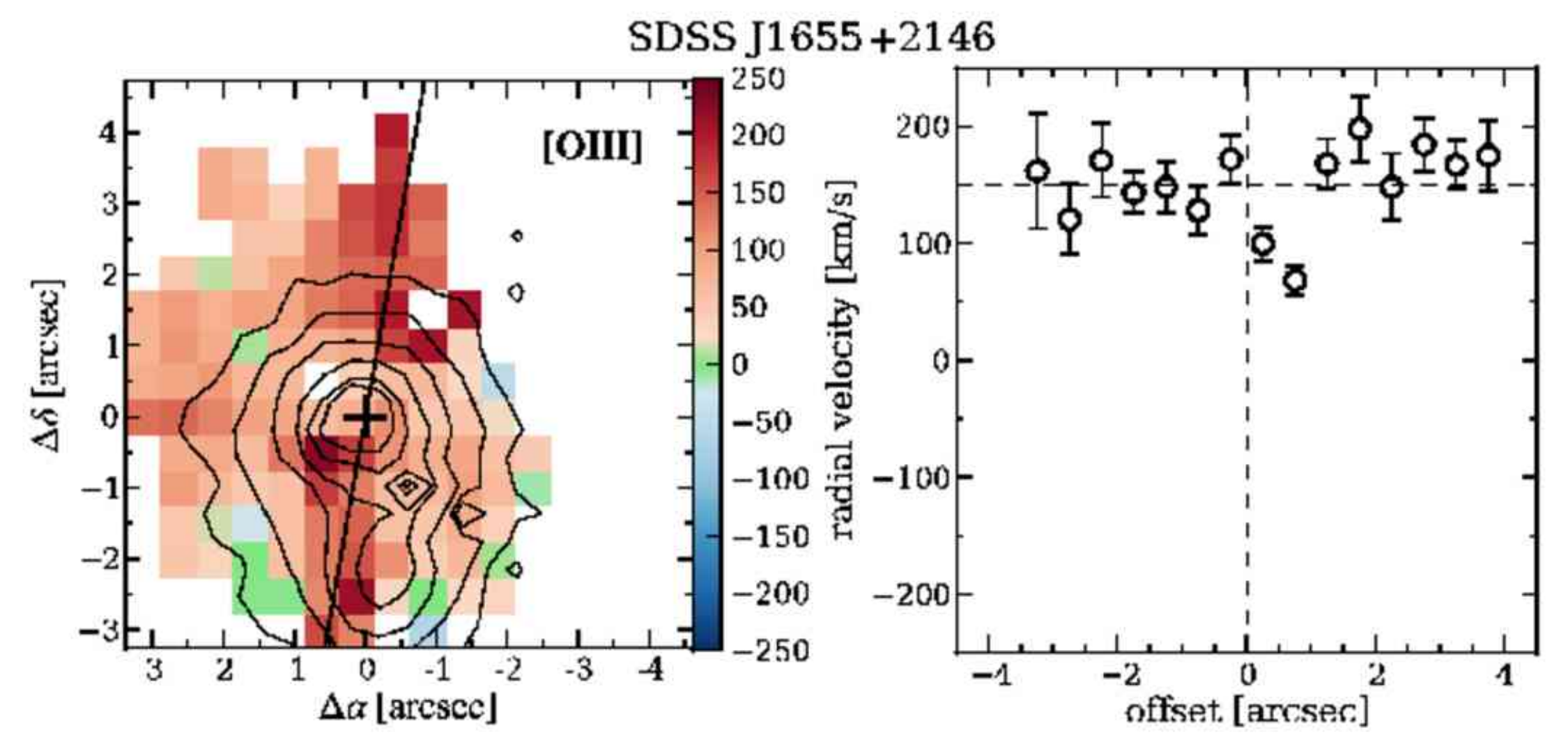}

\caption{Same as Fig.~\ref{fig_pmas:EELR_kin_rot}, presenting the miscellaneous velocity fields of the remaining objects.}
\label{fig_pmas:EELR_kin_pec}
\end{figure*}

We used the narrow \mbox{[\ion{O}{iii}]}\ line in the QSO spectrum to estimate the systemic redshifts of the QSO host galaxies. Previous studies showed that the spatially unresolved \mbox{[\ion{O}{iii}]}\ line is often systematically blueshifted in QSOs \citep[e.g.][]{Eracleous:2004,Boroson:2005,Komossa:2007,Villar-Martin:2011b}, which has been interpreted as a signature for circumnuclear outflows. Whether such outflows commonly extend out to kpc scales is one of the issues we want to address. We present the 2D velocity fields in three categories,  (i) apparently large scale rotational motions (Fig.~\ref{fig_pmas:EELR_kin_rot}), (ii) velocity fields of morphologically identified major mergers (Fig.~\ref{fig_pmas:EELR_kin_merger}), and (iii) miscellaneous velocity fields (Fig.~\ref{fig_pmas:EELR_kin_pec}), which we discuss separately in the following.

\subsubsection{EELRs dominated by rotational motion}
We find signatures of large scale rotational motion around I~Zw~1, PG~0052$+$251, PKS~1545$+$21, PG~1612$+$261, HE~2158$-$0107 and HE~2353$-$0420. The corresponding gas velocity fields are shown in Fig.~\ref{fig_pmas:EELR_kin_rot}. Because the turn-over point in the velocity curve is not reached in most cases, we modelled the synthetic longslit curves with a straight line ($v_r = \Delta v_r\cdot r+v_0$) to search for the major kinematic axis with the steepest velocity gradient $\Delta v_r$ among all $-90\degr<\mathrm{PA}<90\degr$ in steps of 3\degr. The velocity curves along this kinematic major axis are also shown in Fig.~\ref{fig_pmas:EELR_kin_rot} and the corresponding parameters are given in Table~\ref{tab_pmas:kin_axis_rot}. The majority of these systems (4/6) are morphological disc-dominated galaxies for which the kinematic major axis is well aligned with the photometric major axis within $\pm14\degr$ (rms). Thus, we suggest that the ionised gas is following the gravitational motion of the stellar disc in these cases.  

To verify the reliability of our velocity fields, considering that the bright QSO has been subtracted before, we compare our ionised gas kinematics of I~Zw~1 with ${}^{12}$CO(1-0) data on scales of 1\arcsec--14\arcsec\  \citep{Schinnerer:1998}. A radial \mbox{H$\alpha$}\ velocity of $v_r=166\pm22\,\mathrm{km}\,\mathrm{s}^{-1}$ at a distance of 2\arcsec\ is in good agreement with a velocity of $180\,\mathrm{km}\,\mathrm{s}^{-1}$ at 1.6\arcsec\ reported by \citet{Schinnerer:1998} from the CO kinematics. Such a comparison between CO and ionised gas kinematics is only possible for I~Zw~1, because  CO observations with the required spatial resolution and depth do not yet exist for other QSOs in our sample.

\begin{table}\centering
\begin{tiny}
\caption{Kinematic properties of rotational velocity fields.}
\label{tab_pmas:kin_axis_rot}
\begin{tabular}{lcccccc}\hline\hline\noalign{\smallskip}
Name & PA\tablefootmark{a} & $\Delta v_\mathrm{r}$\tablefootmark{b} & $v_0$\tablefootmark{c} & R\tablefootmark{d} & $v_\mathrm{r}(R)$\tablefootmark{e}  \\
     & [\degr] & $[\frac{\mathrm{km}}{\mathrm{s}}]$ & $[\frac{\mathrm{km}}{ \mathrm{s}}]$ & [\arcsec/kpc] & $[\frac{\mathrm{km}}{\mathrm{s}}]$\\\noalign{\smallskip}\hline\noalign{\smallskip}
I Zw 1 & $-28$ & $88$ & $704$ & $3/3.5$ & $166$\\
PG 0052$+$251 & $2$ & $94$ & $134$ & $3/8.1$ & $284$\\
PKS 1545$+$210 & $-46$ & $165$ & $3$ & $2/8.2$ & $299$ \\
PG 1612$+$261 & $6$ & $57$ & $39$ & $3/7.0$ & $119$\\
HE 2158$-$0107 & $-49$ & $17$ & $5$ & $2/6.9$ & $49$ \\
HE 2353$-$0420 & $20$ & $70$ & $-25$ & $2/7.3$ & $138$\\
\noalign{\smallskip}
\hline
\end{tabular}
\tablefoot{
\tablefoottext{a}{Position angle along the highest velocity gradient.}
\tablefoottext{b}{Radial velocity gradient along the major kinematic axis per arcseconds.}
\tablefoottext{c}{Radial velocity zero point that the QSO position with respect to the rest-frame defined by the nuclear  \mbox{[\ion{O}{iii}]}\ line.}
\tablefoottext{d}{Characteristic radius from the centre used for a dynamical mass estimate}
\tablefoottext{e}{Radial velocity at the radius given given in (d).}
}

\end{tiny}
\end{table}

PKS~1545+210 and possibly also HE~2158$-$0107 are bulge-dominated host galaxies for which the inclinations of the putative rotating gas discs are unknown. It is unlikely that the gas kinematics are caused by AGN-driven bipolar outflows, because the radio jet of PKS~1545+210 is oriented perpendicular to the kinematic major axis (see section~\ref{pmas_sect:jet-cloud}), and HE~2158$-$0107 is a RQQ with a low radio luminosity.  Interestingly, an apparent companion galaxy is located $\sim$3\arcsec\ to the West of PKS~1545+210 in the direction of the kinematic major axis and HE~2158$-$0107 features an extended tail of gas . Both case are suggestive for the accretion of gas from companions or the environment itself \citep[see the case of HE~2158$-$0107 in][]{Husemann:2011}, which will govern the rotation axis of the gas.

A direct comparison of the \mbox{H$\alpha$}\ and \mbox{[\ion{O}{iii}]}\ kinematics is only possible for PG~0052+251. We find that the \mbox{[\ion{O}{iii}]}\ kinematics does not match the clear rotational \mbox{H$\alpha$}\ velocity field with an offset of up to $\sim$100\,$\mathrm{km}\,\mathrm{s}^{-1}$. This kinematic separation supports our interpretation of the emission-line ratios in section~\ref{pmas_sect:BPT_diagnostic}, in which the \mbox{H$\alpha$}\ emission originates dominantly from \mbox{\ion{H}{ii}}\ regions in the discs, whereas the \mbox{[\ion{O}{iii}]}\ emission more often represents AGN ionisation cones. 

\subsubsection{Velocity fields of interacting QSO host galaxies}

In Fig.~\ref{fig_pmas:EELR_kin_merger} we present the gas velocity fields of Mrk~1014, SDSS J1131$+$2632 and PKS 2349$-$012, which we identified to be ongoing major mergers based on their optical continuum images. Because of the complex mass distribution and tidal forces, we do not necessarily expect a clear rotational velocity pattern in these systems.

Our PMAS observation of Mrk~1014 covers only the inner 10\,kpc part of the QSO host (cf. Fig.~\ref{fig_pmas:EELR_det}), so that we cannot trace the bulk motion of the ionised gas across the whole galaxy. The \mbox{[\ion{O}{iii}]}\ velocity curve along the EELR major axis appears to be peculiar. The symmetric curve has two sharp peaks in the radial velocity at $\pm2\arcsec$ distance from the nucleus at which the radial motion changes from receding to approaching and vice versa. These features are associated with a region of strong jet-cloud interactions as discussed below in section~\ref{pmas_sect:jet-cloud}, likely leading to non-relaxed bulk motion of the gas.

The EELR of SDSS~J1131+2632 is very elongated, and we can only study the kinematics in a part of the galaxy similar to Mrk~1014. The most striking feature is a giant shell or arc of ionised gas, 4\arcsec\ South-East of the nucleus, offset by $352\pm33\,\mathrm{km}\,\mathrm{s}^{-1}$ from the systemic velocity in the rest-frame. This is a unique feature among all the QSO in our sample and does not correspond to any continuum substructure in the broad-band image of the complex merging system. One possibility is that the gas belongs to a faint companion galaxy illuminated by the AGN radiation. 

A strong velocity gradient is detected in the ionised gas of PKS~2349$-$014. The main axis of this gradient follows the large curved wisps that were identified in the HST images by \citet{Bahcall:1995}. Thus, the gas motion is clearly driven by gravitation. However, the radio-jet axis of this RLQ is oriented North-South, and the radio jet seems to kinematically disturb the ionised gas at the interception point of jet and ENLR.

\subsubsection{Miscellaneous velocity fields}
The radial velocity fields of the remaining four QSO hosts, PG~0026$+$129, SDSS~J0836$+$4426, PKS 1217$+$023 and SDSS~J1655$+$2146,  are either asymmetric or lack a significant velocity gradient across the galaxy. All four host galaxies appear to be bulge-dominated, in which the gas could be kinematically hot and slowly rotating. However, the SDSS images of J1655$+$2146 and J0836$+$4426 do not rule out a disc component, which would need to be studied with higher spatial resolution imaging. We do not find any strong evidence for extreme kinematic features that would indicate non-gravitational motion.

\subsection{Evidence for jet-cloud interactions on kpc scales}\label{pmas_sect:jet-cloud}
We already discussed a close connection between the ENLR and the presence and morphology of radio jets emerging from the nucleus in Section~\ref{sect_pmas:ENLR_prop}. To follow up on this we now address the question whether the kinematics of the ionised gas is also significantly influenced by the jets.
\begin{figure*}
 \includegraphics[width=\textwidth,clip]{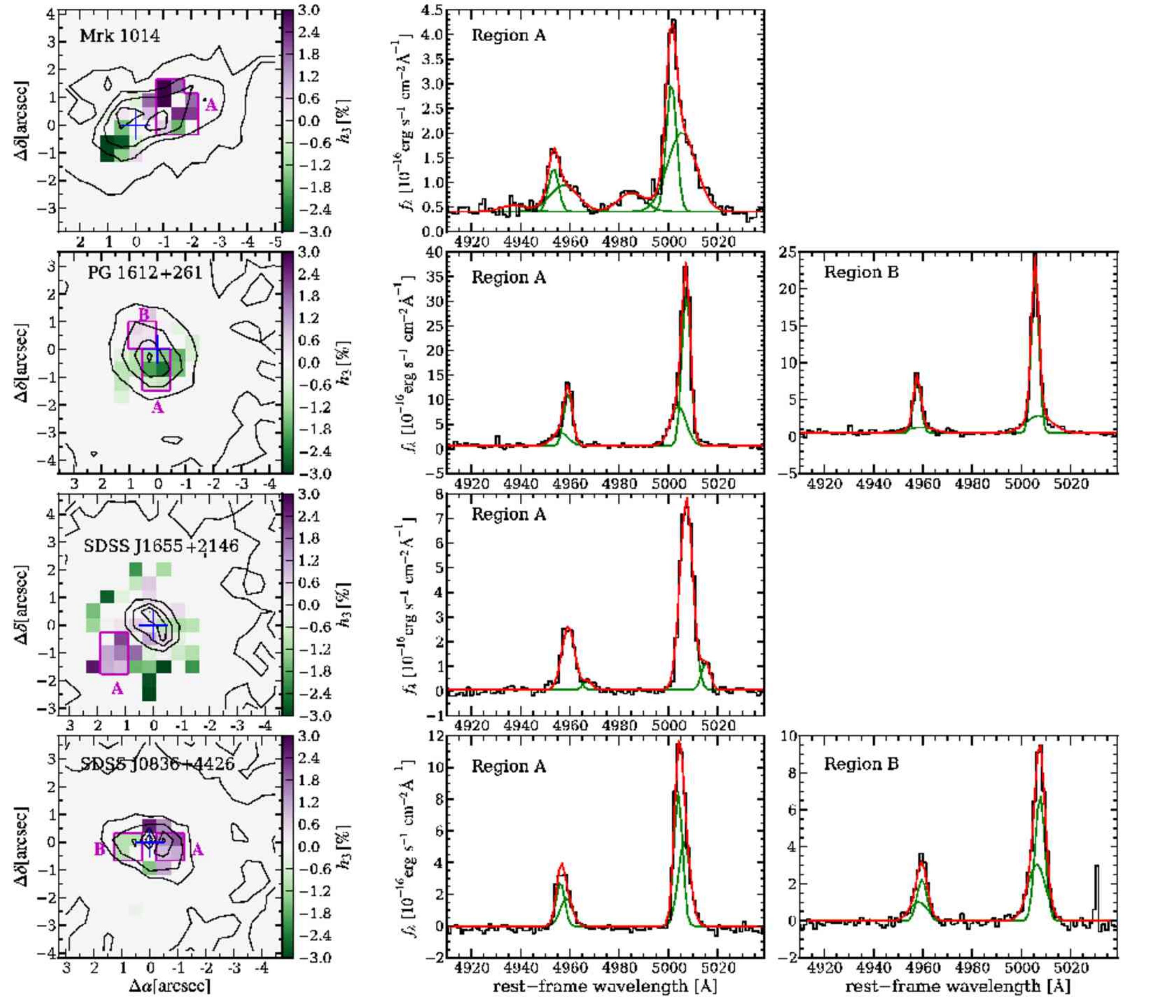}
\caption{\emph{Left panel:} Strength of the third Gauss-Hermite coefficient $h_3$ representing an asymmetry in the line profile found to be significant only in 4 galaxies, i.e. Mrk~1014, PG~1612$+$261, SDSS~J1655$+$2146, and SDSS~J0836$+$4426. The black contours represent the \mbox{[\ion{O}{iii}]}\ surface brightness distribution for comparison and the blue cross marks the QSO position for reference. \emph{Right panel:} Modelling of \mbox{[\ion{O}{iii}]}\,$\lambda 4960,5007$ doublet line in the co-added spectra (of the spaxels in the magenta box on the left) with two Gaussian components. The red line is the model to the measured spectrum shown in black, and the green line represents the individual Gaussian components.}
\label{fig_pmas:EELR_outflows}

\end{figure*}

We detect significantly asymmetric \mbox{[\ion{O}{iii}]}\ line profiles close to the nucleus of Mrk~1014 and PG~1612$+$261 right at the location of their known radio hot spots (Fig.~\ref{fig_pmas:EELR_outflows}). The line width and radial velocity offset are $508\pm84\,\mathrm{km}\,\mathrm{s}^{-1}$ and $-1012\pm48\,\mathrm{km}\,\mathrm{s}^{-1}$ for Mrk~1014 and $424\pm24\,\mathrm{km}\,\mathrm{s}^{-1}$ (FWHM) and $-220\pm20\,\mathrm{km}\,\mathrm{s}^{-1}$ for PG~1612$+$261 at region A, respectively, indicating clearly non-gravitational motions. In addition to the blue-shifted component in Mrk~1014, a broad redshifted component is blended with the \mbox{[\ion{O}{iii}]}\ narrow component. The extended ionised gas kinematics of Mrk~1014 was already studied in detail by means of long-slit \citep{Leipski:2006a} and IFU spectroscopy \citep{Fu:2009}. Our data for this object are fully supporting their results. In addition, SDSS~J0836$+$4426 and SDSS~J1655+2146 similarly show asymmetric \mbox{[\ion{O}{iii}]}\ line profiles close to the nucleus, which we separated into two different components. It is possible that jet-cloud interactions play a significant role in producing these complex kinematic feature as well given their significant integrated radio luminosity of $2\times 10^{24}\,\mathrm{W}\,\mathrm{Hz}^{-1}$ and $4\times 10^{23}\,\mathrm{W}\,\mathrm{Hz}^{-1}$, respectively. Yet, no interferometric radio images are available for these two objects that would be needed to confirm a clear association between radio emission and line asymmetries for a jet-cloud interaction scenario.
 
The EELR of PG~1700$+$518 is almost coincident with the south-east radio hot spot, considering the low spatial resolution of the optical compared to the radio observation. It is tempting to speculate that the radio-jet is  accelerating the gas to  the observed velocities ($\sim800\,\mathrm{km}\mathrm{s}^{-1}$). On the other hand, the EELR may also belong to a companion galaxy or a starburst region as part of the merger, because a knot in the H-band adaptive optics image was identified by \citet{Marquez:2003}. The brightest region of the EELR around the RQQ PG~2130$+$099 is matching with the southern hot spot of the triple radio source. However, we could not find any kinematic impact on the gas as in the other cases.

The \mbox{[\ion{O}{iii}]}\ kinematics of the major merger host of the RLQ PKS 2349$-$012 are mainly quiescent. However, we find redshifted \mbox{[\ion{O}{iii}]}\ emission by $400\pm70\,\mathrm{km}\,\mathrm{s}^{-1}$  with an intrinsic line width of $\sim$700\,$\mathrm{km}\,\mathrm{s}^{-1}$ (FWHM) at the interception point with the radio jet axis (2\arcsec\ North of the nucleus). This highlights the impact of the radio jet on the kinematics at the interception point, but not in the surrounding regions.

The RLQs PKS~1217+023 and PKS~1545+210 reside both in bulge-dominated galaxies. The jet axis of PKS~1217+023 is aligned with the EELR major axis  along which the velocity curve does clearly not agree with rotational pattern. Whether the gas is directly interacting with the jet remains uncertain because of the narrow line width and radial velocity $<$200\,$\mathrm{km}\,\mathrm{s}^{-1}$. An interaction of the jet with the EELR around PKS~1545+210 is most unlikely because the radio jet is aligned with the rotation axis of the gas disc. Such rotating gas discs are common in low-redshift radio galaxies \citep[e.g.][]{Kotanyi:1979,Baum:1992} with Centaurus A being the most prominent example. 

\begin{figure}
\resizebox{\hsize}{!}{\includegraphics[clip]{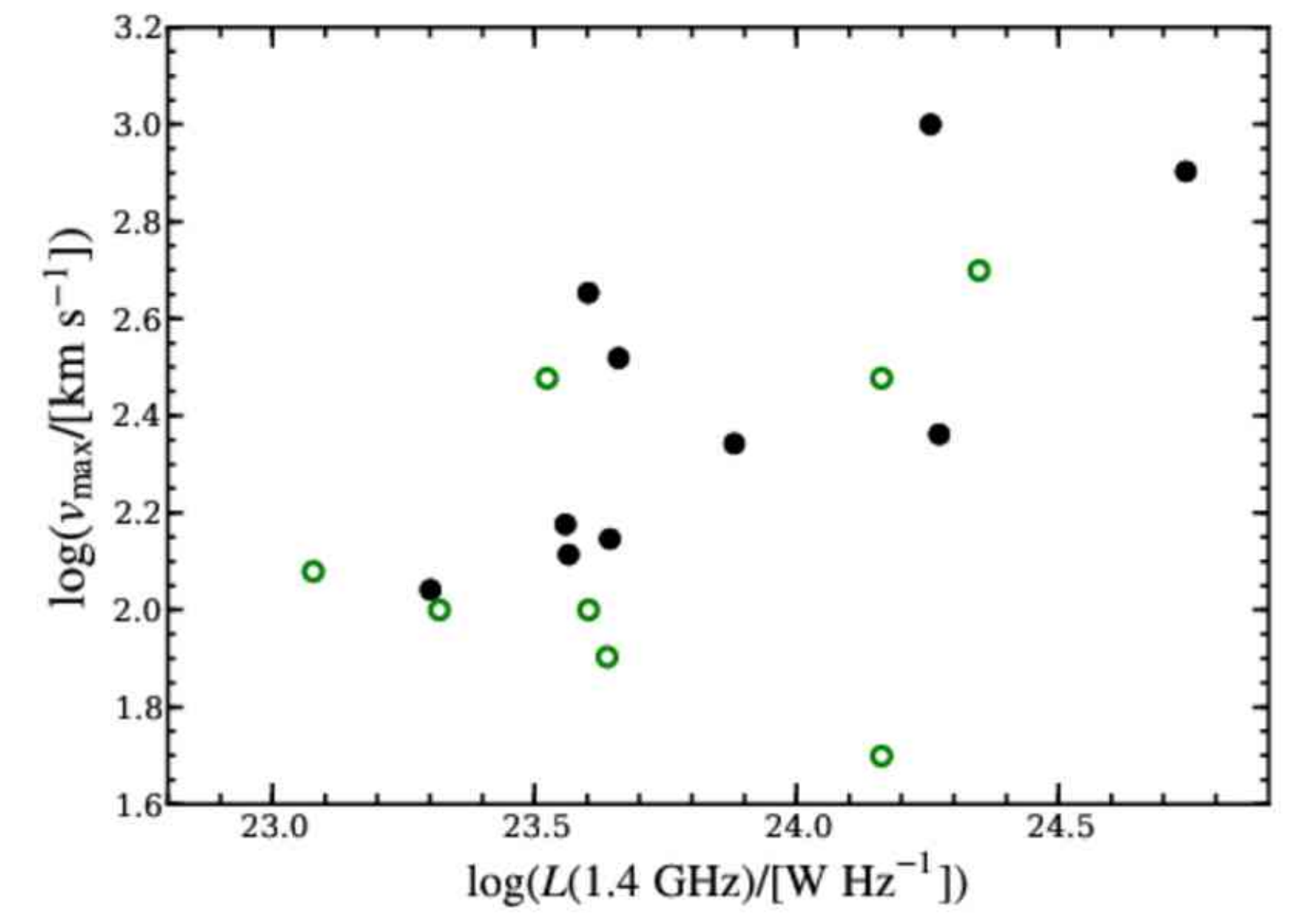}}
\caption{Extended maximum [OIII] radial velocity in the ENLR of RQQs as a function of radio luminosity at 1.4\,GHz. Only objects with detected radio emission are included in the diagram. Data from our IFU study of RQQs are indicated by the black filled symbols and open green symbols are from the long-slit data of type 2 QSOs \citep{Greene:2011}.}
\label{fig_pmas:EELR_radio_vmax}
\end{figure}

To summarize these individual results, in Fig.~\ref{fig_pmas:EELR_radio_vmax} we compare the maximum radial \mbox{[\ion{O}{iii}]}\ EELR velocity offset with respect to the zero-point of the velocity curves against the radio luminosity for all RQQs in our sample and that of \citet{Greene:2011}. A tentative trend is emerging that the radial velocity of the highly ionised gas on kpc scales increases with the radio luminosity, i.e. jet power, even in RQQs.  Despite some contamination with purely rotational motions at low radio luminosity and the unknown geometry and inclination of the gas motion, this supports the notion of a the systematic influence of radio jets on the ionised gas kinematics on host galaxy scales. A quite similar trend was already reported by \citet{Mueller-Sanchez:2011} for sub-kpc scales in lower luminosity Seyfert galaxies based on near-infrared IFS. Nevertheless, our inferred radial velocities are in the majority of cases below the escape velocity, which is of the order of  500--1000\,km\,$\mathrm{s}^{-1}$ \citep[see][]{Greene:2011}. Such a trend with increasing radio luminosity may be expected as the jets will carry mechanical energy and momentum of which a certain fraction will certainly be dissipated in the surrounding gas. Radio-loud AGN are therefore known to exhibit jet-cloud interactions that accelerate the gas to significant velocities \citep[e.g.][]{Baldwin:1980,Holt:2003,Sanchez:2004a,Morganti:2005,Christensen:2006,Nesvadba:2006}. On the other hand, detailed studies of some RQQs \citep[e.g. SDSS J1356+1026,][]{Greene:2012} and ULIRGs \citep[e.g. Mrk 231,][]{Rupke:2011} have shown that a radio jet alone is energetically insufficient to power high-velocity outflows present in these objects. The lack of high velocity ionised gas clouds in the majority of luminous RQQs is not necessarily in contrast with those observations. It could be that radio jets dominate the cloud acceleration even energetically in the majority of cases, but with some particular exceptions. An alternative explanation could be that radio jets significantly increase the effective energy dissipation of the QSO radiation into the ambient medium. We mentioned already in Sect.~\ref{sect:radio_prop} the possibility that a jet might decrease the clumpiness of the surrounding medium so the QSO radiation can act more strongly as proposed by \citet{Hopkins:2010} in their two-stage AGN feedback model. 

\subsection{The incidence of outflows among RQQs: No evidence for AGN feedback?}
AGN feedback has generally become an important ingredient in cosmological simulations of galaxy formation to efficiently stall star formation on short time scales. In these models either a significant amount of gas is expelled from the host galaxy by AGN outflows \citep[e.g.][]{Sanders:1988b, Hopkins:2005} or the gas in the galaxy halo is episodically heated by radio jets to prevent further condensation of cold gas \citep[e.g.][]{Croton:2006,Bower:2006,Best:2006}. Clear observational evidence for the latter process are the intra-cluster X-ray cavities driven by radio jets of the brightest cluster galaxies \citep[see][and references therein]{Cattaneo:2009, Fabian:2012}. On the other hand, observational signatures for AGN driven outflows are still scarce, and their interpretations are a subject of debate.  

Galactic superwinds with outflow velocities of up to 1000\,$\mathrm{km}\,\mathrm{s}^{-1}$ in the ionised gas have been detected in Ultra Luminous Infrared Galaxies (ULIRGs) \citep[e.g.][]{Heckman:2000,Rupke:2005b,Rupke:2011,Harrison:2012, Westmoquette:2012} and massive post-starburst galaxies \citep{Tremonti:2007}. These extraordinary outflows may be driven by the radiation of the central AGN with a comparably low contribution from the starburst. Whether AGN-driven outflows are solely responsible for such outflow velocities is however questioned by recent observation of post-starburst galaxies that display high-velocity outflows without clear evidence for AGN \citep{Diamond-Stanic:2012}. This is consistent with a comparison study of the radial gas velocity in X-ray selected AGN hosts and post-starburst galaxies by \citet{Coil:2011}. They concluded that while typical outflow velocities of $200\,\mathrm{km}\,\mathrm{s}^{-1}$ are seen in both samples, the winds are likely driven by vigorous star formation and not by AGN. Furthermore, ionised gas with radial velocities exceeding more than $500\mathrm{km}\,\mathrm{s}^{-1}$ have been detected mainly around radio-loud AGN \citep[e.g.][]{Nesvadba:2006,Holt:2008,Fu:2009}, which can be attributed to the mechanical energy of jet.

In our sample of 31 low-redshift QSOs we find signatures of  outflowing ($\gtrsim400\,\mathrm{km}\,\mathrm{s}^{-1}$) ionised gas on kpc scales only in 3 objects. In all these cases we identify a radio jet that is most likely driving the outflow. Quiescent gas kinematics, or no ionised gas at all, are characteristic for the rest of the sample.  Similarly, \citet{Greene:2011} found extended quiescent kinematics in the ionised gas of all but one type 2 QSO, but they attempt to use the wings of the \mbox{[\ion{O}{iii}]}\ line to argue for the presence of some level of AGN feedback.  The low outflow detection rates agree with the result of \citet{Krug:2010}, who compared the radial velocity of the \ion{Na}{i}\,D absorption line in the spectra of infrared-faint (SFR$<$10$\,\mathrm{M_{\sun}}\,\mathrm{yr}^{-1}$) Seyfert galaxies at $z<0.05$ and reported an outflow detection rate of 6\% for Seyfert 1 and 18\% for Seyfert 2 galaxies, compared to more than 50\% for starburst galaxies.  

Our observations clearly show a lack of ionised gas outflows  in radio-quiet AGN at low redshift, which may challenge the incorporation of AGN feedback in some theoretical models. However, no ionised gas at all is found around almost half of the QSOs in our sample. These QSOs typically show weak \mbox{[\ion{O}{iii}]}\ emission in their nuclear spectra, which could be interpreted in two ways: (i) The majority of gas has already been expelled from the host galaxy \citep[e.g.][]{Netzer:2004}, (ii) the majority of gas within the host galaxy is not sufficiently ionised by the AGN, or (iii) the galaxies are intrinsically poor of gas. With the advent of new radio/sub-mm telescopes, like the Atacama Large Millimetre Array (ALMA), we have the opportunity to study the amount, distribution, and kinematics of the cold molecular gas  at a sufficient spatial and spectral resolution. Given that the molecular gas is the seed of star formation, it may be the more direct tracer for AGN feedback processes than the ionised gas phase. 

\section{Conclusions}
In this paper we presented a detailed analysis of the extended ionised gas around a large sample of 31 low-redshift QSOs observed with the PMAS integral field spectrograph on the 3.5\,m telescope at the Calar Alto observatory. This is currently the largest sample of low-redshift type 1 QSOs observed with integral field spectroscopy in the optical wavelength regime. In \citet{Husemann:2008} we already studied the presence of extended ionised gas in relation to the spectral properties of the QSO nucleus based on a subsample already observed at that time. With the full sample now in hand, we investigated the ionisation states, luminosities, sizes of the extended narrow line regions (ENLRs), and the kinematics of the ionised gas. In order to separate the unresolved QSO emission from extended host galaxy emission in the IFU data we introduced our new software tool \texttt{QDeblend${}^{\mathrm{3D}}$}. The main results of our analysis can be summarised as follows:

\begin{itemize}
\item We detect extended ionised gas around 19 of 31 QSOs (61\%), of which 13 are ionised mainly by the QSO (the ENLR), 3 are \mbox{\ion{H}{ii}}-like regions produced by ongoing star formation, and the remaining 3 objects show signatures of both types in different extended regions. \\

\item Previous ENLR sizes obtained from HST narrow-band images are underestimated by at least a factor of 2 because of the much shallower surface brightness limits. Nine of sixteen ENLRs in our sample are elongated which means that longslit observations will strongly depend on the slit orientation. In the worst case they will underestimate the size by up to a factor of 6  for individual object, and a factor of 2 in the mean.\\

\item The typical size of the ENLR is $r_{95}=10\pm3\,\mathrm{kpc}$ at a median \mbox{[\ion{O}{iii}]}\ luminosity $\log( L([\mbox{\ion{O}{iii}]})/[\mathrm{erg}\mathrm{s}^{-1}])= 42.7\pm0.15$ of our QSO sample. Recent studies of type 2 QSOs at comparable luminosity by \citet{Greene:2011} and \citet{Fu:2012} obtained very similar ENLR sizes compared to our type 1 QSOs. This implies that the ENLR size-luminosity relation for type 1 and type 2 AGN cannot be as different as implied by the results of \citet{Schmitt:2003b}. The reported discrepancy between their extrapolated ENLR sizes of luminous type 2 and type 1 QSOs is therefore obsolete.\\

\item We find that the QSO continuum luminosity correlates much better with the ENLR size than does the total \mbox{[\ion{O}{iii}]}\ luminosity, while the \mbox{[\ion{O}{iii}]}\ luminosity of the ENLR is totally uncorrelated with the QSO continuum luminosity. Thus, we argue that the \mbox{[\ion{O}{iii}]}\ luminosity is not a good proxy for the intrinsic luminosity of the QSO and that the amount of ionising photons does not determine the luminosity of an ENLR alone. Either the amount and distribution of gas is quite different in the host galaxies or other factors such as the impact of radio jets play a additional role. This is supported by a trend that the ENLR luminosity increases with radio luminosity. We therefore caution to use type 2 AGN to study  the size-luminosity relation given that the \mbox{[\ion{O}{iii}]}\ luminosity is a \emph{secondary} and less accurate proxy for the intrinsic luminosity of the nucleus \citep[e.g.][]{Punsly:2011}. Additionally, type 2 QSOs are often found to be radio-intermediate objects and have a much higher incidence of flat-spectrum radio source \citep{Lal:2010}, which might cause a certain selection effect for type 2 QSOs if the radio jets enhance the \mbox{[\ion{O}{iii}]}\ luminosity relative to the bolometric AGN luminosity.

\item Signatures of presumably rotational gas kinematics are identified in 6 host galaxies of which 4 are disc-dominated. Signatures of non-gravitational kinematics have been found in in 5 cases for specific regions within the EELRs among which 3 show radial velocity greater than $\gtrsim$400$\,\mathrm{km}\,\mathrm{s}^{-1}$. Given the coincidences with extended radio sources, we argue that jet-cloud interactions are the most likely origin for the disturbances in the kinematics in the majority of cases. However, the detection rate and radial velocities of the outflows are much lower than reported for starburst galaxies or radio-loud AGN. \\

\end{itemize}

One of the main results of our study is the low detection rate of gas outflows around low-redshift radio-quiet QSOs (RQQs), which is in agreement with similar results from type 2 QSOs and infrared faint Seyfert galaxies. Taken all together this is evidence that massive AGN-driven bulk outflows of gas are not ubiquitous around RQQs at low redshifts. Signatures of AGN feedback are mainly observed in radio-loud AGN which can expel the gas from the host galaxy or even drive large cavities in the intra-cluster medium. The lack of gas outflow signatures in RQQs may challenge some the theoretical AGN feedback scenarios in which the expulsion of gas is essential to stall  star formation on short time scales. One explanation could be that the QSOs  investigated here, while among the most luminous at these low redshifts, are still too faint compared to their more ultra-luminous high-redshift counterparts, to display the same strong signature of vigorous outflows in the ionised gas \citep{CanoDiaz:2012,Maiolino:2012}. But our RQQs may also be in late stages of their evolution, after most of the gas has either been heated or was already expelled from the host galaxies considering the presumed time delay between the peak of starburst and AGN activity of around a few 100\,Myr \citep{Li:2008b,Wild:2010}. On the other hand, the ENLR sizes, luminosities and kinematics provide evidence for the general importance of radio jets to increase the covering factor of the ambient gas for the ionising photons of the nucleus in a pre-stage of feedback. 
To understand AGN feedback processes in RQQs therefore still remains an open issue.

\begin{acknowledgements}
We thank the referee  Jenny Greene for a prompt report with many valuable comments that clearly improved the quality of the article. Additionally, we thank Vardha Nicola Bennert for kindly providing us their analysed HST narrow-band images, Christian Leipski for sending us their reduced radio maps, and Hai Fu for making their unpublished ENLR size measurements available to us. BH and LW acknowledge financial support by the DFG Priority Program 1177 ``Witnesses of Cosmic History: Formation and evolution of black holes, galaxies and their environment'', grant Wi 1369/22. KJ is funded through the DFG Emmy Noether-Program, grant JA 1114/3-1.

Part of the data used in the paper originate from  observation made with the NASA/ESA Hubble Space Telescope, and obtained from the Hubble Legacy Archive, which is a collaboration between the Space Telescope Science Institute (STScI/NASA), the Space Telescope European Coordinating Facility (ST-ECF/ESA) and the Canadian Astronomy Data Centre (CADC/NRC/CSA).

Funding for the SDSS and SDSS-II was provided by the Alfred P. Sloan Foundation, the Participating Institutions, the National Science Foundation, the U.S. Department of Energy, the National Aeronautics and Space Administration, the Japanese Monbukagakusho, the Max Planck Society, and the Higher Education Funding Council for England. The SDSS was managed by the Astrophysical Research Consortium for the Participating Institutions.

This research has made use of the NASA/IPAC Extragalactic Database (NED) which is operated by the Jet Propulsion Laboratory, California Institute of Technology, under contract with the National Aeronautics and Space Administration. 

For the preparation of this paper we have made used of the cosmology calculator ``CosmoCalc'' \citep{Wright:2006}
\end{acknowledgements}
\bibliographystyle{aa}
\bibliography{references}
\end{document}